\documentclass[
    paper=A4,
    twoside,
    BCOR=15mm,
    DIV=14,
    headsepline,
    parskip=half,
    titlepage,
    openright,
    11pt,
    abstractoff,
    ngerman,
    english
]{scrbook}

\usepackage[T1]{fontenc}
\usepackage[utf8]{inputenc}
\usepackage[english]{babel}
\usepackage[stretch=20]{microtype}

\usepackage{marvosym}
\usepackage{lmodern}
\usepackage{textcomp,gensymb}
\usepackage{amsmath}
\usepackage{amssymb}
\usepackage{graphicx}
\usepackage{caption}
\usepackage{subfigure}
\usepackage{hyperref}
\usepackage{booktabs}
\usepackage[svgnames,table,hyperref]{xcolor}
\usepackage{placeins}
\usepackage{wasysym}
\usepackage{wrapfig}
\usepackage{stmaryrd}
\usepackage{pdfpages}
\usepackage{multirow}
\usepackage{comment}
\usepackage{nicefrac}
\usepackage{units}
\usepackage[numbers,square]{natbib}
\usepackage{xspace}

\addtokomafont{caption}{\small\itshape}
\addtokomafont{captionlabel}{\bfseries}
\addtokomafont{sectioning}{\rmfamily}

\DeclareGraphicsExtensions{.pdf,.png,.jpg}

\newcommand{\methane}{\ensuremath{\text{CH}_4}\xspace}
\newcommand{\carbondioxide}{\ensuremath{\text{CO}_2}\xspace}
\newcommand{\Sr}{\ensuremath{\text{Sr}^{90}}\xspace}
\newcommand{\Fe}{\ensuremath{\text{Fe}^{55}}\xspace}
\newcommand{\Ytt}{\ensuremath{\text{Y}^{90}}\xspace}

\newcommand{\uETp}{\ensuremath{\nicefrac{V}{\text{cm}}\nicefrac{K}{\text{mbar}}}\xspace}

\newcommand{\hptpc}{\mbox{\emph{HPTPC}}\xspace}
\newcommand{\hpgmc}{\mbox{\emph{HPGMC}}\xspace}
\newcommand{\magboltz}{\mbox{\texttt{\textsc{MagBoltz}}}\xspace}
\newcommand{\agros}{\mbox{\texttt{\textsc{Agros2D}}}\xspace}
\newcommand{\garfield}{\mbox{\texttt{\textsc{Garfield++}}}\xspace}
\newcommand{\ltspice}{\mbox{\texttt{LTSPice}}\xspace}
\newcommand{\scipy}{\mbox{\texttt{SciPy}}\xspace}

\newcommand{\Rpara}{{\ensuremath{\text{R}_{\text{para}}}}\xspace}
\newcommand{\Cpara}{{\ensuremath{\text{C}_{\text{para}}}}\xspace}
\newcommand{\TOhm}{\ensuremath{\text{T}\Ohm}\xspace}
\newcommand{\Ohm}{\ensuremath{\Omega}\xspace}

\newcommand{\ETp}{\ensuremath{\nicefrac{\text{ET}}{\text{p}}}\xspace}
\newcommand{\Ep}{\ensuremath{\nicefrac{\text{E}}{\text{p}}}\xspace}
\newcommand{\pE}{\ensuremath{\nicefrac{\text{p}}{\text{E}}}\xspace}
\newcommand{\vd}{\ensuremath{v_d}\xspace}
\newcommand{\alphap}{\ensuremath{\nicefrac{\alpha}{p}}\xspace}
\newcommand{\gammase}{\ensuremath{\gamma_{\text{se}}}\xspace}

\newcommand{\clabel}[1]{\texttt{#1}\xspace}  %
\newcommand{\HRule}{\rule{\linewidth}{0.5mm}}

\begin{document}
    \begin{titlepage}
        \center
        \textsc{\LARGE RWTH Aachen University}\\[1.5cm]
        \textsc{\Large Masterarbeit in Physik \\ angefertigt im III. Physikalischen Institut B}\\[0.5cm]

        \HRule \\[0.4cm]
        { \huge \bfseries Design of a Gas Monitoring Chamber for High Pressure Applications}
        \HRule \\[1.5cm]

        \begin{minipage}{0.4\textwidth}
            \begin{flushleft} \large
            \emph{von:}\\
            Philip \textsc{Hamacher-Baumann}
            \end{flushleft}
        \end{minipage}
        ~
        \begin{minipage}{0.4\textwidth}
            \begin{flushright} \large
            \emph{bei:} \\
            PD Dr. Stefan \textsc{Roth}\\
            \end{flushright}
        \end{minipage}\\[4cm]
        ~
        \begin{center}
            \large{
                vorgelegt der
                \\
            }
            \textsc{
                \large{
                    Fakult\"{a}t f\"{u}r Mathematik, Informatik und Naturwissenschaften \\
                }
            }
        \end{center}

        \vfill
        {\large September 2017}
    \end{titlepage}

    \tableofcontents
    \chapter{Introduction}
    Gaseous detectors have been used for decades by particle physics experiments at beamlines, as subdetectors or standalone experiments.
    Some can provide insight on the open question of the origin of the observed matter-antimatter asymmetry in the universe.
    One experiment, that searches for an answer to this question is the Tokai To Kamioka (T2K) experiment in Japan.
    This is done by searching for differences in matter-antimatter processes, so-called Charge-Parity violation (CPV), in the leptonic sector.
    Experience from its run time so far have shown, that a high pressure gaseous detector could benefit the active analyses.
    This work presents the construction of a high pressure gaseous detector and testing of its subsystems.

\section{The T2K Experiment}
    In T2K, a $\nu_\mu$ neutrino beam produced at Japan Proton Accelerator Research Complex (JPARC) is sent across Japan and measured at a near and far detector location.
    It is also possible to run in anti-neutrino ($\bar{\nu}_\mu$) mode.
    T2K measures parameters of neutrino oscillation and probes CPV by comparing $\nu_\mu$ and $\bar{\nu}_\mu$-run data.

\begin{figure}[!h]
        \centering
        \includegraphics[width=.7\linewidth]{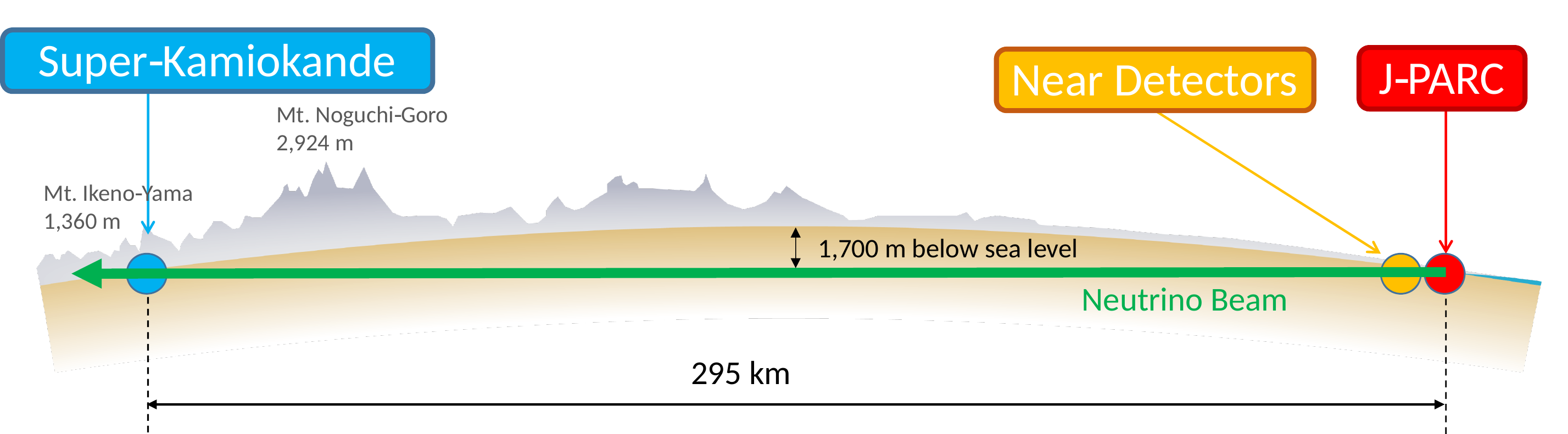}
        \caption[Cross-sectional view of the T2K neutrino beam]{
            The neutrino beam produced at JPARC travels through ND280 towards SK.
            On its path, it crosses Japan from east to west in \unit[295]{km}.
        }
        \label{fig:t2k_crossection}
    \end{figure}

\begin{figure}[!h]
        \centering
        \includegraphics[width=.7\linewidth]{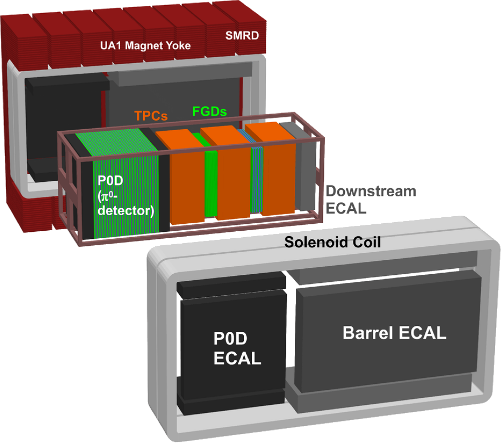}
        \caption[Opened ND280 with subdetectors visible]{
            ND280 contains three large TPCs that can provide measurements for low momentum cross-sections of protons.
        }
        \label{fig:ND280_explosion}
    \end{figure}

    Figure~\ref{fig:t2k_crossection} shows the full neutrino path from JPARC, through Near Detector 280 (ND280) and finally arriving at Super-Kamiokande (SK).
    SK is a water Cherenkov detector located in the Kamioka mine at the central west coast area of Japan.
    Part of ND280 are three large TPCs filled filled with $\text{Ar}(\unit[95]{\%})$, $\text{CF}_4(\unit[3]{\%})$ and $\text{iC}_4\text{H}_{10}(\unit[2]{\%})$, see figure~\ref{fig:ND280_explosion}.
    With them, it is already possible to extract cross-sections of neutrinos on argon.
    However, statistics are very low and the gas admixtures necessary for stable TPC operation are also included in the measurement.

\section{Motivation for High Pressure Detectors}
    Currently, one of the dominating systematic uncertainties of the T2K experiment's oscillation and CPV results is due to the uncertainties on $\nu$-cross-sections, especially at energies below \unit[250]{MeV}~\citep{Abe2017}.
    Future experiments like the Deep Underground Neutrino Experiment and Hyper-Kamiokande assume a significant reduction of these uncertainties in their physics goals.
    That is why a technology is needed, that provides increased interaction statistics with low detection thresholds to improve the understanding of processes in the low momentum region.
    High pressure gaseous detectors can fulfill these requirements.

\begin{figure}[h]
        \centering
        \includegraphics[width=.7\linewidth]{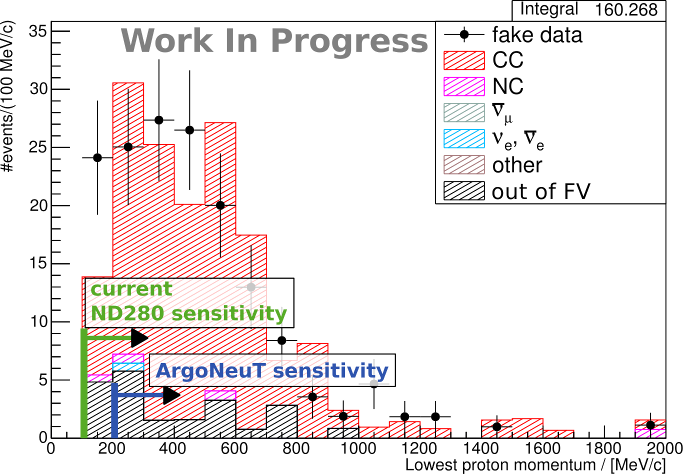}
        \caption[Comparison of NEUT and GENIE Monte Carlo predictions]{
            Momentum distribution of protons exiting an argon nucleus after neutrino interaction.
            Simulated with NEUT (histograms) and GENIE (fake data), Monte Carlo predictions differ especially in the low momentum region.
            Taken from~\citep{NuINT17:Koch}.
        }
        \label{fig:neut_genie_comp}
    \end{figure}

    The momentum threshold for a well reconstructed proton in a gaseous TPC is expected to be at \unit[60]{MeV}, compared to \unit[200]{MeV} for a liquid argon TPC~\citep{NuINT17:Koch}.
    This increased sensitivity range can be used to improve neutrino interaction generators, like NEUT and GENIE, see figure~\ref{fig:neut_genie_comp}.
    Additionally, increasing the number of target atoms by increasing internal detector pressure, also increases statistics compared to atmospheric pressure gaseous detectors.

\FloatBarrier
\section{Gaseous Detectors at Atmospheric Pressure}
    In the past, atmospheric pressure detectors have been very versatile and successful in high energy experiments.
    Some of their characteristics that contribute to their versatility are:
    \begin{itemize}
        \item Low material budget that causes only minimal energy loss of through-going particles.
        \item Gas amplification makes readout of small signals easy.
        \item Energy loss and, assuming a magnetic field, momentum reconstruction allow identification of particles.
        \item For Time Projection Chambers [TPC], a large volume can be covered with few read-out channels.
    \end{itemize}

\begin{figure}[h]
        \centering
        \includegraphics[width=.7\linewidth]{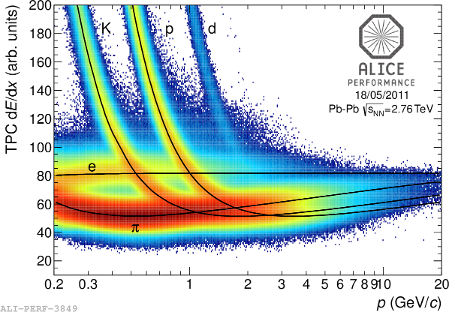}
        \caption[Particle identification with their energy loss and momentum]{
            The charged particle type can be extracted by comparing energy loss per unit length with the measured momentum.
            Taken from~\citep{Christiansen2013}.
        }
        \label{fig:ALICE_PID}
    \end{figure}

    Often, gaseous detectors in beamline experiments are of the TPC-type with a magnetic field for momentum measurements of particles.
    Energy loss can be calculated from the measured charge along the track.
    It is possible to extract the particle type from a $\nicefrac{\text{d}E}{\text{d}x}$-$p$-plot, see the example in figure~\ref{fig:ALICE_PID}.
    For some areas and high momentum, a separation is not possible with this strategy.

    High pressure gaseous detectors are not interesting for most non-neutrino experiments.
    Increasing the material budget is often not desirable, especially considering that a high pressure TPC would need a thick wall, e.g. made of steel.

\section{Principle of a Gas Monitoring Chamber}
    Track reconstruction in a TPC relies heavily on the precisely known drift velocity of ionization tracks inside the gas.
    Furthermore, particle identification is done by calculating the deposited energy in the gas from measurement of the gas-amplified charge of a track.
    Since both gain and drift velocity depend on environment parameters (i.e. $T$, $p$) and the precise gas composition, which all constantly vary, a system is needed that continuously monitors the gaseous detector.
    Such a system is called a gas monitoring chamber or GMC for short.

\begin{figure}[h]
        \centering
        \subfigure[Adapted from~\citep{Terhorst2008}]{
            \includegraphics[width=.4\linewidth]{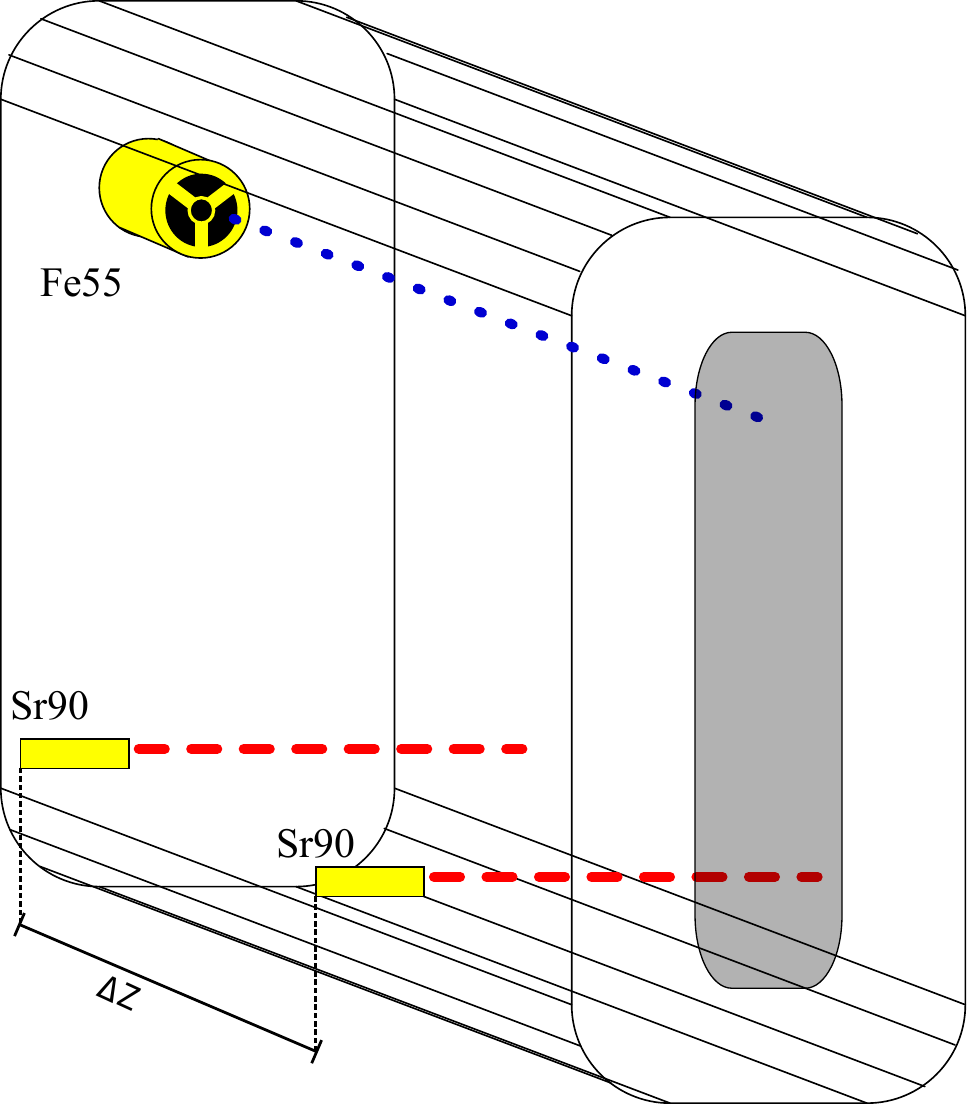}
            \label{fig:GMC_scetch}
        }%
        \subfigure[]{
            \includegraphics[width=.6\linewidth]{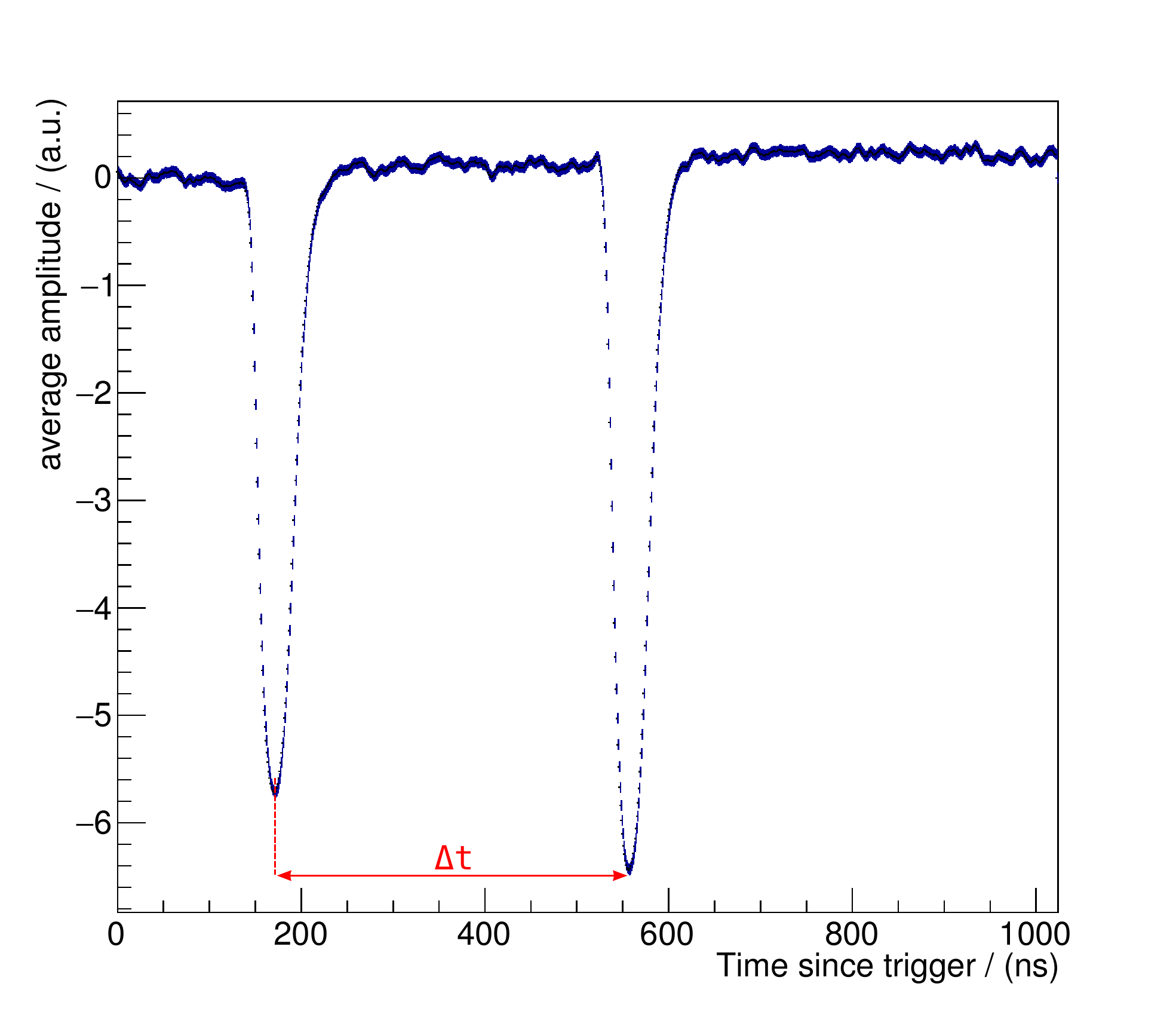}
            \label{fig:GMC_signal}
        }
        \caption[Conceptual view of a GMC]{
            (a)~A GMC can be used to continuously calibrate a running gaseous detector.
            Tracks created at known positions (\Sr) can be used to measure drift properties and known energy deposition (\Fe) for a gain calculation.
            (b)~Timing measurement between two track positions.
        }
    \end{figure}

    A GMC is similar to a conventional TPC, but in this case not the tracks are reconstructed using known gas properties but the gas properties are reconstructed using known track positions.
    In figure~\ref{fig:GMC_scetch} a sketch of such a system is shown.
    Collimated, radioactive sources at known positions, here \Sr, create tracks with distance $\Delta z$.
    Electrons from the decay of \Sr that traversed the gas can exit the chamber and start a time measurement.
    Their tracks left in the gas are drifted towards a readout plane using the same drift field as in the main detector.
    The arrival time difference between the two positions is used for the drift velocity measurement using
    \begin{align}
        \vd=\frac{\Delta z}{\Delta t}\,,
    \end{align}
    see \ref{fig:GMC_signal} for such a time measurement.
    This eliminates many of the present systematic uncertainties.
    For example, from the unknown field inhomogeneities at the edges, especially in the anode region with the readout plane and amplification voltage or the timing measurement.
    \Fe sources, that produce x-ray photons which are absorbed in the gas and produce an electron cloud with well defined charge content, are used for gain measurement.
    \Sr electrons are not always fully absorbed and are not mono-energetic, making a gain measurement difficult.

\FloatBarrier
    \chapter{Drifting Electrons in Gases}
\label{ch:vdtheo}
\section{First Ionization Event}
\begin{figure}[h]
        \centering
        \includegraphics[width=\linewidth]{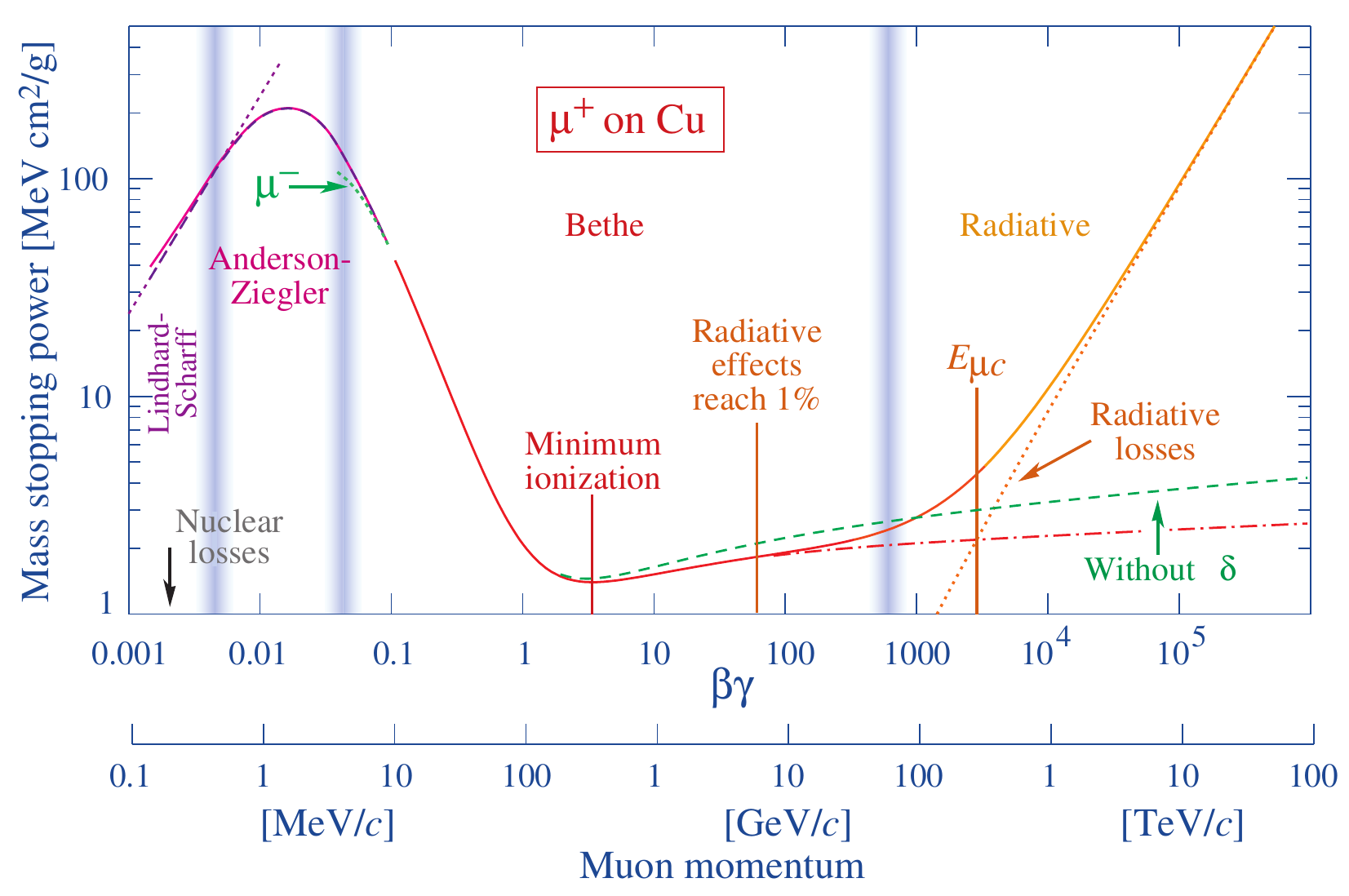}
        \caption[Energy loss of muons in matter]{
            Energy loss of muons per unit length in pure copper.
            The shape of this curve is very similar for other materials.
            Taken from~\citep[fig.~33.1]{PDGReview2016}.
        }
        \label{fig:dEdx_muCu}
    \end{figure}

    When charged, heavy\footnote{heavier than electrons} particles traverse any material, they lose energy by interaction.
    At intermediate energies, for muons $\mathcal{O}(\unit[1]{GeV})$, this loss is almost entirely due to interactions with electrons and their electric field inside the target material.
    How much energy is lost strongly depends on atomic properties of the target material and the kinematics of the incident particle.
    Figure~\ref{fig:dEdx_muCu} shows the energy loss per unit length in copper for anti-muons~\citep[fig.~33.1]{PDGReview2016}.
    The shape of the curve is very similar for most materials.

    Atomic properties that influence the energy loss rate are the maximum possible energy transfer $W_\text{max}$ per collision with a shell electron and their excitation energies.
    Those values are dependent on the atomic nucleus mass $A$ and proton number $Z$.
    For intermediate energies, the exact shell structure can be approximated with a single mean excitation energy $I$.
    Bethe does this in the formulation of the average energy loss per unit length:
    \begin{align}
        \left<-\frac{\text{d}E}{\text{d}x}\right>=Kz^2\frac{Z}{A}\frac{1}{\beta^2}\left[\frac{1}{2}\ln\left(\frac{2m_\text{e}c^2\beta^2\gamma^2W_{\text{max}}}{I^2}\right)-\beta^2-\frac{\delta(\beta\gamma)}{2}\right]\,.
        \label{eq:bethe_formula}
    \end{align}
    The ionizing field of the incident particle could also be that of a multiply charged particle, e.g. an $\alpha$ particle, so the incident particle's charge number $z$ also has to be taken into account.
    Electromagnetic fields are propagated with the speed of light, so for highly energetic particles approaching the speed of light, their ionizing fields are distorted.
    This is introduced through the term $\delta(\beta\gamma)$ in Bethe's formula.
    Remaining constants are absorbed into $K$.\citep[ch.~33]{PDGReview2016}

    A fraction of the deposited energy is used to ionize the material, so free electron-ion pairs are produced along the incident particle's trajectory.
    On a statistical basis, an average energy needed for the creation of a single electron-ion pair can be measured, which also includes secondary production mechanisms.
    For this, completely stopped particles from radioactive sources were used in~\citep[tab.~1.3]{Blum2008}.
    Table~\ref{tab:W_values} lists some prominent examples.
    Singly and doubly charged particles differ in energy loss according to equation~\ref{eq:bethe_formula}, so $\beta$ and $\alpha$ radiation are distinguished with $W_\beta$ and $W_\alpha$ respectively.

    \begin{table}[h]
        \centering
        \caption{
            Energy $W$ spent on average for the creation of an electron-ion pair~\citep[tab.~1.3]{Blum2008}
        }
        \begin{tabular}{lcclc}
            Gas & $W_\alpha$ (eV) & $W_\beta$ (eV) & Gas mixture & $W_\alpha$\\
            \hline
            \hline
            He              & 46.0 & 42.3 & Ar:$\text{CH}_4$ 97:3           & 26.0  \\
            Ar              & 26.4 & 26.3 & Ar:$\text{C}_3\text{H}_8$ 98:2  & 23.5  \\
            Xe              & 21.7 & 21.9 &                                 &       \\
            $\text{CH}_4$   & 29.1 & 27.1 &                                 &       \\
            \hline
        \end{tabular}
        \label{tab:W_values}
    \end{table}

    When going from pure gases to mixtures, more ionization processes become possible.
    In the most direct case, one electron is directly stripped from the atom by the incident particle $x^\mp$:
    \begin{equation*}
        x^\mp + A\rightarrow {x^\mp}'+e^-+A^+\,.
    \end{equation*}
    In some cases, the particle excites one atom or molecule instead of ionizing it, which then performs a radiation-less transfer to a second atom or molecule.
    If the ionization threshold of the receiver is lower than the transferred energy, ionization follows:
    \begin{align*}
        {x^\mp}+A &\rightarrow {x^\mp}'+A^*\\
        A^*+B &\rightarrow A+B^++e^-\,.
    \end{align*}
    This process is also performed in the strong fields at amplification regions.
    The probability with which $A^*$ transfers its surplus energy to $B$, is called the Penning transfer probability.
    A precise simulation is not possible, but an estimation can be computed.
    Simulation results have to be tuned with measurements to reach higher precisions.

\FloatBarrier
\section{Drift of Electrons in Gases}
    In gaseous detectors (e.g. TPCs), the electrons created along tracks are drifted towards the readout plane with an electric field.
    The velocity of this drift is mainly influenced by the electric field and gas properties.
    Magnetic fields also contribute to the drift velocity, but are not employed in GMCs in general.
    It is a reasonable assumption, that the final drift velocity in the gas is reached instantaneously.
    That is because the drift is a statistical effect of electrons moving much faster than the actual drift velocity.
    They are scattering almost isotropically on gas atoms or molecules~\citep[ch.~2.2]{Blum2008}.
    Unfortunately, a functional description is not available.
    Monte Carlo simulation tools such as \magboltz exist that can simulate electron drift velocity with sufficient precision for predefined parameters~\citep{Biagi1999}.

    Even though a full parametrization might not be known, the dependency the drift velocity curve follows can be derived rather simple.
    The drift velocity depends on the energy gained by electrons in the drift field $E_\text{drift}$ along the path $\lambda_\text{path}$
    \begin{align}
        \vd=v_d(E_\text{drift}\cdot\lambda_\text{path})\,,
    \end{align}
    where $\lambda$ is the mean free path between collisions in equilibrium.
    For a description of $\lambda$, the ideal gas law can be used; at least for lower, atmospheric pressures:
    \begin{align}
        \lambda_\text{path}\propto \frac{1}{n}=\frac{k_\text{B}T}{pV}\,.
    \end{align}
    Putting it all together, the drift velocity shows a functional dependency of
    \begin{align}
        \vd=\vd(\ETp)
    \end{align}
    for moderate pressures.
    This quick derivation completely ignores important influences such as energy dependent cross-sections or non-pure gas mixtures.
    Nevertheless, this dependency was experimentally found to be in good agreement with data and simulations done in~\citep[ch.~4.3]{Wrobel2011}.

\begin{figure}[hb]
        \centering
        \includegraphics[width=.7\linewidth, angle=-90]{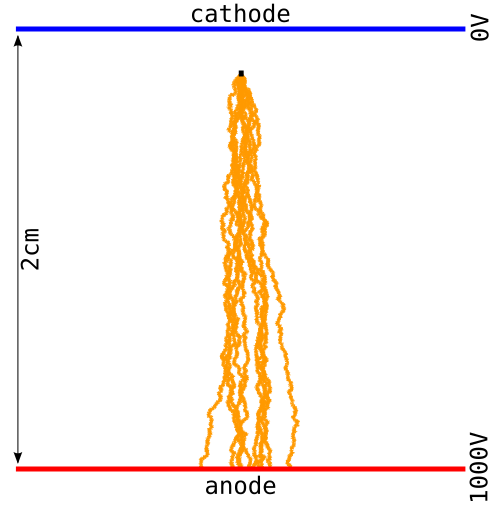}
        \caption[Microscopic view of electrons in a homogeneous field]{
            Simulated microscopic view of \unit[10]{} electrons in a homogeneous field of \unit[500]{V/cm}.
            The isotropic scattering with an effective travelled distance after many collisions is visible.
            Simulated with \garfield.
        }
        \label{fig:micro_view_electron_avalanche}
    \end{figure}

    From the microscopical image of isotropically scattering electrons on gas atoms, it is clear, that an electron cloud is enlarged when drifting, see figure~\ref{fig:micro_view_electron_avalanche} for a simulated example.
    This effect is called diffusion and can be separated into a longitudinal and transversal contribution.

\begin{figure}[h]
        \centering
        \includegraphics[width=.7\linewidth]{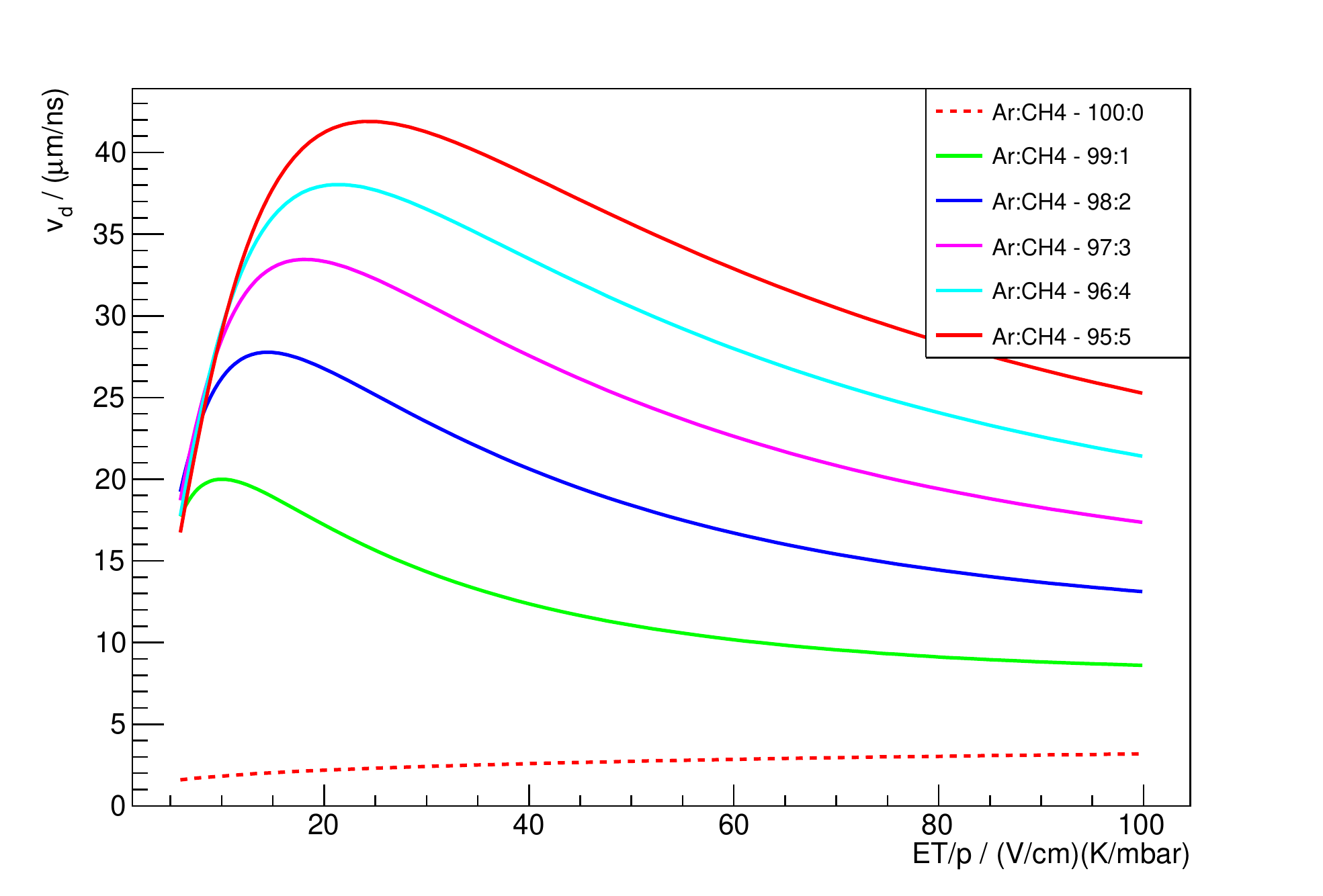}
        \caption[Drift curves of argon with methane additions]{
            Drift velocity curves for argon with additions of methane.
            Adding methane to argon increases the maximum drift velocity.
            Simulated with \magboltz.
        }
        \label{fig:vd_curves_sim}
    \end{figure}

    Most used drifting gas mixtures are not a pure gas but a mixture of a noble gas and some sort of quencher, e.g. methane.
    Already small amounts of quencher can change the drift velocity significantly as can be seen in figure~\ref{fig:vd_curves_sim}.
    Additionally quenchers are also necessary to absorb photons from the gas, that are created by the incident particles and in gas amplification regions.
    The absorbed photons are mostly converted to heat.
    Depending on the gas photon absorption capability, the $\gamma$'s can produce additional free electrons far away from the original location.
    In the worst case, for tracking detectors, all of the detector gas is ionized by these photons.

\section{Gas Amplification of Electrons}
    After passing through the drift space, the electrons need to be detected in a readout plane.
    The charge carried by the arriving cloud is too small to be detected by the readout electronics, hence the arriving electrons are usually amplified by strong electric fields.
    In these field, the acceleration of electrons is sufficient for additional ionization of the gas.
    The newly freed electrons can ionize the gas again.
    This creates an avalanche-like built-up of separated charge, which is also referred to as an electron avalanche or (avalanche) gas amplification.

    To quantify the gain, the traversed electrical field needs to be integrated.
    The first Townsend coefficient $\alpha$ describes how many electrons are added through ionization for every infinitesimal step $\text{d}s$ taken towards the anode.
    It has to be a function of the electron's kinetic energy and start at zero for particles below the ionization threshold.
    Since the kinetic energy of the electrons comes from the gain of energy along a step $\text{d}E=E\text{d}s$, $\alpha$ can be assumed to be a function of the amplification field instead.
    Another dependency is the gas density $\rho$, that is inversely proportional to the average collision length.

    Then, the number of electrons $N$ in the amplification region are increased every step by
    \begin{align}
        \text{d}N=N\alpha(E_\text{amp}, \rho)\text{d}s\,.
    \end{align}
    Integration yields
    \begin{align}
        N(E_\text{amp}) &= N_0\exp\left( \int\limits_{s_0}^{s(\text{end})} \alpha(E(s), \rho)ds \right)\\%
        &= N_0\exp\left( \int\limits_{E_0}^{E(\text{end})} \frac{\alpha(E, \rho)}{\nicefrac{\text{d}E}{\text{d}s}}\text{d}E \right)
    \end{align}
    for the final number of electrons inducing a signal on a wire for example.

    Since electron-ion pairs are produced with a strongly energy dependent cross-section, no global formula for $\alpha(E,\rho)$ exists.
    It has to be determined by simulation or measurement.

\FloatBarrier

    \chapter{Construction of the \hpgmc}
    The goal of this work is to build a gas monitoring chamber that can be operated with highly pressurized gases, a High Pressure Monitoring Chamber (\hpgmc).
    However, increasing the pressure also means reducing the value of \ETp and thus is hindering the comparison to well measured ranges of atmospheric pressure detectors.
    To reach values of \ETp that are comparable again, the electric field has to be increased as well, which means a very high cathode voltage.
    Countering by heating is not feasible as the temperature is incorporated in Kelvin, i.e. an increase from \unit[1]{bara} to \unit[10]{bara} would need an absolute temperature of \unit[3000]{K}.

    Those high voltages need to be properly insulated to protect the interior from sparks and to keep the \hpgmc's outside potential free.
    It is possible to decrease the overall drift distance and thus decrease the needed voltage.
    However, the impact of a constant error on the drift distance measurement can be reduced by increasing the drift distance.
    Diffusion is the limiting factor here as it smears out the signal depending on the drift length.
    Furthermore, the goal is to design a chamber that is safe to operate from atmospheric pressure and maximum cathode voltage up to maximum pressure and maximum cathode voltage.
    The gas is taken from a pressurized bottle with a pressure regulator directly on the bottle with an output of up to \unit[10]{barg}.
    Safety measures have to be taken against overpressure through maloperation or failure of the regulator.
    It needs to be taken into account that some useful quenching gases are flammable or should not be inhaled.

    Since this detector aims to measure in experimentally relatively uncharted area, the \hpgmc is a modular system, that allows removal and easy exchange of components.
    It is also a demonstrator for some new components that were not used before.
    In addition, the whole system should not constrain the possible gases that can be measured.
    The materials should be chemically inert and not contaminate the gases, i.e. low outgassing materials with little water attachment.

    The collected requirements are:
    \begin{enumerate}
        \item Gas system operation safety given at all times
        \begin{itemize}
            \item overpressure protection
            \item consider flammable gases
        \end{itemize}
        \item Long drift distance
        \begin{itemize}
            \item very high cathode voltages
            \item sufficient electric insulation at atmospheric pressure
        \end{itemize}
        \item Modular setup
        \begin{itemize}
            \item easy exchange of detector parts
        \end{itemize}
        \item Inside of chamber chemically inert
        \item Low outgassing materials
    \end{enumerate}

\section{Materials Selection Guidelines}
    General considerations regarding the construction materials have to be taken to assure gas purity.
    Impurities in the drift gas change the gas properties and thus drift and gain characteristics.
    Additionally, this makes measurements incomparable with other experiments.
    The main contributors to impurities are water attachment, outgassing of construction materials and inherent contaminations of supply gas.

\begin{figure}[h]
        \centering
        \includegraphics[width=.7\linewidth]{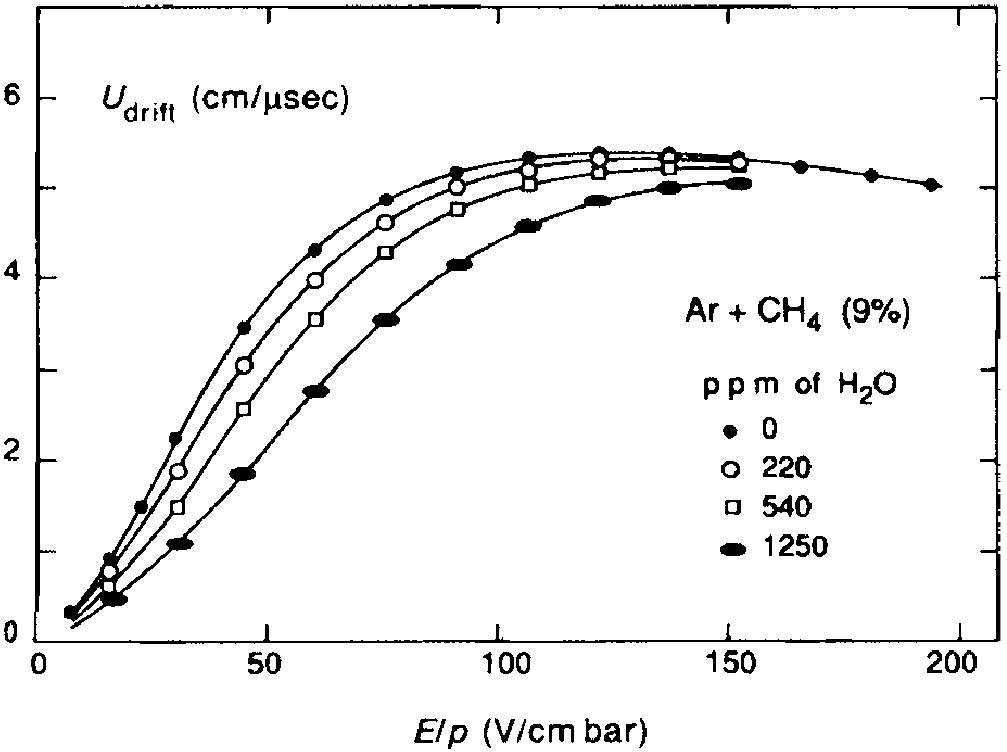}
        \caption[Drift velocities of Ar:\methane 91:9 with small water impurities]{
            Small impurities of water in Ar:\methane 91:9 can have a significant impact on drift velocity.
            With increasing water content, the drift becomes slower for constant \ETp.
            Figure taken from~\citep[fig.~12.5a]{Blum2008}.
        }
        \label{fig:water_drift_impact}
    \end{figure}

    Water attachment on detector parts is already caused by air humidity.
    Porous materials, such as oxidized aluminium surfaces, are more prone to attachment than smooth surfaces.
    This water is usually slowly released during the first weeks to months of operation into the dry gas.
    Since the molecular structure of water contains oxygen, which is highly electronegative, electrons are attached to it and are not available for gas amplification anymore, i.e. are lost during drift.
    An impact on drift speeds can also be measured, see figure~\ref{fig:water_drift_impact} and~\citep[ch.~12.3.3]{Blum2008}.
    To reduce water attachment, materials with microscopically smooth surfaces should be favoured over porous ones.
    A smooth surface can be achieved by polishing.

    Outgassing of construction materials is partly due to desorption but also the material itself or a volatile compound in it, for example colouring or paint.
    This introduces foreign matter into the drift gas after construction and during operation.
    Especially hydrocarbons are known to cause degradation, also called ageing in this context, of amplification geometries, through polymerization~\citep{Zarubin1989}.
    As a guidance for material selection, NASA has compiled a large database of raw materials and compounds, such as tape, glues or insulated cables and published it in~\citep{url:NASA_outgassing_2017}.
    Some examples can be found in table~\ref{tab:examples_outgas}.
    \texttt{TML} and \texttt{CVCM} are the values that describe the outgassing of the tested materials.
    \texttt{TML} is mass loss after baking at \unit[398]{K} in percent and \texttt{CVCM} the mass of collected volatile and condensable material in percent of the starting mass of the sample.
    For low outgassing materials, it is advised by the database's authors to select materials with $\text{\texttt{TML}}<\unit[1.0]{\%}$ and $\text{\texttt{CVCM}}<\unit[0.1]{\%}$.
    In the case of a high pressure environment, outgassing is suppressed by the counteracting pressure of the gas.

    \begin{table}
        \centering
        \caption{
            Examples from used outgassing database~\citep{url:NASA_outgassing_2017}
        }
        \begin{tabular}{lrrl}
            Material & \texttt{TML} / \% & \texttt{CVCM} / \%& Application\\
            \hline \hline
            Bendix connector green glass insert & 0.00 & 0.00 & connector insulation \\
            PFA convoluted tubing & 0.02 & 0.00 & tubing \\
            Silastic 675 silicone & 0.41 & 0.04 & seal-gasket \\
            Delrin white acetal (POM) & 0.78 & 0.10 & mould compound \\
            Black Kapton tape DM 151 & 0.95 & 0.00 & tape \\
            PVC thermocouple wire insulation & 21.46 & 7.23 & wire insulation\\
            Apiezon N hydrocarbon grease & 100.00 & 0.03 & lubricant grease \\
            \hline
        \end{tabular}
        \label{tab:examples_outgas}
    \end{table}

    Contamination of the supply gas or decomposition by radiation alter the gas mixture and thus characteristics.
    Irradiation of molecular gas can dissociate it, which often creates radicals and can lead to polymerization which accelerates ageing.
    Constantly providing gas throughput flushes out dissociated drift gas components, but some may still oxidize the inside of the detector.

\FloatBarrier
\section{High Voltage Considerations}
    Resulting from the available power supplies, the \hpgmc has to handle cathode voltages of up to \unit[30]{kV}.
    Sufficient insulation of the cathode, but also field cage and resistor chain, inside the gas have to be considered during the design phase.
    Paschen's law can be used to calculate a minimal breakthrough gap with given static voltage, medium and pressure.

\subsection{Paschen's Law}
    The derivation of Paschen's law assumes infinite, planar surfaces as cathode and anode with distance $d$ and an electric field only along one coordinate.
    Demanding conservation of total charge by setting the fluxes $\Gamma_{\text{ions}}(\text{anode})-\Gamma_{\text{ions}}(\text{cath.})\overset{!}{=}\Gamma_{\text{elec.}}(\text{cath.})-\Gamma_{\text{elec.}}(\text{anode})$ leads to Paschen's law as derived in~\citep{book:discharges}:
    \begin{align}
        V_\text{b}(p, d) = \frac{Bpd}{\ln\left(Apd\right) - \ln\left[\ln\left(1+\frac{1}{\gammase}\right)\right]}\,.
        \label{eq:paschen_law}
    \end{align}
    Here, $V_b$ is the breakthrough voltage, \emph{p} gas pressure, \emph{d} insulation gap, \gammase the secondary electron emission coefficient and \emph{A}, \emph{B} parameters.
    These parameters are specific to the used gas, pressure and electric field and can be assumed constant within a certain range of pressure reduced electric field \Ep.
    A change in temperature would affect \emph{A} and \emph{B}, but for neither heated nor cooled gases, the temperature can be assumed to be constant at e.g. \unit[300]{K}.
    Some example values can be found in table~\ref{tab:AB_examples}.
    \begin{table}[h]
        \centering
        \caption{
            Some values for \emph{A} and \emph{B}, taken from~\citep[tab.~14.1]{book:discharges}
        }
        \begin{tabular}{lccc}
            & E/p & A & B \\
            Gas & ($\nicefrac{\text{V}}{\text{cm}\,\text{mbar}}$) & ($\nicefrac{\text{1}}{\text{cm}\,\text{mbar}}$) & ($\nicefrac{\text{V}}{\text{cm}\,\text{mbar}}$) \\
            \hline
            Ar            & 150\dots800 & 15.33 & 235 \\
            He            &  40\dots300 & 3.73  & 103 \\
            $\text{CF}_4$ &  30\dots280 & 14.67 & 284 \\
            \hline
        \end{tabular}
        \label{tab:AB_examples}
     \end{table}

    The secondary electron emission coefficient \gammase is defined as $\Gamma_e\left(\text{cath.}\right)=\gammase\Gamma_i\left(\text{cath.}\right)$.
    It describes the impact of ions onto the cathode that can result in free electrons.
    Therefore, \gammase is a function of gas (ions), cathode material and, albeit weakly, the kinetic energy of ions reaching the cathode.
    Values are typically $\ll1$, peaking for alkali materials as cathode in conjunction with noble gases ~\citep[tab.~9.1]{book:discharges}.
    Because of their double-logarithmic entry into equation~\ref{eq:paschen_law}, their influence is negligible.

    Paschen's law can be understood not as a function of independent \emph{(p, d)}, but as a function of \emph{(pd)}.
    Under the assumption that \mbox{$\ln(Apd)\approx \ln(pd)$} and \mbox{$Bpd\approx pd$}, Paschen's law reduces to \mbox{$V_b\approx\nicefrac{pd}{\ln(pd)}$}.
    This shows, that insulation can be achieved both by larger distance and higher pressure.

\begin{figure}[h]
        \centering
        \includegraphics[width=.7\linewidth]{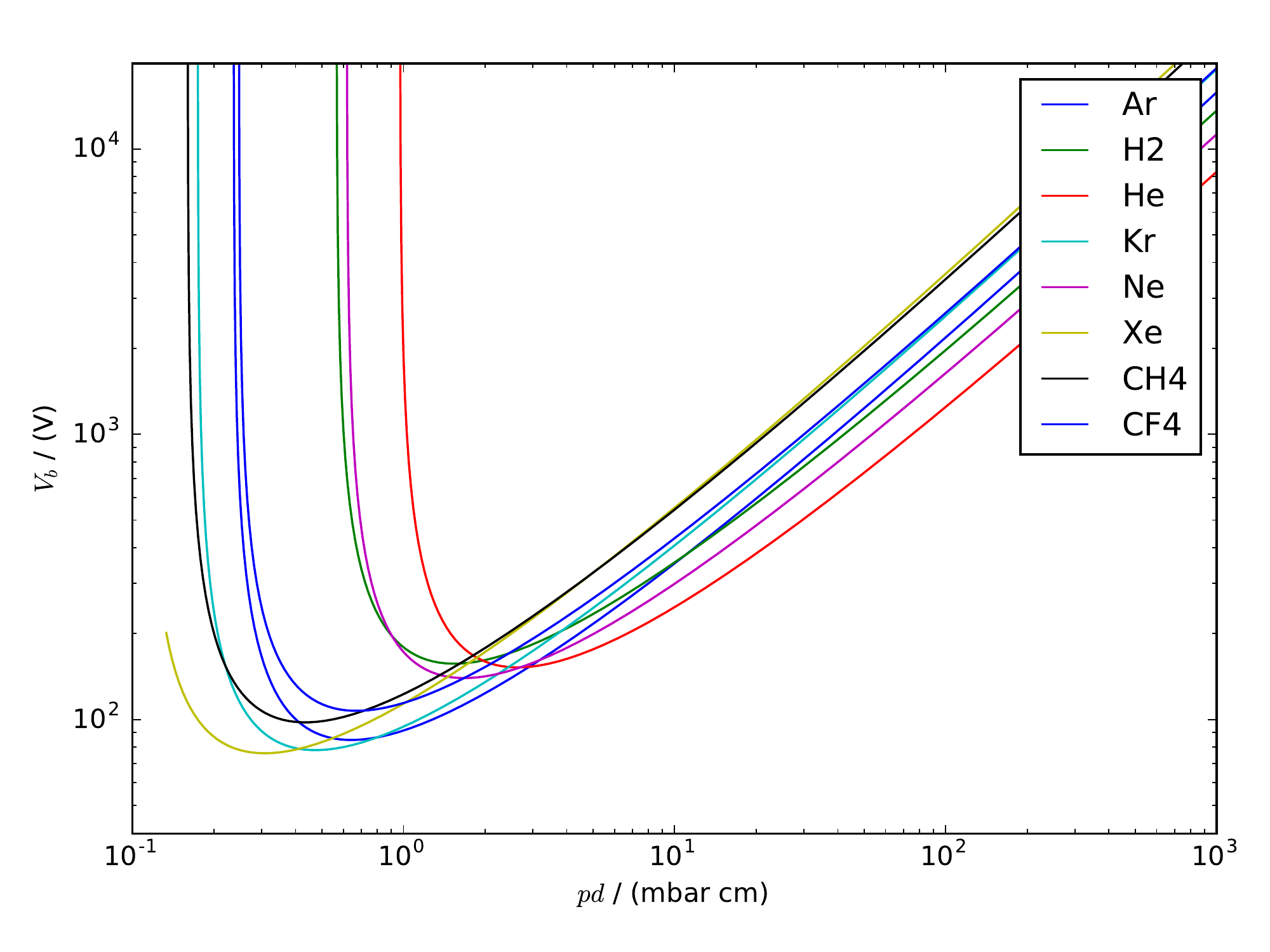}
        \caption[Example Paschen curves]{
            Paschen curves for most noble gases and some molecular gases commonly used in gaseous detectors.
            Calculated with values for \emph{A} and \emph{B} taken from~\citep[tab.~14.1]{book:discharges}.
        }
        \label{fig:example_paschen_curves}
    \end{figure}

    A collection of representative Paschen curves is shown in figure~\ref{fig:example_paschen_curves}.
    The dip and very steep ascend to smaller values of $pd$ is caused when the mean free path of electrons in a gas nears and exceeds the distance between anode and cathode.
    There is a $pd$ under which breakdown can not occur because no gas amplification is possible and electrons can freely travel from cathode to anode.

\subsection{Experimental Verification}
    One of the assumptions for Paschen's law was an infinite, flat anode and cathode.
    This assumption can never be met in experiment.
    Any deviation in shape will result in a locally increased electric field and cause breakdown of an insulation gap prior to the prediction.

\begin{figure}
        \begin{center}
            \subfigure[]{
                \includegraphics[width=.45\linewidth]{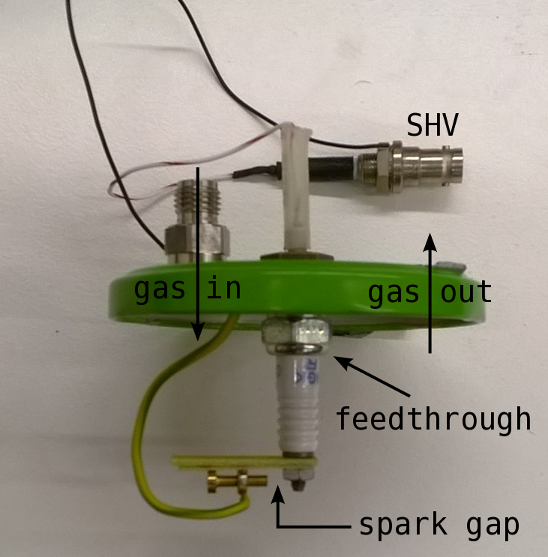}
                \label{fig:sparkgap_lid}
            }
            \subfigure[]{
                \includegraphics[width=.45\linewidth]{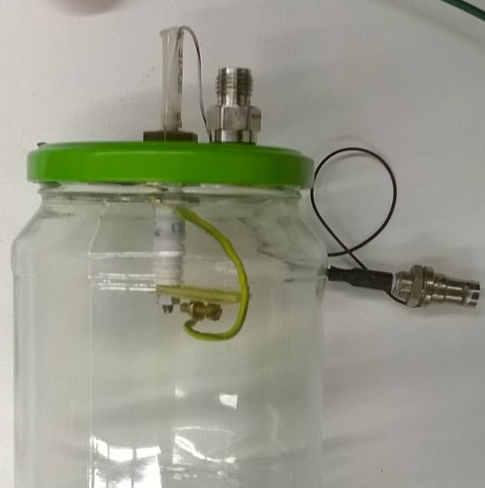}
                \label{fig:sparkgap_full}
            }
        \end{center}
        \caption[Test gas volume containing a visible spark gap]{
            (a)~Flange with feedthroughs and gas connectors.
            The spark gaps width can be changed with the attached screw.
            The cathode surface was removed for this picture.
            (b)~Setup with gas containment attached.
        }
        \label{fig:sparkgap}
    \end{figure}

\begin{figure}
        \centering
        \includegraphics[width=.7\linewidth]{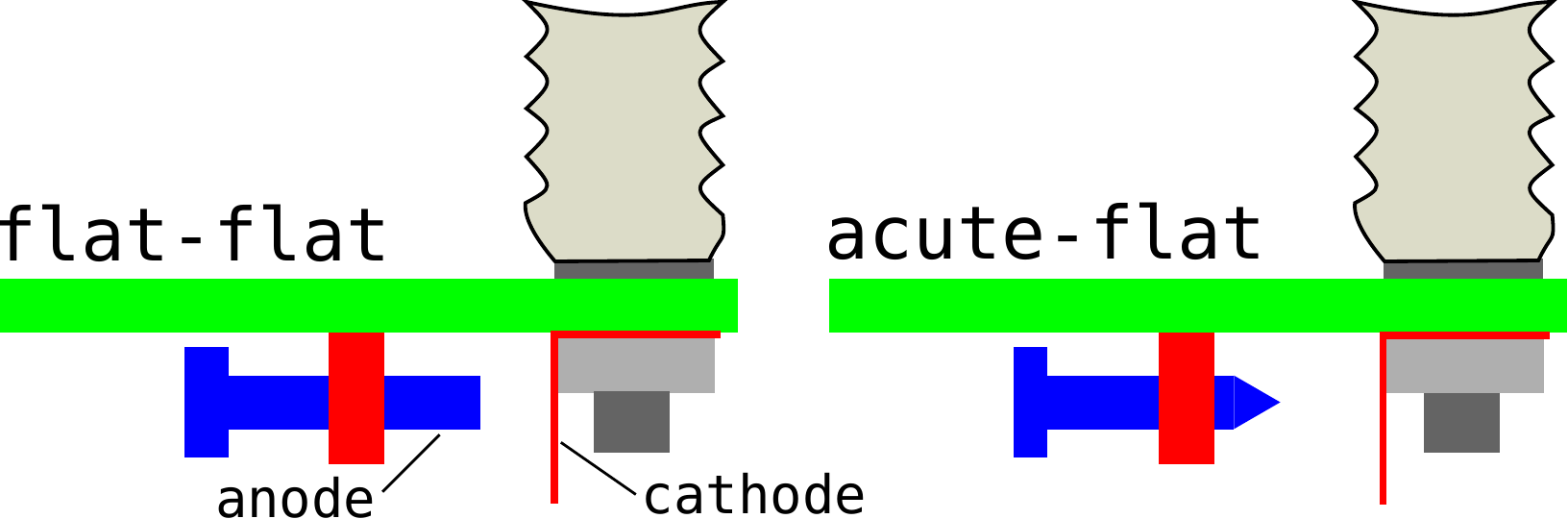}
        \caption[Geometries used in breakthrough tests]{
            Different anode and cathode geometries were used to estimate the effect of non-perfectly flat geometries on the value of the breakthrough voltage.
        }
    \label{fig:geometries_gurkenglas}
    \end{figure}

    For a rough verification of the prediction and estimation of deviations, a small testing gas volume was build, see figure~\ref{fig:sparkgap}.
    It contains a cathode and anode separated by a variable width gas gap.
    The volume is filled with either argon or helium.
    Voltage is then slowly increased until a breakthrough is observed and the corresponding voltage is noted.
    After each measurement, the volume is thoroughly vented to release fumes released during breakthrough.
    Different geometries have been tested by modifying or exchanging the anode or cathode, see figure~\ref{fig:geometries_gurkenglas}.
    As representatives were chosen a flat and acute or tipped geometry.
    An overestimation of the breakthrough voltage is expected for the tipped geometries, while a combination were both anode and cathode have a flat surface should match the prediction better.

\begin{figure}[h]
        \centering
        \includegraphics[width=.7\linewidth]{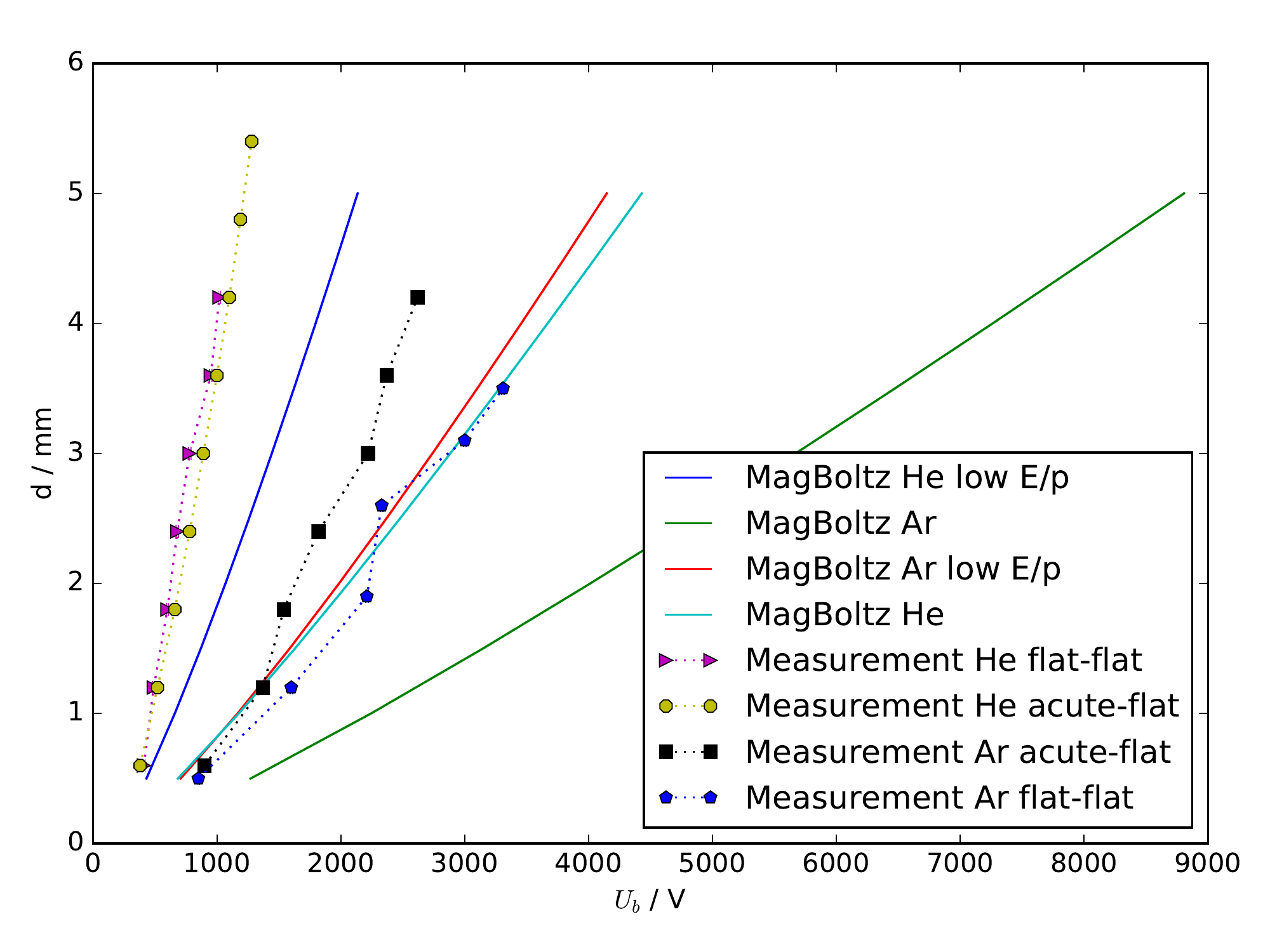}
        \caption[Experimental and theoretical Paschen curves]{
            Comparison of computed Paschen curves with experimental curves.
            The impact of taken assumption in the theory is evaluated by trying different geometries.
        }
        \label{fig:paschen_exp_curves}
    \end{figure}

    The results of this experimental series can be seen in figure~\ref{fig:paschen_exp_curves}.
    The measurement curves show similar behaviour within the same gas.
    For argon the \emph{acute-flat} geometry combination shows the expected reduction in breakthrough voltage compared to the \emph{flat-flat} measurement series.
    Some theory curves however show a distinct deviation.
    In one case of helium, the curve even fits argon seemingly better.
    Most concerning is the far right curve of predicted argon breakthrough gaps, that severely underestimates the appropriate insulation gap.
    The reason can be found with the parameters \emph{A} and \emph{B}, that are only valid within a certain range of \Ep.
    Curves with the addendum \emph{low E/p} used values for those parameters that were calculated for a lower range of \Ep.
    Those fit the experiment's \Ep more closely and thus show a better prediction of the insulation gap.
    Nevertheless, it is only possible to measure and calculate the magnitude, but not a more precise value, of the breakthrough voltage this way.

\subsection{Extrapolation and Safety Zones}
    \label{ssec:extrapolation_safety_zones}
    It has become evident, that extrapolation of Paschen's law can not be naively performed to prospected operation voltages and pressures of the \hpgmc.
    For a trustworthy extrapolation, a piecewise recalculation of \emph{A} and \emph{B} is needed.
    Their origin lies with an approximate parametrization of the influence of increased pressure on the first Townsend coefficient.
    This parametrization is given by:
    \begin{align}
        \frac{\alpha}{p}=A\cdot\exp{(-B\frac{p}{E})}\,.
        \label{eq:alpha_A_B}
    \end{align}

    \begin{figure}[h]
        \centering
        \includegraphics[width=.7\linewidth]{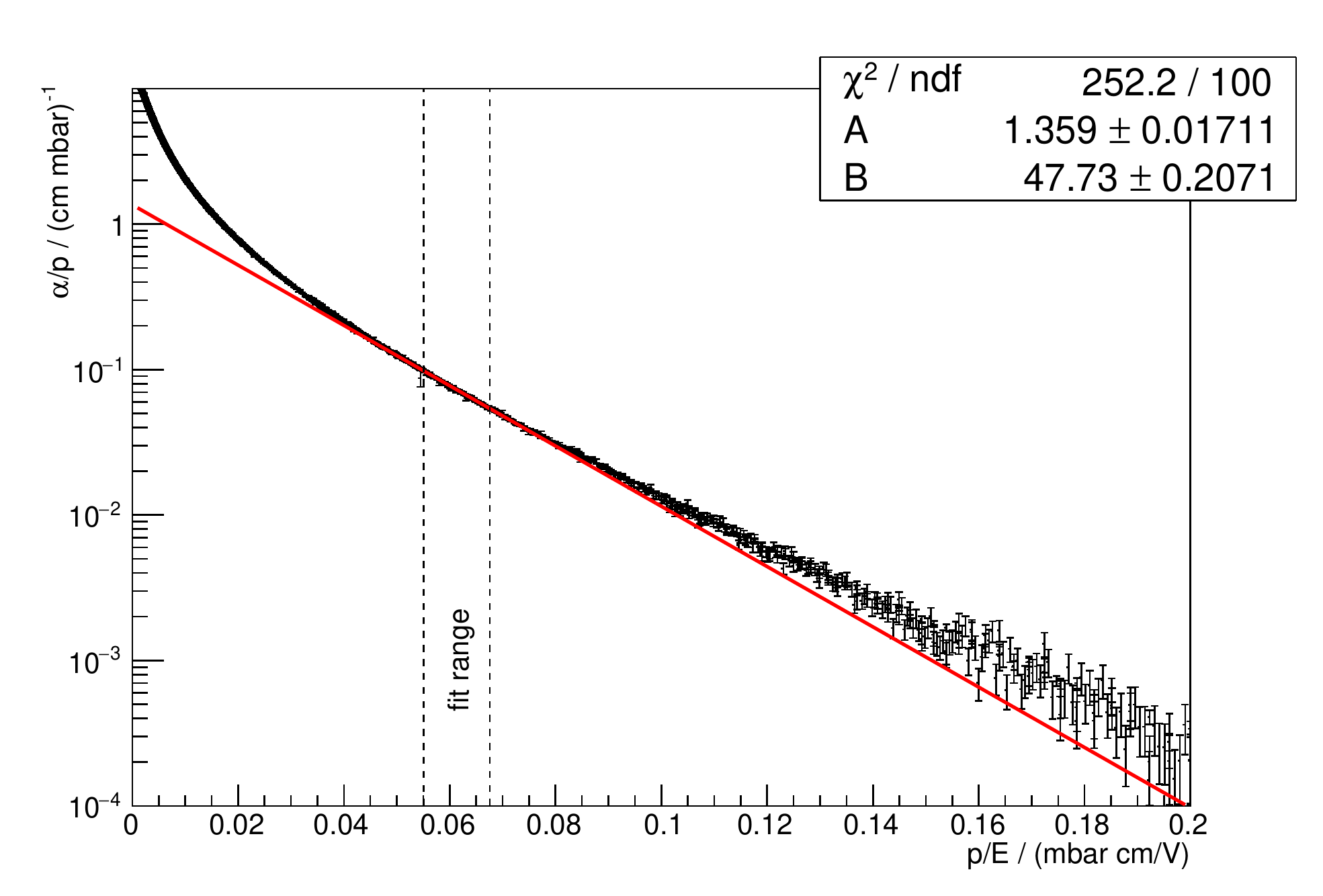}
        \caption[Example of \alphap versus \pE curve for argon]{
            Exemplary curve of \alphap versus \pE for argon.
            The plotted fit of equation~\ref{eq:alpha_A_B} shows good agreement locally, but significant deviation on a larger \pE range.
        }
        \label{fig:alpha_p_curve}
    \end{figure}

    With \magboltz, it is possible to simulate $\nicefrac{\alpha}{p}$ versus \Ep for different, pure gases.
    Mixed gases can unfortunately not be simulated reliably, because the Penning effect is not known for all gas mixtures to high precision.
    The gain can vary quite drastically with a change in the Penning transfer probability, so a fit to $\nicefrac{\alpha}{p}$ can not be trusted~\citep[fig.~6]{Andronica2004}.

    The simulated $\nicefrac{\alpha}{p}$ curves are then fitted with equation~\ref{eq:alpha_A_B} in an interval of 100 points around the requested \Ep.
    Figure~\ref{fig:alpha_p_curve} shows a representative fit in a subrange around \unit[0.06]{mbar~cm/V}.
    A side effect of this is, that the breakthrough voltage depends implicitly on the electric field, which is derived from applied voltage and geometry.
    For the design of the \hpgmc, the breakthrough voltage for a given insulation gap distance versus applied voltage has to be known.

\begin{figure}[!h]
        \begin{center}
            \subfigure[Atmospheric pressure]{
                \includegraphics[width=.5\linewidth]{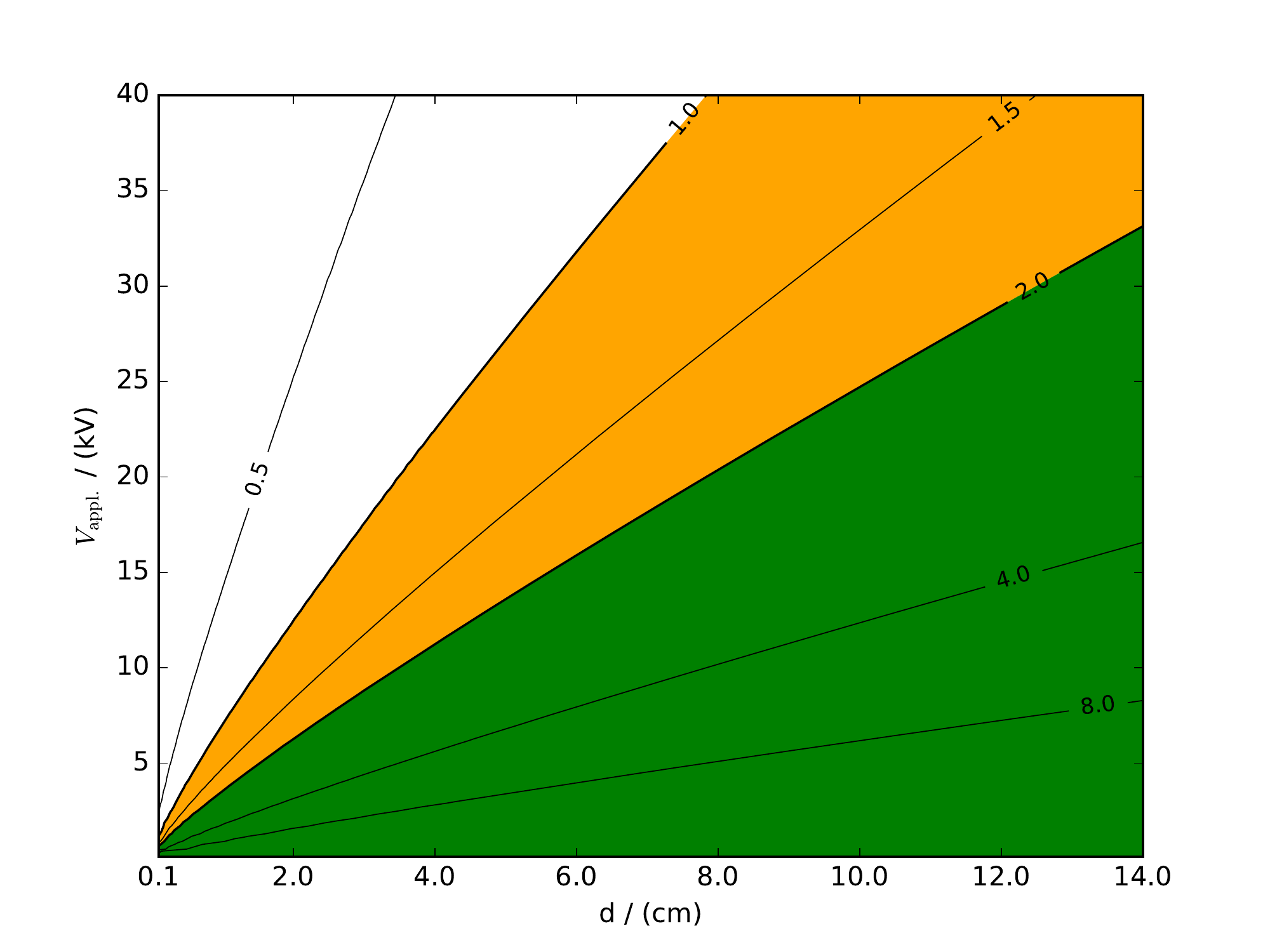}
                \label{fig:breakdown_safety_zones_1_bar}
            }%
            \subfigure[Pressurized to 11 bara]{
                \includegraphics[width=.5\linewidth]{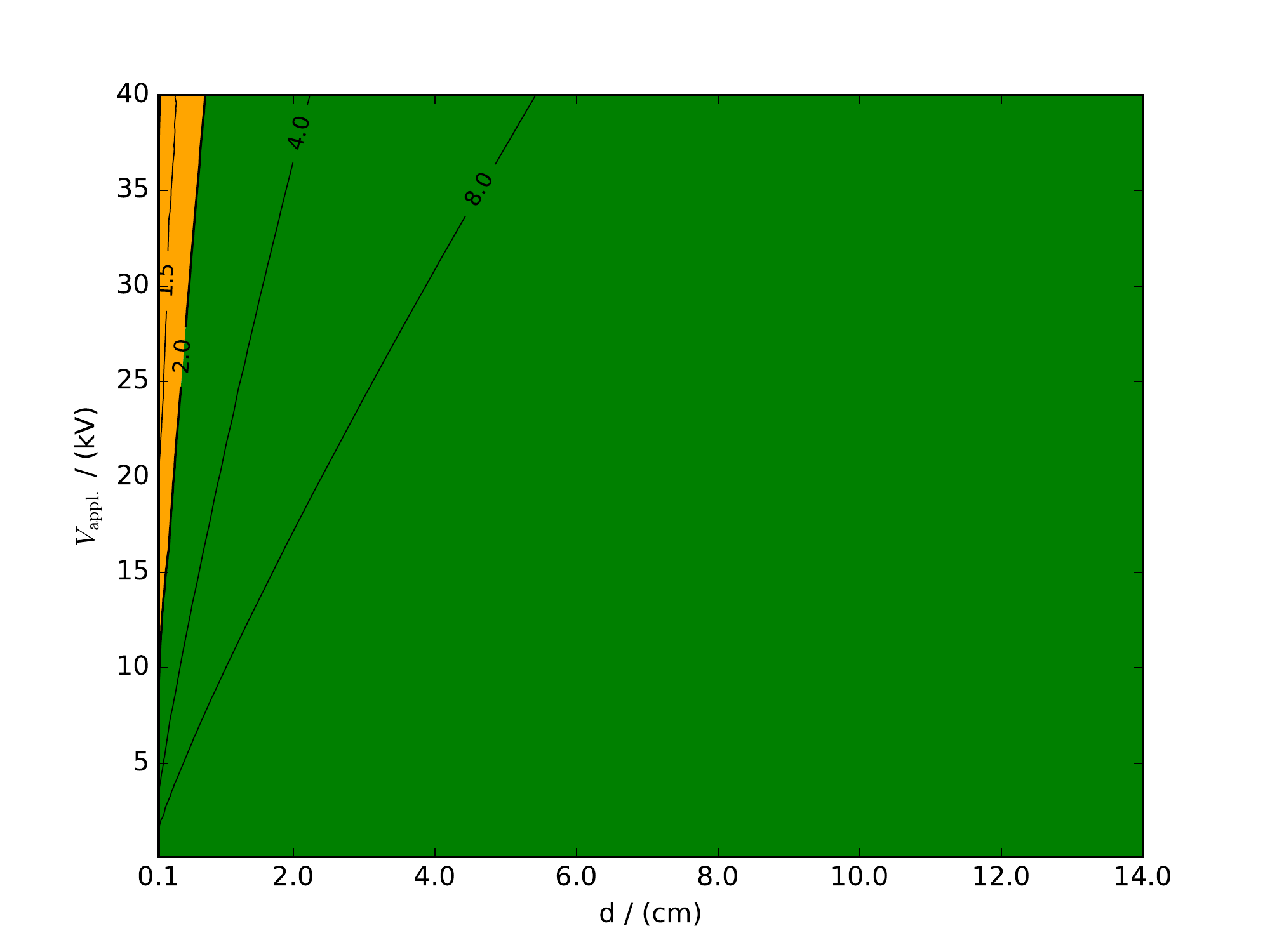}
                \label{fig:breakdown_safety_zones_11_bar}
            }
        \end{center}
        \caption[Contours of breakthrough voltage and applied voltage for argon]{
            The quotient of breakthrough voltage and applied voltage can be used to define safety zones of a setup for which insulation gaps and applied voltage is given.
            When the quotient is below 1, breakthrough will happen before operation voltage can be reached.
            For anywhere above 1, stable operation is theoretically possible, but a safety factor of 2 should be taken.
        }
        \label{fig:safety_zones}
    \end{figure}

    Paschen's law from equation~\ref{eq:paschen_law} is only exact at voltages that are just high enough to cause breakthrough, because of the dependency of \emph{A} and \emph{B} of \emph{E}.
    A prediction of $V_\text{b}$ that uses values for \emph{A} and \emph{B} computed with fields $E=\nicefrac{V_\text{appl.}}{d}$ is not guaranteed to have the same prediction if \emph{A} and \emph{B} are computed at the predicted $V_\text{b}$.
    The impact of this shortcoming is estimated by the verification at low voltages and looking at the general trend of $V_\text{b}$ with increasing $V_\text{appl.}$.
    Plotting the quotient $\nicefrac{V_\text{b}}{V_\text{appl.}}$ gives this trend and shows contours where operation is save, i.e. $\nicefrac{V_\text{b}}{V_\text{appl.}}>1$.
    To account for geometric effects and non-perfect cleanliness, that can reduce the breakthrough voltage, areas in figure~\ref{fig:breakdown_safety_zones_1_bar} where the predicted breakthrough voltage is at least twice that of the applied voltage are deemed as safe.
    The benchmark gas is argon, as it is the main component of the gas mixtures used in some prominent large scale experiments with gaseous detectors (e.g. T2K, ALICE).
    It is also readily available and has a very thorough implementation in \magboltz.
    Figure~\ref{fig:breakdown_safety_zones_11_bar} shows the same plot for argon at \unit[11]{bara} pressure and makes it clear, that for high pressures insulation is comparatively easy to achieve.

    A \unit[30]{kV} cathode voltage would need to have an argon filled gap of at least \unit[13]{cm} to fulfill the set safety standard for electrostatic insulation.
    This insulating capability of a material is referred to as dielectric strength and usually given in units of \unit{kV/mm}.

\FloatBarrier
\subsection{The Field Cage}
    The central part of the \hpgmc is the field cage surrounding the drift volume.
    A homogeneous drift field is generated inside the cage by putting the cage's field strips on decreasing voltages by connecting them with resistors.
    Packing the field strips tighter and reducing their width increases field homogeneity, however, adjacent strips need to be sufficiently insulated against each other.
    Breakdown between one field strip pair, i.e. reducing one resistor value to \unit[0]{\Ohm}, doubles the voltage difference for the next strip towards the anode.
    If this strip, or its resistor, experiences breakdown as a result, a chain reaction occurs, tripping the HV supply and, in the worst case, damage the \hpgmc.
    The used resistor chain consists of 19 high voltage resistors of $\unit[10]{\text{M}\Ohm}$ each.
    They are rated up to \unit[2]{kV} for continuous operation and absolute maximum \unit[3]{kV} surge.
    The maximum possible cathode voltage of \unit[30]{kV} means a maximum voltage difference of \unit[1579]{V} for each resistor, so regular operation is well within the parameters for continuous operation.

\begin{figure}[h]
        \centering
        \includegraphics[width=\linewidth]{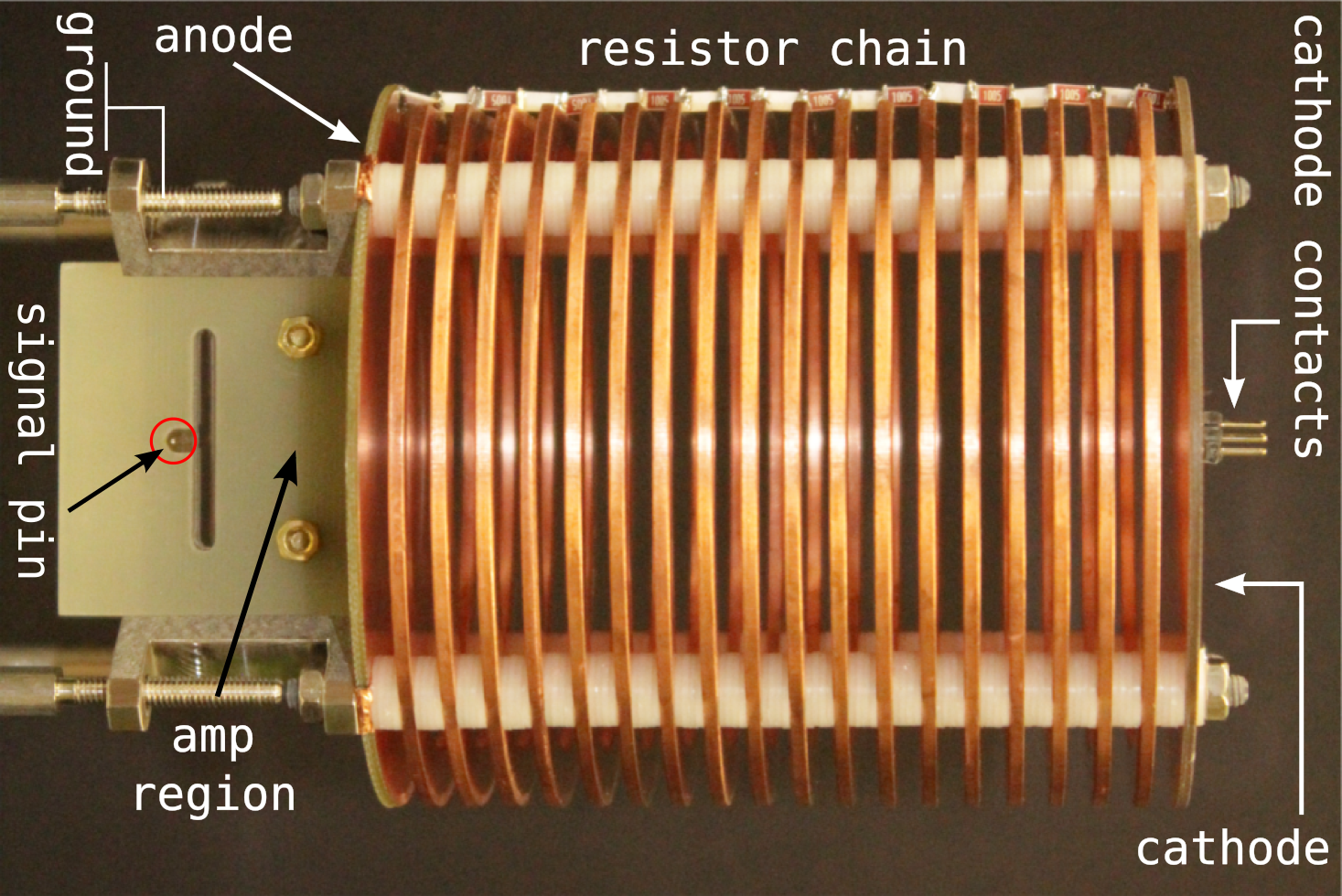}
        \caption[The assembled field cage]{
            The assembled field cage with resistor chain.
            Electrical contact to ground is through the cage's feet.
            The attached part on the anode houses the amplification region.
        }
        \label{fig:field_cage}
    \end{figure}

    A rigid design of the cage counters the risk of deformation during assembly, which could disconnect resistors and thus affect the drift field.
    Using copper rings stacked cylindrically with gas gaps for insulation provides this rigidness.
    With a gas gap, the mechanical contact of the strips is also reduced, eliminating many possible paths for creep currents that deteriorate material and can be the precursor of a breakdown.
    Between cathode and anode planes, 18 copper rings are stacked with a gap of $\unit[3.2]{mm}$ for insulation, see figure~\ref{fig:field_cage}.
    Consulting figures~\ref{fig:breakdown_safety_zones_1_bar} and~\ref{fig:paschen_exp_curves}, a gap of about \unit[3]{mm} is sufficient for at least \unit[2]{kV}.

    In this configuration, a maximum \ETp of $\approx\unit[900]{\uETp}$ is reachable at atmospheric pressure and a temperature of \unit[300]{K}.
    Ramping up the pressure to the maximum value of \unit[11]{bara} leads to an \mbox{$\ETp\approx\unit[84]{\uETp}$}, which covers many working points of already deployed gaseous detectors.
    The T2K TPCs for example operate at approximately \unit[78]{\uETp}.

\FloatBarrier
\section{Pressure Regulations}
    Guidelines and regulations for pressurized vessels in the EU are described by directive 2014/68/EU~\citep{DIRECTIVE2014/68/EU}.
    Since the directive uses the unit bar as synonym for barg, this section will follow this wording.
    Following the classification scheme, the \hpgmc is a pressure vessel for gases of Group 1: Flammable gases below \unit[200]{bar}.
    While a drift gas with minor additions of flammable gases might not be flammable in itself, any admixture of a flammable gas is enough to classify the complete gas mixture as flammable in the sense of the directive.

\begin{figure}[h]
        \centering
        \includegraphics[width=.7\linewidth]{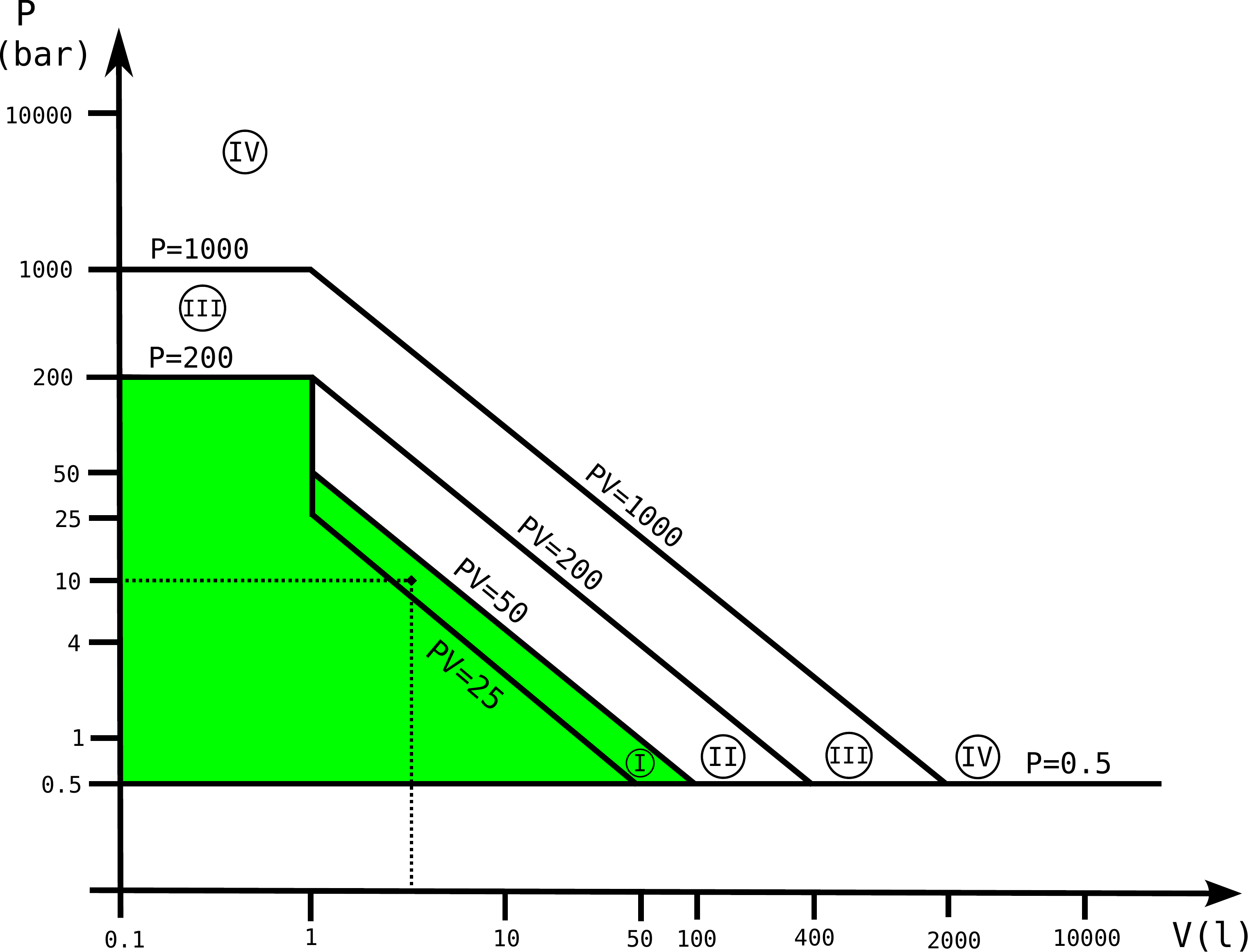}
        \caption[Definition of pressure equipment categories by $pV$ contours]{
            Definition of categories by $pV$ contours for flammable gases.
            Anything equal to or below category \texttt{I} is considered feasible.
        }
        \label{fig:category_contours}
    \end{figure}

    The pressure equipment category a vessel falls into depends on inner volume, pressure and type of contained gas.
    A category dictates necessary safety measurements and quality control requirements and can be read off from the roman numerals in figure~\ref{fig:category_contours}.
    Pressures of less than \unit[0.5]{bar} are not covered by the directive.

    The aim is to design the \hpgmc as a category \texttt{I} pressure equipment.
    In this category, only internal production control is necessary.
    Most testing of single components can be skipped when using standard industry components that are already tested and certified.
    When aiming for an arbitrarily set target overpressure of \unit[10]{bar}, this limits the maximum inner volume of the \hpgmc to $\unit[5]{l}$ to fall into category \texttt{I}.
    The result from section~\ref{ssec:extrapolation_safety_zones} was to accommodate large insulation gaps of up to \unit[13]{cm} between field cage and chamber walls.
    However, doing this would lead to a large volume and increases the pressure equipment category to at least category \texttt{II}.

    A reduction of the needed insulation gap can be achieved through usage of non-conductive materials with high dielectric strength.
    Most commonly used insulators are thermoplastics, which are often also highly outgassing materials.
    A sample of PVC wire insulation is reported in the outgassing database used to have a $\text{\texttt{TML}}=\unit[15.49]{\%}$.
    A material that does not only have excellent insulation characteristics and is chemically inert to nearly any substance, but also shows very little outgassing is PTFE\footnote{Polytetrafluoroethylene, commonly known by one of its brand names Teflon}, or one of its enhancements, PFA\footnote{Perfluoroalkoxy}.
    With a dielectric strength of \unit[80]{kV/mm} for pure PFA, a few millimetres will be sufficient for insulation~\citep{datasheet:DuPont_PFA}.
    This value already takes into account material deficiencies such as pores or accidental surface damaging.
    Notable is also the very low water attachment of less than \unit[0.03]{\%}.

\begin{figure}[h]
        \centering
        \includegraphics[width=.7\linewidth]{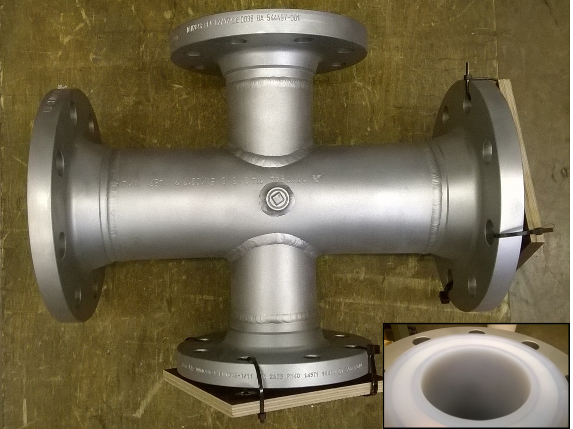}
        \caption[Standard industry double-tee piping lined with PFA]{
            Chemical industry standard double-tee piping lined with \unit[4]{mm} PFA.
            The lining's theoretical, integrated dielectric strength is \unit[320]{kV}.
            The hull is made from high grade stainless steel and rated for \unit[25]{bar} overpressure.
        }
        \label{fig:double_tee_only}
    \end{figure}

    The chamber of the \hpgmc was chosen to be a double-tee piping piece made from stainless steel (1.4571) lined with \unit[4]{mm} PFA.
    It is a standard component for chemical industries and pressure rated and certified up to \unit[25]{bar} (PN25) with an inner volume of about \unit[4.52]{l}.
    According to the manufacturer, the only materials that must not come in contact with the inside lining, assuming room temperature, are liquid alkalines, elemental fluorine and some halides~\citep{BAUM2017}.
    Figure~\ref{fig:double_tee_only} shows the piece shortly after arrival.
    The crossing, smaller tees have a diameter of \unit[81]{mm} (DN80) and the main pipe one of \unit[116]{mm} (DN100).
    Every side can be connected using the flanges described in {DIN EN 1092-1}~\citep{DIN1092}.
    Blind flanges with electrical feedthroughs and bores for gas in and outlet on the four sides of the double-tee will be used to close off the piece.

    The blind flanges for the DN80 sides are regular, off-the-shelf stainless steel blind flanges.
    The feedthroughs on them are through-hole, i.e. not attached by screws and should not significantly weaken the material.
    Since the DIN standard does not differentiate between PN25 and PN40 for DN80 or DN100, the used blind flanges already come with a generous safety margin of \unit[15]{bar} compared to the PN25 rating of the vessel itself.
    However, for the anode and cathode flange, some feedthroughs are flange-mounted themselves.
    That means, the blind flanges need to have threaded blind holes on the outside that weaken the material along a larger circle than single through hole mounting would.
    For safety reasons, blind flanges were built that have a thickness of a standard DN100 blind flange plus the depth of the blind holes which comes to a total of about \unit[55]{mm}.

\FloatBarrier
\section{Generation of Drift Electrons}
    For the generation of the primary ionization electrons in the gas within the field cage, two strontium (\Sr) sources are used.

\subsection{Protection of \Sr Sources}
    As a $\beta$-emitter, shielding of radiation is not difficult.
    The inner PFA lining is already enough shielding even for maximum energy electrons.
    From this follows, that the sources have to be installed inside the pressure vessel to generate any tracks in the gas.
    This also means, that the sources are exposed to the high voltage inside.
    Great care has to be taken to ensure no sparks can reach the sources and liberate material.

\begin{figure}[h]
        \centering
        \includegraphics[width=.7\linewidth]{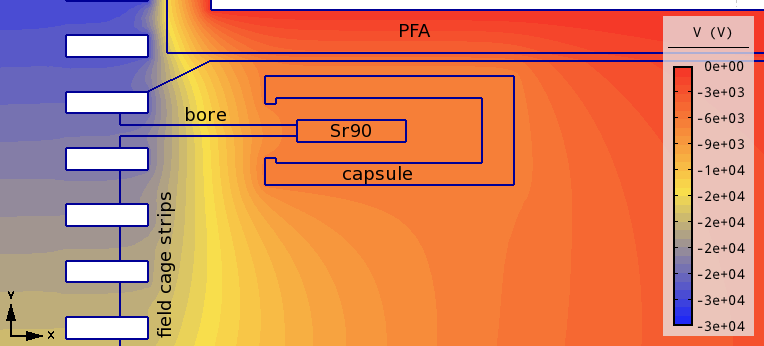}
        \caption[Simulated electrostatic fields near the far-side \Sr source holder]{
                Electrostatic simulation of fields in the proximity of one \Sr source holder.
                The other holder is closer to the anode and experiences far smaller potential gradients.
                Electrons exit into the drift volume through a small \unit[1]{mm} diameter bore.
                Each contour step corresponds to a change in electric potential of \unit[1]{kV}.
        }
        \label{fig:sr90_fieldsim}
    \end{figure}

    The first protection measure is to embed the sources in a non-conductive retainer machined from POM\footnote{Polyoxymethylen}, a robust plastic.
    Furthermore, the \Sr pellet is encased in a brass capsule that acts as a free floating Faraday cage.
    The decay electrons can exit through a small bore of \unit[1]{mm} diameter and enter the drift volume by passing between the field strips.
    A piece of aluminized Mylar foil acts as an additional safeguard between the source and the exit channel towards the drift volume.
    The described geometry, without foil, is depicted in figure~\ref{fig:sr90_fieldsim} and pictures of the machined parts can be found in section~\ref{apx:sources}.
    By designing a safe setup for the \Sr capsule closer to the cathode, the one towards the anode is also safe since the voltage decreases towards the anode.

    The program \agros was used to simulate the interior of the \hpgmc with the described geometry and electrostatic fields generated by the field cage~\citep{Agros2D}.
    It only works on two-dimensional problems where a third spatial dimension can be added only for cylindrical symmetries or if the problem is homogeneous along the third dimension, i.e. planar symmetry.
    For the \hpgmc, the planar symmetry option was chosen as an estimator for the real geometry.

\begin{figure}[h]
        \centering
        \includegraphics[width=.7\linewidth]{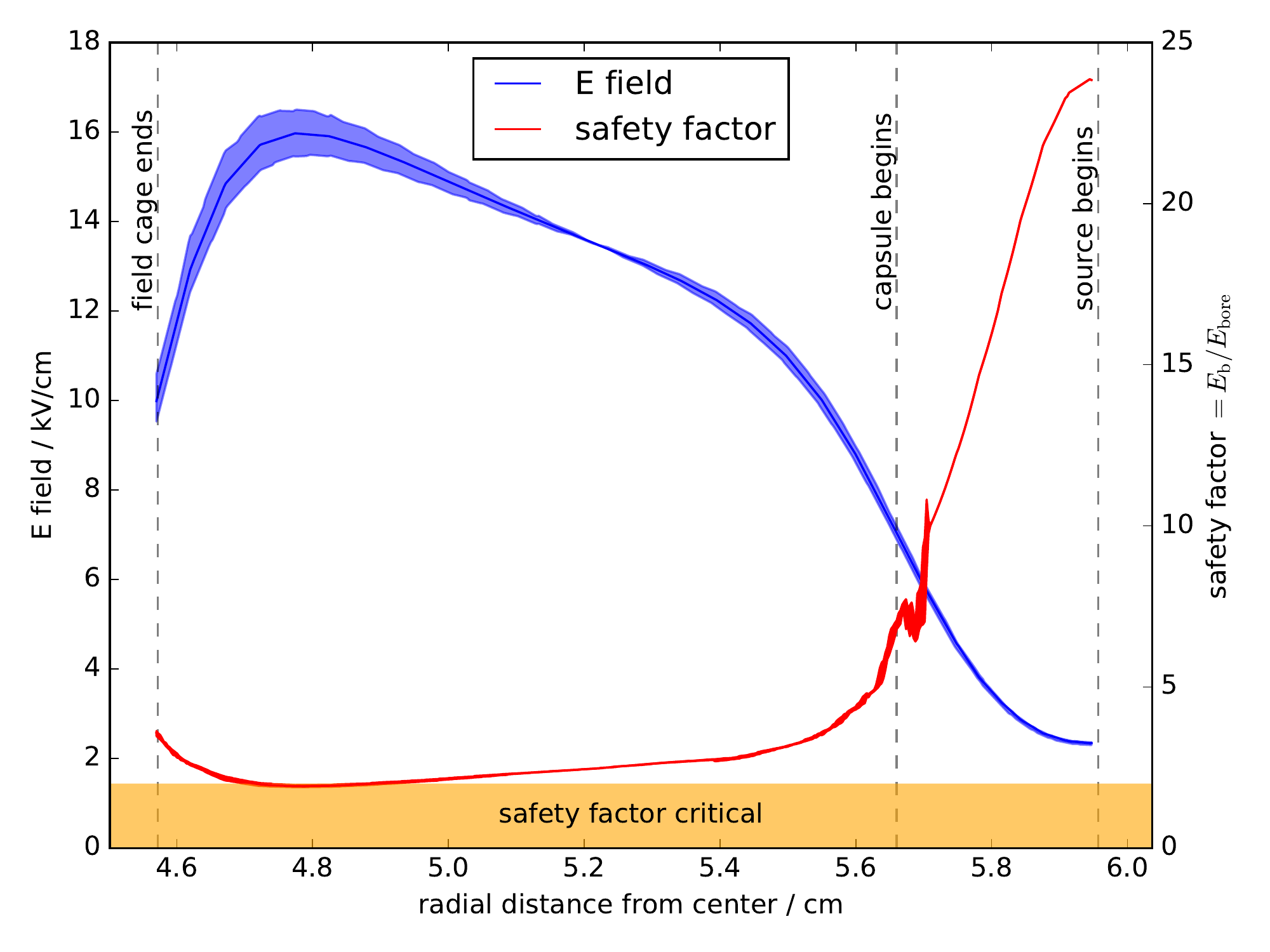}
        \caption[Electrical field and safety factor along bore towards \Sr capsules]{
            The electric field along the bore towards the \Sr sources in comparison to the calculated ratio of breakthrough field with the local electric field.
            If the safety factor exceeds~2, usage is considered safe.
            Simulation artifacts around the position where the capsules begin are of no concern for safety.
        }
        \label{fig:safety_factor_channel}
    \end{figure}

    The electric field inside the bore towards the \Sr capsules is taken along the central line and compared to the predicted breakthrough field calculated by equation~\ref{eq:paschen_law} divided by the used step size $d$.
    A safety factor is calculated analogous to section~\ref{ssec:extrapolation_safety_zones}.
    The bore is a circular hole that has a very smooth surface, so uncertainties from imperfect geometries are not expected to have a large impact.
    The main deviation from equation~\ref{eq:paschen_law} comes from free electrons inside the bore generated by the passing high energy $\beta$-electrons.
    Following the field gradient, the electrons drift onto the Mylar foil and must not be accelerated enough for gas amplification and breakdown.
    A safety factor of at least \unit[2]{} is required again.
    Figure~\ref{fig:safety_factor_channel} shows the calculated safety factors and the simulated electric field used for calculation along the radial coordinate with origin in the centre of the field cage.
    The error-bands are calculated by taking the electric field along two upwards and two downwards shifted paths inside the bore and taking the maximal and minimal resulting safety factors as the band's limits.

    A discontinuity around \unit[5.7]{cm} is a simulation artifact that does not affect safety of operation in that region.
    The electric fields are smooth in this region and thus the safety factor should also be smooth and will not dip below a factor of \unit[2]{}.
    Around this coordinate, the real setup will have an aluminized Mylar foil that was not included for simulation.
    The foil will be held in place only by friction.
    Attaching it by gluing would risk rupture when changing the operation pressure too quickly.

    The lowest safety factor is at $\approx\unit[2]{}$ around \unit[4.8]{cm} from the centre of the field cage.
    For a sustained gas amplification, more than just a local undercutting has to take place, so the simulated geometry can be considered safe for use.

\FloatBarrier
\subsection{Range of Electrons inside the Chamber}
    The daughter isotope of \Sr, \Ytt, is also a pure $\beta$-emitter with a maximum energy of $Q_{\text{max}}=\unit[2.283]{MeV}$ in contrast to \Sr's maximum energy of \unit[0.546]{MeV}~\citep[tab.~37.1]{PDGReview2016}.
    The complete decay chain is
    \begin{align*}
        {}^{90}_{38}{\text{Sr}}\overset{\beta}{\rightarrow} {}^{90}_{39}{\text{Y}} \overset{\beta}{\rightarrow} {}^{90}_{40}{\text{Zr}}.
    \end{align*}
    The half-life of the Yttrium isotope is much shorter than that of \Sr ($\unit[2.7]{d}\ll\unit[28.5]{y}$), so the main contribution of the used \Sr sources to ionization in the gas volume actually comes from Yttrium $\beta$-radiation.
    The most probable energy of an $\beta$-radiation electron can be estimated to be about $\nicefrac{1}{3}$ of $Q_{\text{max}}$.
    To assure sufficient statistics, the design takes the maximum range of electrons of \mbox{$\nicefrac{Q_{\text{max}}}{3}\approx\unit[700]{keV}$} as basis for construction.

    For calculation of attenuation of electrons in gas at different pressures, the NIST database \texttt{ESTAR} was used~\citep{ESTAR}.
    It allows to select many commonly used elements and materials in radiation detection or detector construction and generates stopping power tables for the specified compound.
    There is also the possibility to get a more accurate value for the density of pressurized gases than by using the ideal gas proportionalities.

\begin{figure}[h]
        \centering
        \includegraphics[width=.8\linewidth]{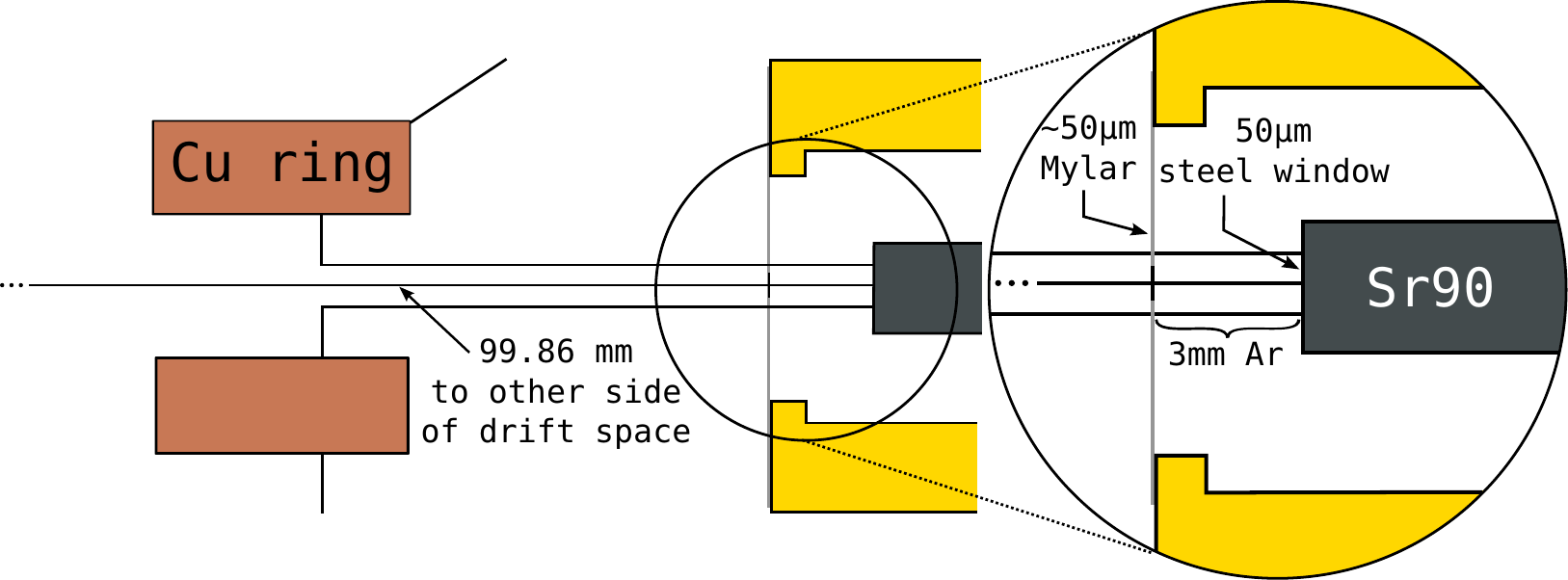}
        \caption[Passed matter of \Sr electrons before entering scintillating fibers]{
            Path of electrons emitted from \Sr sources and the passed material depths on their way to the trigger's scintillating fibres.
        }
        \label{fig:passed_matter_budget}
    \end{figure}

    In pure argon at \unit[1]{bara}, the penetration depth of a \unit[700]{keV} electron is calculated to be \unit[234.9]{cm} and $\text{d}(\unit[11]{bar})=\unit[21.2]{cm}$ both with a maximal uncertainty of \unit[3]{\%} from the stopping power tables of \texttt{ESTAR}.
    After traversing the inner part of the field cage, the particles need to have enough energy left to start a measurement.
    For this, the electrons have to enter a scintillating fibre and produce a sufficient amount of scintillation light to be picked up by SiPMs.
    Furthermore, the original \Sr source is not open, but behind a $\unit[50]{\mu m}$ stainless steel window, see figure~\ref{fig:passed_matter_budget}.
    In order to leave enough energy with the electron after traversing the detection volume, the field cage diameter is restricted to below \unit[100]{mm}.
    Table~\ref{tab:material_integrated} shows all materials that shield the electrons before reaching the scintillating fibres, their thicknesses and the equivalent depth in \unit[11]{bar} argon.

    \begin{table}
        \centering
        \caption{
           Matter traversed by $\beta$-electrons emitted from the \Sr sources and equivalent depth in \unit[11]{bara} argon, see also figure~\ref{fig:passed_matter_budget}
        }
        \begin{tabular}{lll}
            Material & Thickness & \unit[11]{bara} Ar equiv. \\
            \hline
            \hline
            Stainless steel & $\unit[50]{\mu m}$ & \unit[23]{mm} \\
            Argon & \unit[3]{mm} & \unit[3]{mm} \\
            Mylar foil& $\approx\unit[50]{\mu m}$ & $\approx\unit[5]{mm}$ \\ %
            Argon & \unit[99.86]{mm} & \unit[99.86]{mm} \\
            \hline
            & Total & \unit[130.86]{mm} \\
        \end{tabular}
        \label{tab:material_integrated}
    \end{table}

\FloatBarrier
\section{Amplification Region}
    The amplification region of the \hpgmc consists of a regular, off-the-shelf razor blade.
    Strong electric fields are build up close to the tip of the blade due to its strong spatial gradient.
    Measurements done using a microscope show, that the blade is only about $\unit[(5.0\pm1.4)]{\mu m}$ thick at the tip.
    Compared to typically used wires with a diameter of $\mathcal{O}(\unit[10]{\mu m})$, it is expected that a sufficiently strong electric field for gas amplification can be produced.
    An advantage over wires is the robustness of a blade to overvoltages and easy installation.
    Single sparks can destroy wires, and re-installation would need depressurizing of the chamber.

\begin{figure}[h]
        \centering
        \includegraphics[width=.7\linewidth]{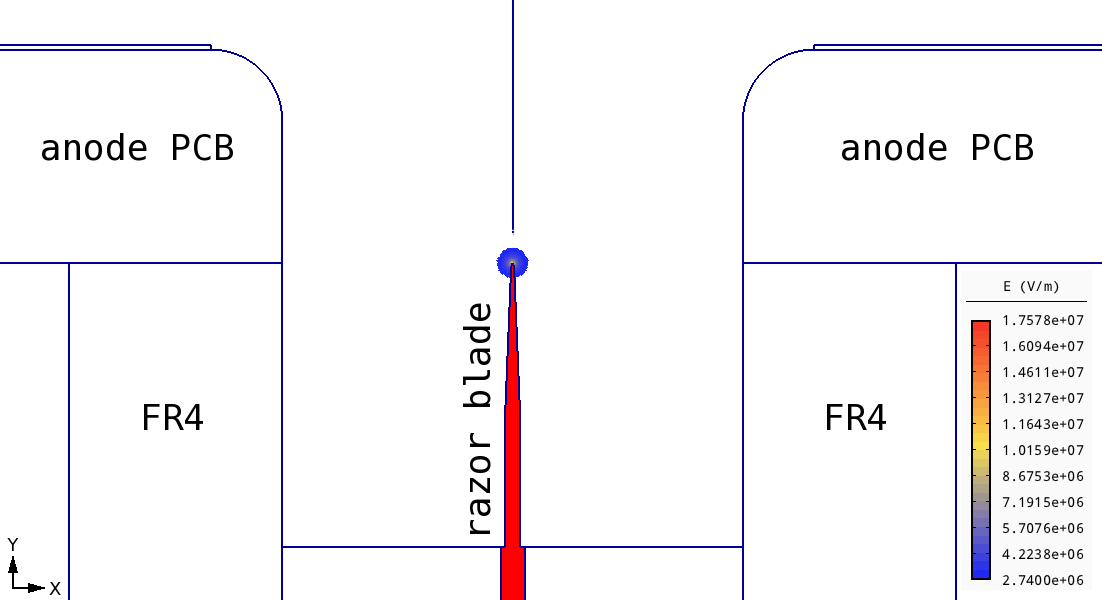}
        \caption[Geometry of amplification region with electric fields]{
            Simulation of amplification region with razor blade.
            Red areas are connected to the amplification voltage source.
            Contours show electric fields above $\unit[2.74\cdot10^6]{V/m}$, that are sufficient for gas amplification.
        }
        \label{fig:amg_region_simu}
    \end{figure}

    A simulation done with \agros, again under the assumption of planar symmetry, shows that the electric field is indeed strong enough for gas amplification.
    Figure~\ref{fig:amg_region_simu} shows the central blade sandwiched between two PCBs for electric contact and two for shielding within a pure argon atmosphere.
    Red areas are put on amplification voltage, here \unit[2.5]{kV}.
    Blue contours at the tip of the blade start at an electric field of $\unit[2.74\cdot10^6]{V/m}$, which is the field strength used for gas amplification in the current T2K readout planes~\citep{Abgrall2011}.
    Higher pressure increases the dielectric permittivity of argon only slightly by about \unit[0.5]{\%} for \unit[11]{bara}~\citep{Schmidt2002}.
    This does not change the electrostatic properties of argon significantly enough to affect the creation of an amplification electric field.

\subsection{Determination of Parasitic Influences}
    From here, it is already possible to extract an estimate for the capacitance of the amplification region.
    \agros calculates the electrostatic energy content $W_e$ of all simulated volumes.
    Utilizing
    \begin{align}
        W_e = \frac{1}{2} C U^2_{\text{ref}}\,,
        \label{eq:estat_energy}
    \end{align}
    one can calculate the capacitance with an assumption for a reference voltage $U_\text{ref}$.
    This voltage is assumed to be $V_{\text{amp}}$ for the volumes below the anode.
    The reference voltage for the upper drift region is not so easily determined.
    A first estimate can be the drift voltage applied to the space directly above the anode.
    Using $V_\text{amp}=\unit[2500]{V}$ and $V_\text{drift}=\unit[-300]{V}$ gives $C_\text{amp.reg.}=\unit[72.3]{pF}$.

\begin{figure}[h]
        \centering
        \includegraphics[width=.7\textwidth]{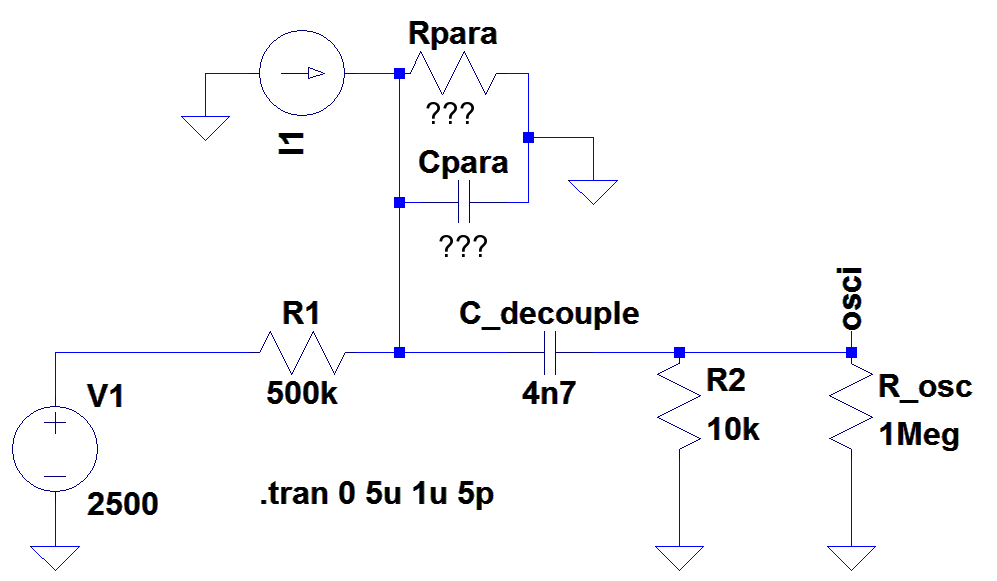}
        \caption[\ltspice schematic used for parasitic determination]{
            The \ltspice schematic used to ascertain the parasitic components introduced in the amplification region.
            \Rpara and \Cpara can be more accurately determined by comparing simulated signals to measured ones.
            As input, a time dependent current simulated with \garfield is used.
        }
        \label{fig:schematic_amp_region}
    \end{figure}

    For a more detailed simulation, one can assume the amplification region to be a current source with a pulsed signal a few \unit[100]{ns} wide and with a parasitic capacitance and resistance, i.e. a \clabel{R} and \clabel{C} element tied to ground, see figure~\ref{fig:schematic_amp_region}.
    In the circuit shown, $C_\text{amp.reg.}$ is equal to \Cpara.
    The values of \Cpara and \Rpara influences the shape of the gas amplified signal.
    For example, choosing $\Cpara=\unit[1]{fF}$, which is unrealistically small, and $\Rpara=\unit[1]{\TOhm}$ makes the signal very fast, i.e. it follows minute fluctuations of the incoming signal.
    Increasing $\tau=\Cpara\Rpara$ gives a signal with sharp turn-on and exponentially decaying turn-off flanks.
    Amplifier-ICs are limited by an IC specific gain-bandwidth-product.
    High amplifications can be achieved, but at a cost of bandwidth, meaning time resolution decreases.
    Knowing reliable parasitic values helps choosing suitable amplifier stages by providing input for electronics simulation such as \ltspice\citep{Engelhardt2016}.

\subsubsection{Experimental Setup for Verification}
    \label{sec:parasitics}
\begin{figure}[h]
        \centering
        \includegraphics[width=.7\linewidth]{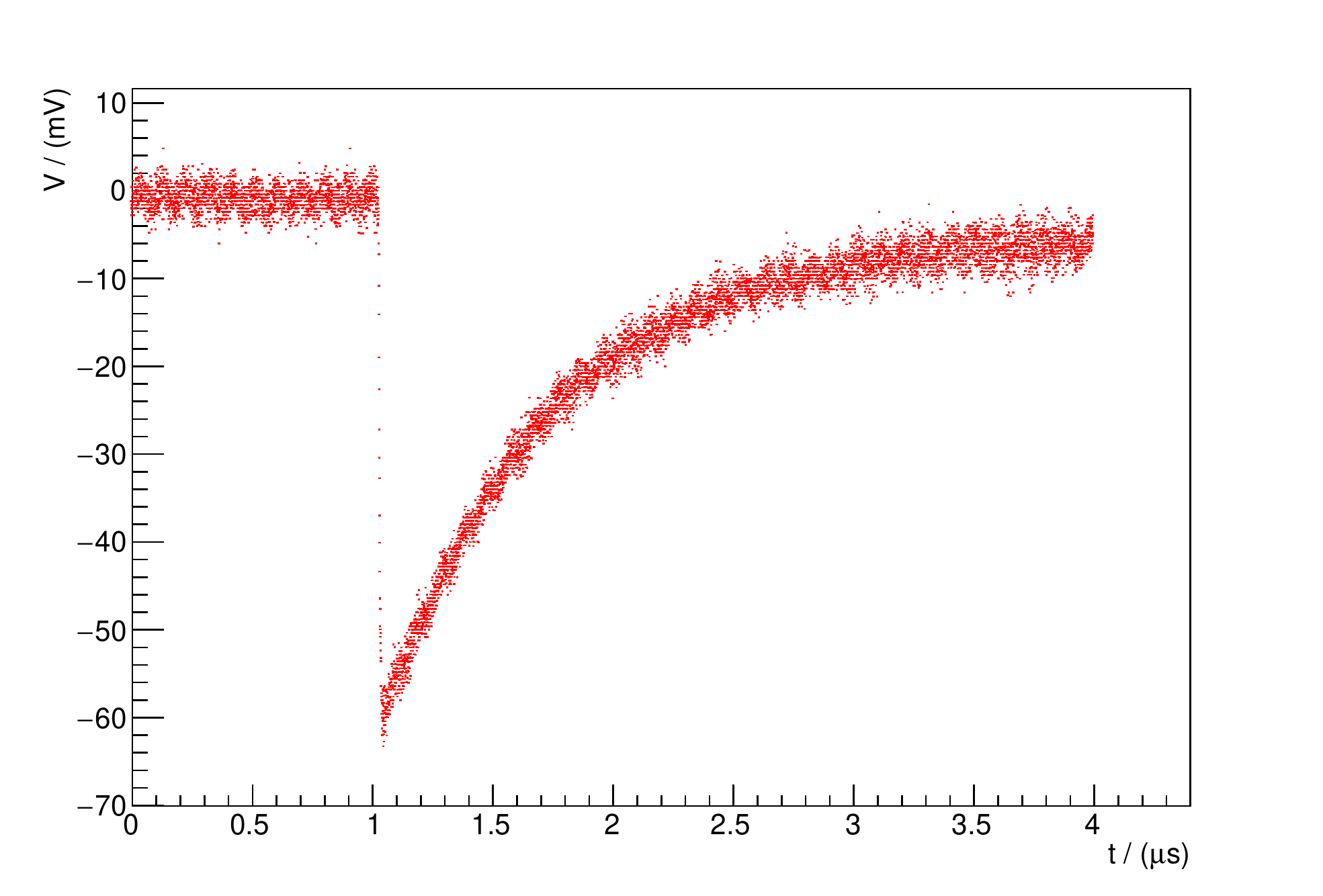}
        \caption[Single waveform measured with mock-up chamber]{
            Gas amplification signal with the razor blade as amplification geometry and a \Fe for creation of free electrons inside the gas.
        }
        \label{fig:osci_example_wf}
    \end{figure}

\begin{figure}[!h]
        \begin{center}
            \subfigure[]{
                \includegraphics[width=.45\linewidth]{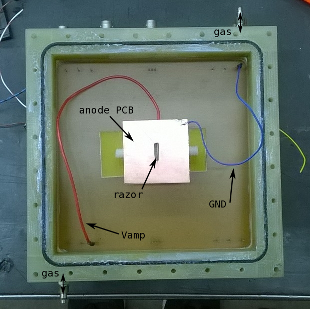}
                \label{fig:open_mockup}
            }
            \subfigure[]{
                \includegraphics[width=.45\linewidth]{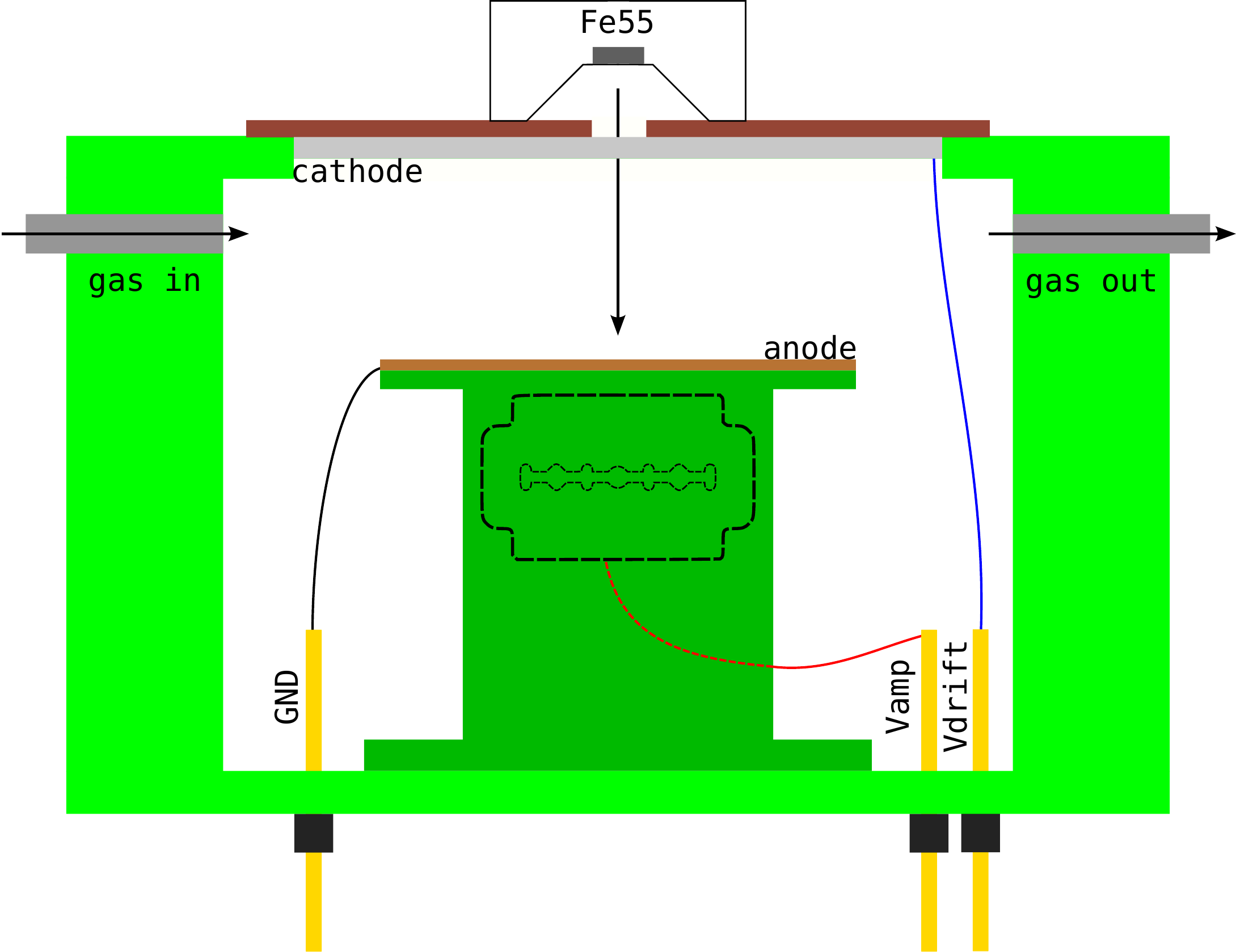}
                \label{fig:mockup_xsec}
            }
        \end{center}
        \caption[Test chamber for amplification region]{
            (a)~Open test chamber with mock-up amplification region in the middle.
            The razor blade is below the slit in the square anode.\\
            (b)~Cross sectional view of the test chamber with the test amplification region using a razor blade.
            The cathode is an aluminized Mylar foil that is electrically contacted for application of a drift voltage.
            }
        \label{fig:mockup_chamber}
    \end{figure}

    Figure~\ref{fig:open_mockup} shows a mock-up amplification region that was build to verify the predictions from \agros simulations.
    It is possible to see signals from electrons produced by a \Fe source placed above the amplification region as shown in figure~\ref{fig:mockup_xsec}.
    The \Fe source is placed directly above the amplification region that sits beneath a section where the chamber wall is made of a thin Mylar foil.
    This is the window where the low energy $\gamma$'s of \Fe pass into the gas.
    As gas, a mixture of Ar:\methane:\carbondioxide 90:7:3 was used with an amplification voltage of $V_\text{amp}=\unit[2500]{V}$ and a drift voltage of $V_\text{drift}=\unit[-300]{V}$.
    Readout was done using an oscilloscope with \unit[1]{M\Ohm} termination, an exemplary waveform is shown in figure~\ref{fig:osci_example_wf}.

\FloatBarrier
\subsubsection{Simulation of the Electron Avalanche Signal}
\begin{figure}[h]
        \centering
        \includegraphics[width=.7\linewidth]{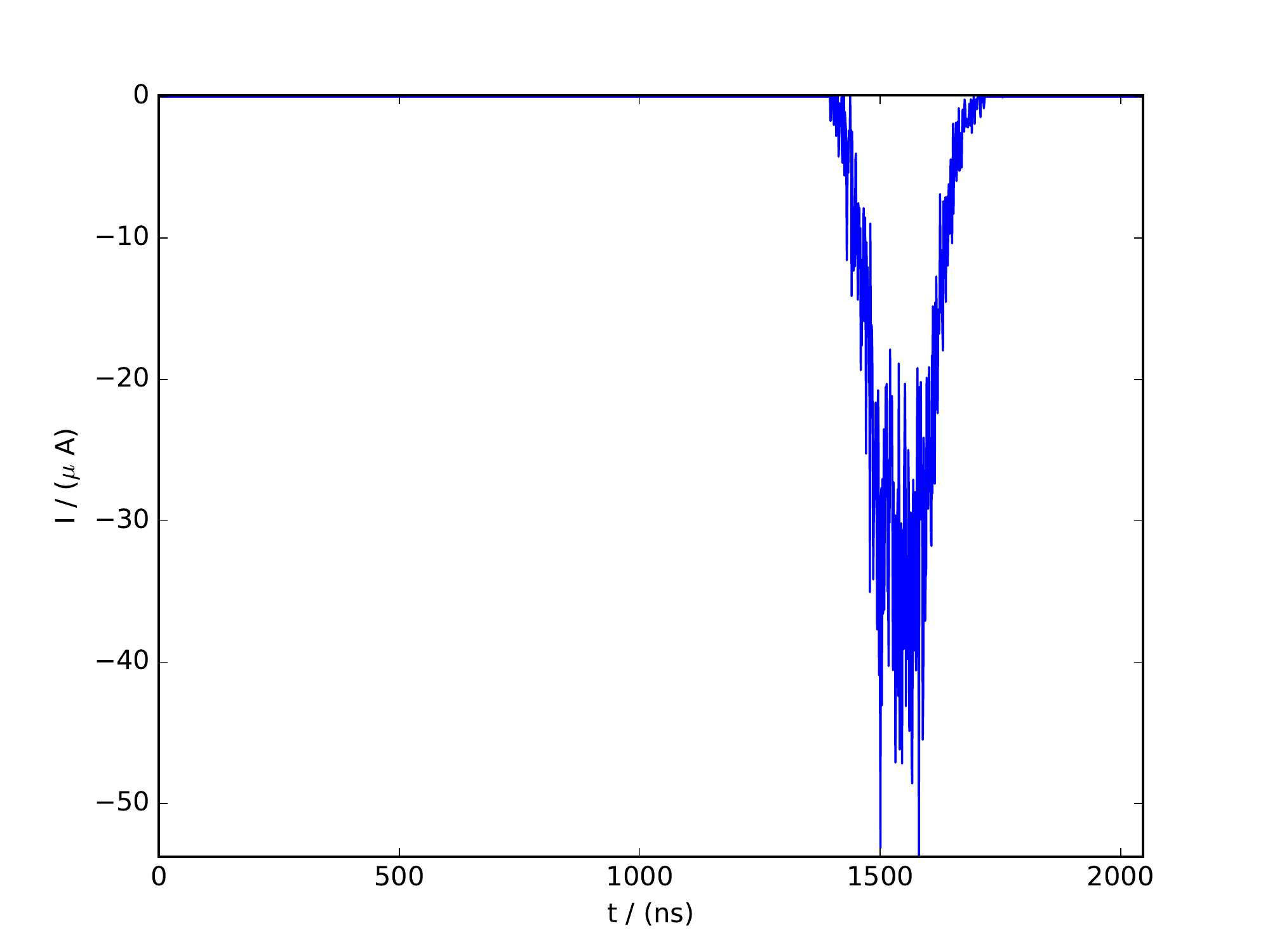}
        \caption[Simulated signal with \garfield]{
            Waveform of a superposition of 220 electrons amplified on a wire using \garfield.
            Electrons were started at $t=\unit[0]{ns}$.
        }
        \label{fig:simulated_signal_on_wire}
    \end{figure}

    \garfield is used to first drift, then amplify single electrons on one wire~\citep{Schindler2017}.
    The razor blade geometry is substituted with a wire for the simulation.
    Figure~\ref{fig:amg_region_simu} shows that the amplification field of the razor is rather circular, so in terms of traversed field gradient shape it is similar to a wire.
    At the start of a simulation run, a single electron is placed at a fixed distance to the wire.
    Since one \Fe $\gamma$ can produce about 220 electrons in argon with its $W_\beta=\unit[26.3]{eV}$, 220 runs are added to form a full signal, see table~\ref{tab:W_values}.
    Such a signal induced on the wire is shown in figure~\ref{fig:simulated_signal_on_wire}.

    The generated signal on the wire is then fed into a \ltspice simulation as a time-dependent current source using the circuit from figure~\ref{fig:schematic_amp_region}.

    The estimation for \Cpara can be made more precise and \Rpara can be determined by a parameter scan with \ltspice.
    In addition to \Rpara and \Cpara as free parameters, a time shift and a global scale parameter were introduced.
    The shift parameter compensates for timing differences in trigger alignment.
    Artificially triggering the smooth, noise free simulation signal is more reliable than a noisy oscilloscope signal.
    Additionally, the measured signal's rise time is shorter than the simulated one's which causes a time difference between the signals.
    A scale parameter accounts for differences in magnitude caused by, for example, the razor blade geometry instead of the simulated wire and gain changes stemming from a deviation in the gas mixture.
    Both scale and shift can not change the decay constant $\tau$ of the signal's tail.
    The scan was performed in predefined ranges and step sizes following table~\ref{tab:raster_ranges_steps}.
    \begin{table}[h]
        \centering
        \caption{
            Raster scan ranges and densities
        }
        \begin{tabular}{lcc}
            Parameter & Range & Step size or steps and spacing \\
            \hline
            \Rpara  & $\unit[1]{\TOhm}\,\dots\unit[10]{\TOhm}$  & \unit[1]{\TOhm} \\
            \Cpara  & $\unit[50]{pF}\dots\unit[90]{pF}$         & \unit[0.5]{pF} \\
            scale   & $0.01\dots3.16$                           & 150, logarithmic \\
            shift   & $\unit[-100]{ns}\dots\unit[+100]{ns}$     & 50, linear \\
            \hline
        \end{tabular}
        \label{tab:raster_ranges_steps}
    \end{table}

\begin{figure}[h]
        \centering
        \includegraphics[width=.7\textwidth]{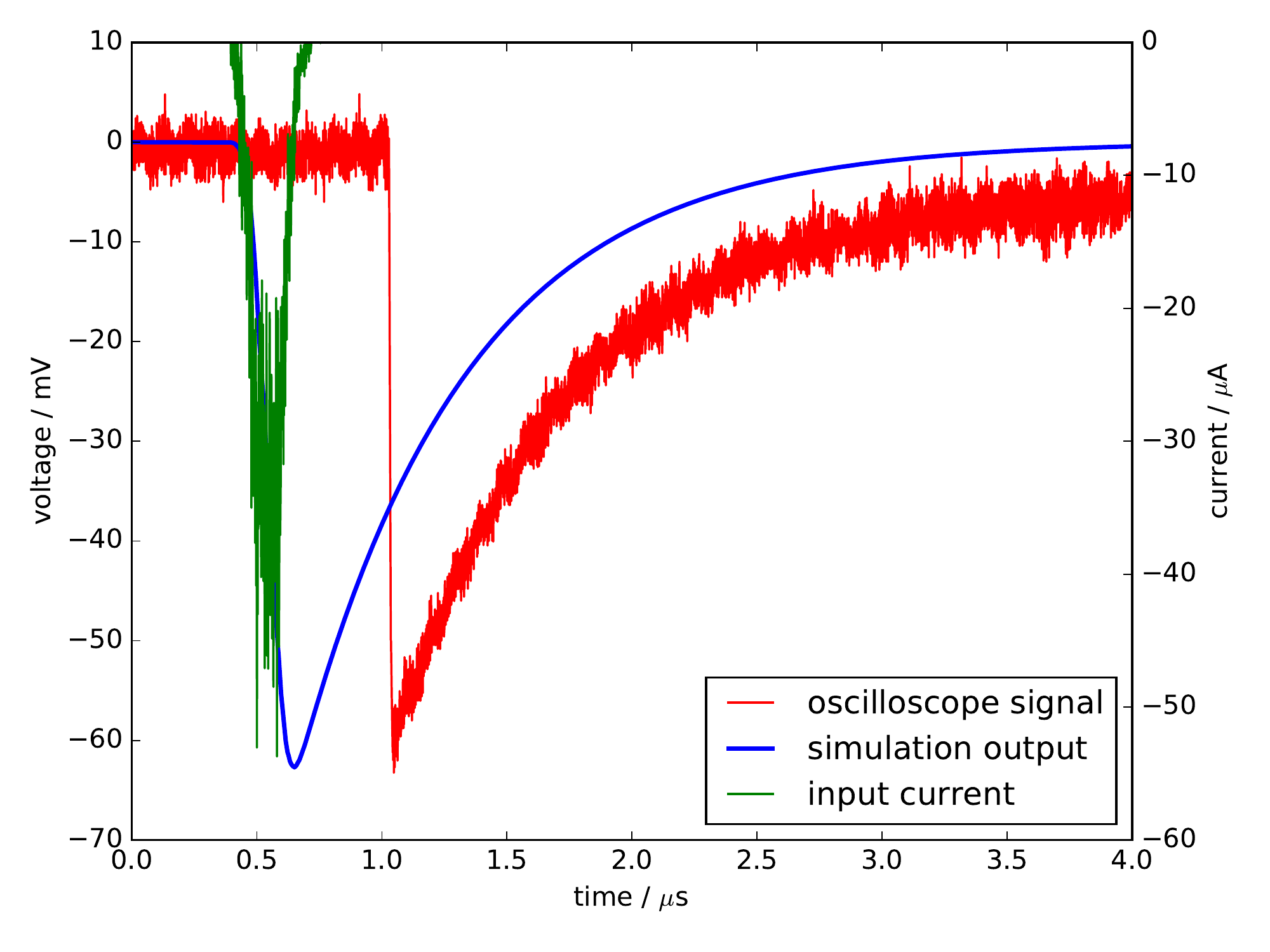}
        \caption[Comparison of simulated signal with a measures curve]{
            The simulated current from \garfield is converted to a voltage signal in \ltspice.
            The resulting response over a \unit[1]{M\Ohm} resistor can be compared to a measured signal to determine parasitic influences.
        }
        \label{fig:singla_sim_compare}
    \end{figure}

    Simulated and measured curves are matched using a $\chi^2$ score calculated by
    \begin{align}
        \chi^2 = \sum^{}_{i} \frac{\left(V_\text{sim., i}-V_\text{meas., i}\right)^2}{\sigma_i^2} = \sum^{}_{i} \frac{\left(V_\text{sim., i}-V_\text{meas., i}\right)^2}{\sigma_\text{baseline}^2}
        \label{eq:def_chi2}
    \end{align}
    with the standard deviation of the baseline as constant uncertainty on every sampling point of the measured data.
    The best matching combination of free parameters is collected and stored with the corresponding $\chi^2$ value and degrees of freedom ndf.
    Figure~\ref{fig:singla_sim_compare} shows the simulated current and response voltage drop along with a measured signal.

\begin{figure}[h]
        \centering
        \includegraphics[width=.7\textwidth]{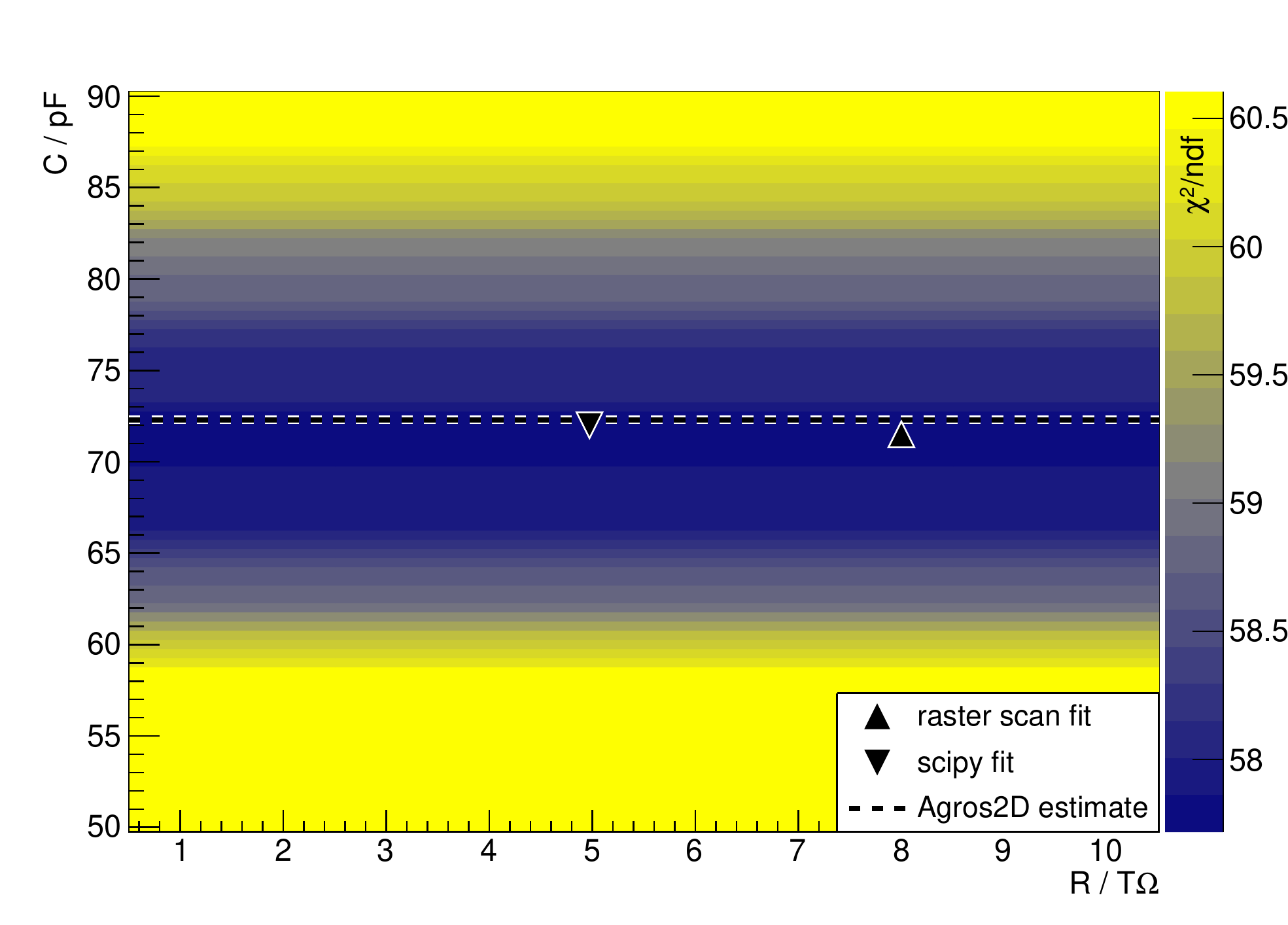}
        \caption[$\chi^2$ distribution for a single waveform]{
            Computed $\chi^2$ distribution for predefined parameters of a single waveform with \Cpara plotted versus \Rpara.
            \Rpara has no distinct shape whereas \Cpara shows a prominent valley shape.
            Also shown are the best fitting point for a \scipy fit and the \Cpara estimate from \agros.
        }
        \label{fig:single_scan}
    \end{figure}

    Usually, one expects to find a gradient towards the best matching combination of the free parameters in the $\chi^2$ distribution.
    It is evident from figure~\ref{fig:single_scan}, that the parasitic resistance is not dominating the $\chi^2$ distribution's shape.
    \Rpara provides a path to ground, but \clabel{R1} is much smaller and the series resistance of the ideal voltage source \clabel{V1} is zero.
    They are the main path through which a current can reach ground.
    Including \Rpara in the simulation gives the possibility to factor in leakage currents in the order of \unit[1]{nA}, which are present in the real system.
    Excluding it leads to increased $\chi^2$ values, but retains the shape along \Cpara.
    A parasitic resistor value of $\Rpara=\unit[5]{\TOhm}$ is used hereafter.

    As for the parasitic capacitance, a distinct valley shaped $\chi^2$ distribution is found.
    Sampling to higher and lower values for \Cpara shows a continuation of this trend.
    The best match between measurement and simulation is found with $\Cpara=\unit[71.5]{pF}$, $\Rpara=\unit[8]{\TOhm}$, $\text{shift}=\unit[-2]{ns}$ and $\text{scale}=\unit[0.22]{}$ with a \mbox{$\chi^2/\text{ndf}\approx57.7$}.
    For the second best match, the same values are found with the exception of $\Rpara=\unit[4]{\TOhm}$ and a $\chi^2/\text{ndf}$ that is only 0.001 larger.
    The smaller-than-one scale factor shows, that the signal is overestimated by the simulation chain.
    Electrons incident on a wire with sufficient voltage for amplification will always be amplified.
    On a razor, where amplification fields are only present at the tip, there exist paths for electrons onto the blade where the field gradient is not sufficient for amplification.
    The large $\chi^2$ comes from shape differences that can not be produced by the assumed parasitic influences, e.g. the different behaviour of the falling flank of the measured and simulated signals.

    The form of $\chi^2$ along \Cpara already suggests that available fitting routines can be used to determine parameters without the restriction of a predefined grid, e.g. the curve fitting routines implemented in the scientific Python library \scipy as \verb|optimize.curve_fit|~\citep{Jones2001}.
    They also use a $\chi^2$ minimization goal as defined by equation~\ref{eq:def_chi2}.
    The initial parameters used were $\Rpara=\unit[5]{\TOhm}$, $\Cpara=\unit[72]{pF}$, $\text{scale}=1.0$ and $\text{shift}=\unit[0]{ns}$ as estimated by \agros and the grid scan.
    The \scipy fitting yields \mbox{$\Rpara\approx\unit[4.97]{\TOhm}$}, \mbox{$\Cpara\approx\unit[72.00]{pF}$}, \mbox{$\text{scale}\approx0.13$} and \mbox{$\text{shift}\approx\unit[-0.11]{ns}$}.
    It was not possible to determine parameter uncertainties because the covariance matrix computed did not have a full rank, in which case the uncertainties are returned to be infinite.
    Fitting with either $\Rpara\overset{!}{=}\unit[5]{\TOhm}$, $\text{shift}\overset{!}{=}\unit[0]{ns}$ or $\text{scale}\overset{!}{=}\unit[1]{}$ reproduces best fit parameters that deviate little from the full-fit \Cpara and \Rpara, albeit with larger $\chi^2_{min}$ values.
    The deviation from the grid search of parameters lies with the relatively flat $\chi^2$ distribution in \Rpara where the fitting algorithm can easily get stuck in local minima.
    Nevertheless, both methods are in agreement about the general value of the parasitic influences.
    More precise values would not lead to further improvements concerning the design of a suitable amplifier for the gas amplification region any more.

\begin{figure}[h]
        \centering
        \includegraphics[width=.7\textwidth]{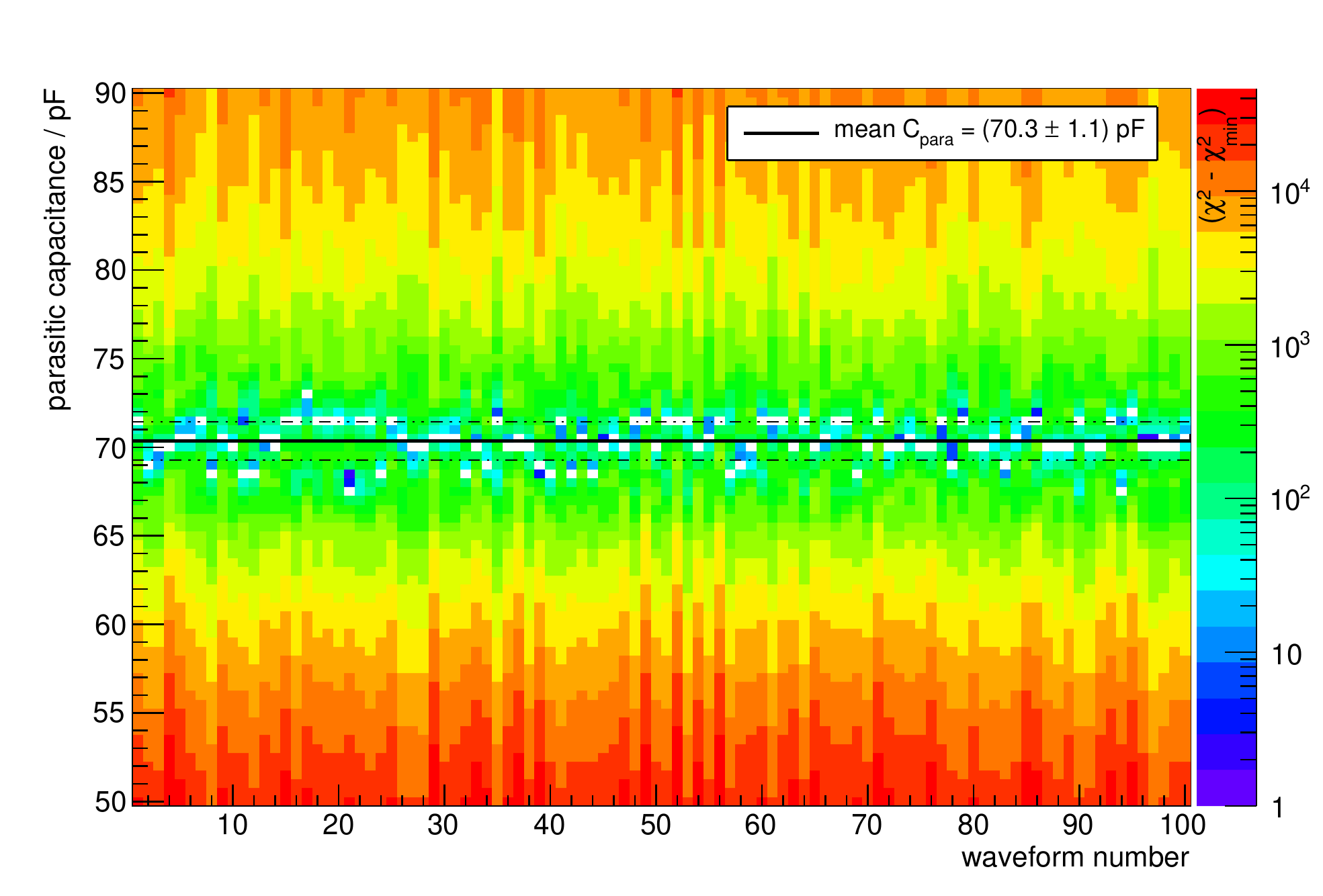}
        \caption[Matching 100 additional waveforms]{
            Redoing the matching procedure for 100 additionally recorded waveforms.
            The shape of the $\chi^2$ distribution is very similar and only differs in an individual offset value $\chi^2_{\text{min}}$.
            The mean of the individually matched \Cpara with standard deviation of the means is taken as final value and uncertainty of $\Cpara=\unit[(70.3\pm1.1)]{pF}$.
            White bins are the minimum for a single waveform.
        }
        \label{fig:stability_check}
    \end{figure}

    The reproducibility is checked by taking 100 waveforms and rerunning the grid search for the best matching parameters for every individual waveform.
    Figure~\ref{fig:stability_check} shows the $\chi^2_\text{i}-\chi^2_\text{i, min}$ values for varying $\Cpara_\text{i}$ where every column is a new waveform.
    Between the waveforms, the value of $\chi^2_\text{i, min}$ differs, so for comparison only the difference is plotted for every waveform.
    The final $\Cpara=\unit[(70.3\pm1.1)]{pF}$ is determined by taking the mean of the individual best matching $\Cpara_\text{i}$ with uncertainties calculated from the variance of $\Cpara_\text{i}$.

\FloatBarrier
\subsection{An Amplifier for the Electron Avalanche Signal}
    With the information gathered in the previous sections on the shape of the signal and the parasitic influences, it is now possible to set up a simulation in \ltspice, that also includes the parametrized amplification region from figure~\ref{fig:schematic_amp_region}.
    The input current simulated with \garfield is scaled down with the ascertained scaling factor of \unit[0.22]{}.

\begin{figure}[h]
        \centering
        \includegraphics[width=\textwidth]{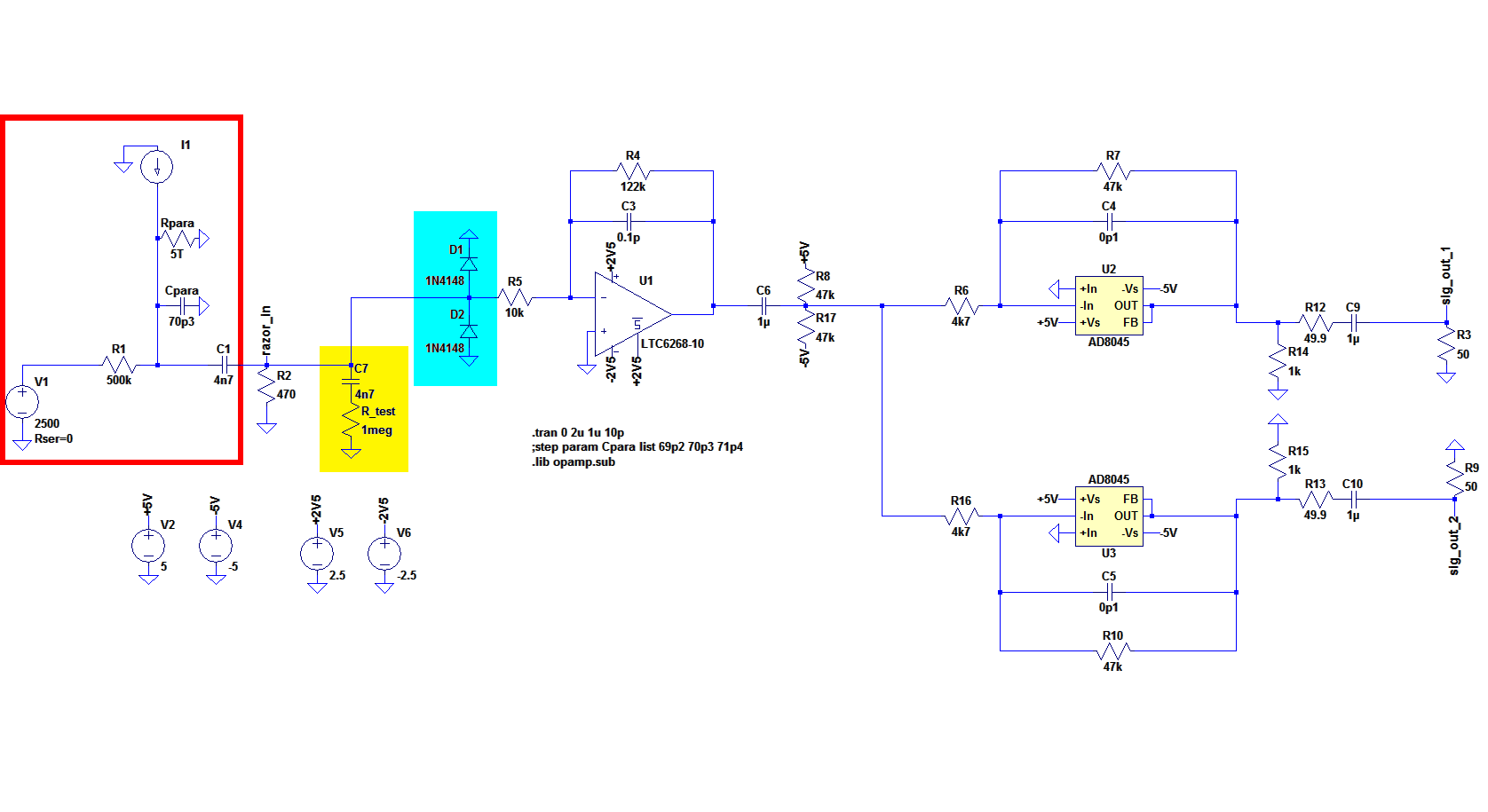}
        \caption[Amplification for electron avalanche induced signal]{
            Amplification circuit for the avalanche induced signal.
            Red marked is the high voltage system.
            In yellow the test input and in blue the input protection diodes are marked.
        }
        \label{fig:circuit_razor_amp}
    \end{figure}

    The amplifying stages were selected to match the requirements (e.g. timing, gain) of the signals from gas amplifications.
    Figure~\ref{fig:circuit_razor_amp} shows the resulting circuit with three highlighted areas.
    In red is the high voltage sub-circuit with \clabel{C1} the HV decoupling capacitance.
    The yellow area shows a point where a test input signal can be injected, here terminated with a \unit[1]{M\Ohm} resistor.
    For input protection, two anti-parallel diodes are connected to ground in the blue marked region.
    They limit signals to $\unit[\pm700]{mV}$.
    This is a signal height that far exceeds the expected, unamplified signals during regular operation.
    After the first amplification stage, the signal is duplicated for output of two identical signals.
    One signal path could be used for triggering while the other serves as input for a flash ADC board.

\begin{figure}[h]
        \centering
        \includegraphics[width=\textwidth]{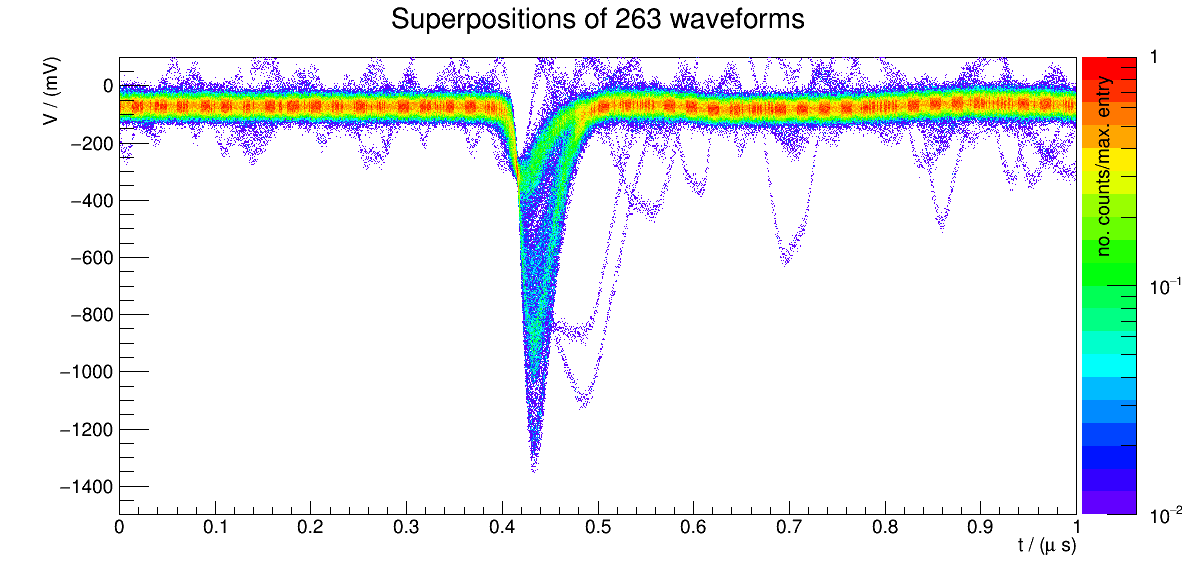}
        \caption[Multiple waveforms taken with the developed amplifier]{
            Normalized superposition of a few hundred waveforms measured with the new built amplifier circuit.
            The mock-up setup from figure~\ref{fig:mockup_xsec} was used again.
        }
        \label{fig:razoramp_signals}
    \end{figure}

    The built amplifier circuit was tested in the mock-up chamber from figure~\ref{fig:mockup_chamber}.
    A total of 263 waveforms were taken with a amplification voltage of only $V_\text{amp}=\unit[1000]{V}$ and a drift voltage of $V_\text{drift}=\unit[-300]{V}$.
    In figure~\ref{fig:razoramp_signals}, all measured signals are shown together.
    The newly designed amplifier is very well suited for picking up small signals generated by amplification voltages less than half of those used before in section~\ref{sec:parasitics}.
    Some noise can be seen left and right of the signal pulse.
    Those events are likely to disappear once the amplifier moves to the inside of the \hpgmc with better shielding and improved signal paths.
    When using signals from a signal generator with shielded cables, no saturated pulses were observed, see figure~\ref{fig:razoramp_signals_generator} for these waveforms.

\FloatBarrier
\section{Implementation of a Triggering System}
\begin{figure}[h]
        \centering
        \includegraphics[width=.7\textwidth]{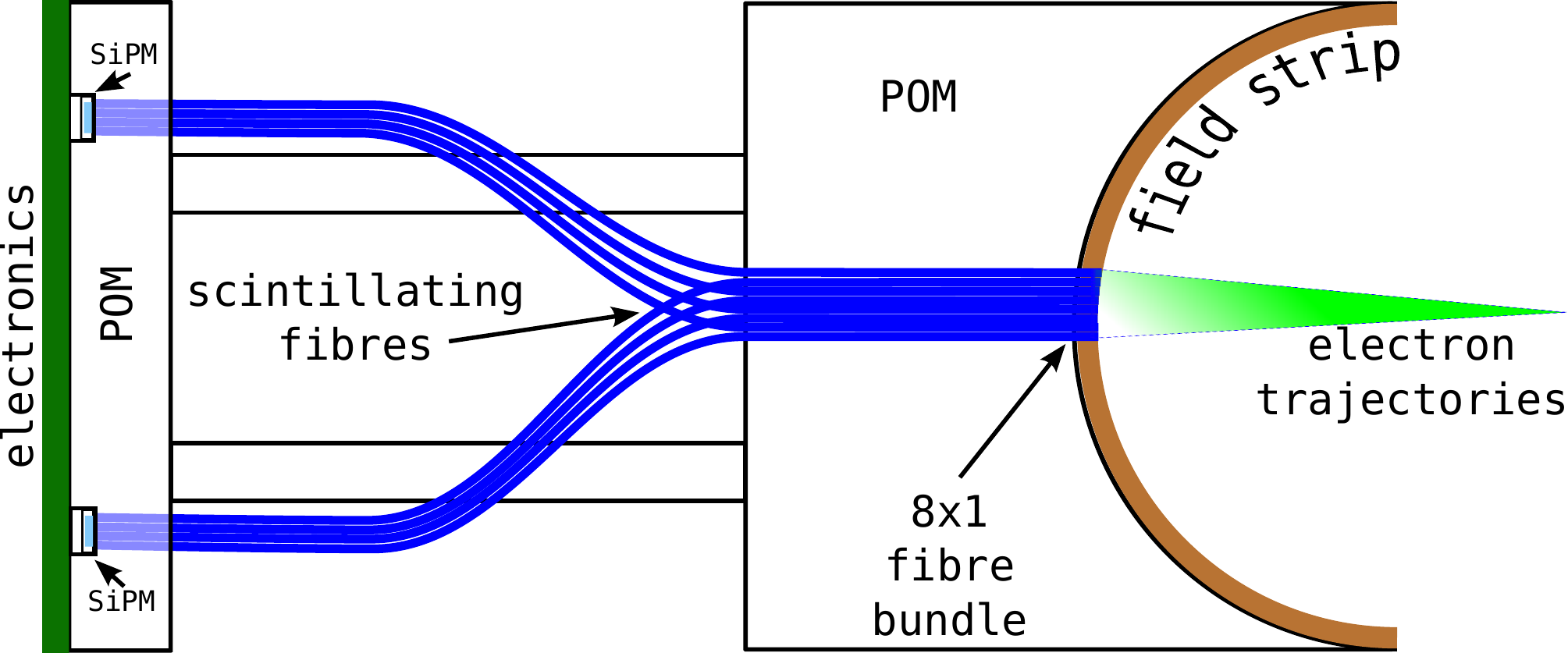}
        \caption[Sketch of the trigger setup]{
            The triggering setup consists of eight fibres mapped in pairs of four onto two SiPMs.
            Interleaving the fibres enables one to use two SiPMs to trigger in coincidence.
            Shown is only one of the two trigger positions.
            The second trigger position would be located below the drawing into the plane.
        }
        \label{fig:trigger_side_scetch}
    \end{figure}

    Timing between the creation of secondary electrons in the drift space and arming the electronics at the gas amplification region is done by a triggering system.
    It lies opposite of the \Sr source holders and consists of four SiPMs and two scintillating fibre bundles of eight fibres each.
    These $8\times1$ fibres are split into bundles of four by mapping them in an alternating pattern to the two SiPMs, see figure~\ref{fig:trigger_side_scetch}.
    Pictures of the machined parts that are described in this section can be found in section~\ref{apx:trigger}.

\subsection{Noise Reduction and Testing}
    SiPMs are able to detect single photons with discrete peaks in an output spectrum for integer multiples of single photons, i.e. photon counting capability.
    However, a dark count rate of $\unit[30]{\text{kcps}/\text{mm}^2}$, with a tail extending to the higher photon count region, introduces the need for a higher threshold to separate noise from signal~\citep{SensLCSeries}.
    Lowering the value for this threshold can be achieved by setting up two SiPMs in coincidence.
    The statistically independent noise signals are unlikely to occur in the same time window.

    However, electrons reaching the scintillating fibres are quickly stopped and are thus unlikely to deposit energy in more than one fibre.
    The photons from a single illuminated fibre need to have a measurable crosstalk with adjacent fibres for the coincidence setup to improve the signal yield.

\begin{figure}[h]
        \centering
        \includegraphics[width=.7\textwidth]{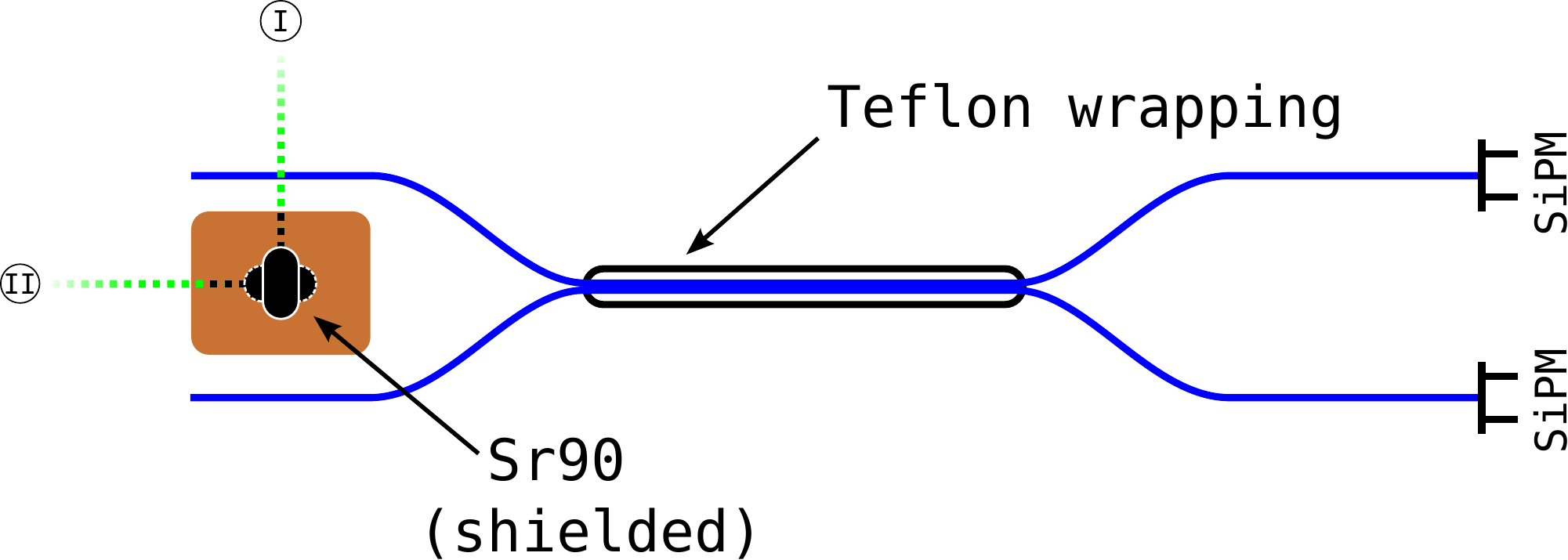}
        \caption[Crosstalk testing experiment]{
            Setup to determine the effectiveness of a crosstalk-based coincidence system for triggering the \hpgmc.
            The \Sr source either irradiates one fibre or none.
        }
        \label{fig:crosstalk_test}
    \end{figure}

    Figure~\ref{fig:crosstalk_test} shows an experimental setup to test whether the amount of optical crosstalk is sufficient for triggering in coincidence.
    Turning the source holder towards position I irradiates only one of the two fibres and position II points away from both fibres.
    This gives an no-signal-hypothesis to test the effectiveness of the source's shielding.
    Accidental count rates are measured by removing the source entirely.

    Count rates are measured over a \unit[10]{s} interval with a coincidence window of \mbox{$\Delta t\approx\unit[11]{ns}$} with a threshold of $V_\text{tresh}=\unit[-130]{\text{mV}}$.
    With this threshold, but without coincidence triggering, the readout electronics still registered events with an $\mathcal{O}(\unit[10]{kHz})$ rate due to the high activity and the close proximity of the \Sr source.
    Listed in table~\ref{tab:cps_trg_test} are the results of the three measurements with the test setup placed in a box for background light reduction.
    There is an increase in the count rates by placing the source in position II compared to no source in the setup, however the count rates for position I are still significantly higher.
    Adding optical grease between fibre ends and SiPMs is meant to reduce light loss and thus increases the count rates in this experiment.
    Sanding the sides of the scintillating fibres that are in contact, to remove the cladding and introduce a diffuse light transfer, showed no further increase in crosstalk.
    \begin{table}
        \centering
        \caption{
            Count rates measured for different source alignments
        }
        \begin{tabular}{l|c}
            configuration & counts \\
            \hline
            \hline
            no source                   & $\unit[(0.6\pm0.3)]{cps}$ \\
            position II                 & $\unit[(3.8\pm0.5)]{cps}$ \\
            position I                  & $\unit[(15.4\pm0.8)]{cps}$ \\
            \quad add optical grease   & $\unit[(23.0\pm0.8)]{cps}$ \\
            \quad add sanded sides     & $\unit[(22\pm1)]{cps}$ \\
            \hline
        \end{tabular}
        \label{tab:cps_trg_test}
    \end{table}

    With the final trigger electronics, another two channels are added.
    As can be seen from figure~\ref{fig:trigger_side_scetch}, the scintillating fibres lie next to each other without optical shielding for some distance.
    If there is too much crosstalk between the upper and lower fibre bundles, it could become impossible to determine whether a signal recorded at the razor blade belongs to the near or far drift distance.
    This cross-talk can be measured by placing a source in front of only one of the $8\times1$ fibre windows in the POM blocks.
    The irradiated site is triggered in coincidence and then a coincidence on the other side's channels is searched.
    The so measured fake double trigger probability was very low with a $V_\text{tresh}=\unit[-30]{\text{mV}}$ threshold:
    \begin{align}
        \frac{
                \overbrace{(\text{CH3} \land \text{CH4}) \land (\text{CH1} \land \text{CH2})}^{\unit[1]{\mu s}}
             }{(%
                \underbrace{\text{CH3} \land \text{CH4}}_{\Delta t = \unit[11]{ns}}%
             )} = \unit[2]{\permil}\,.
    \end{align}
    The coincidence time window between two channels is again $\Delta t=\unit[11]{ns}$.
    The cross-talk coincidence between near (e.g. $(\text{CH3} \land \text{CH4}))$ and far (then $(\text{CH1} \land \text{CH2})$) sides are observed within a \unit[1024]{ns} window.
    This is the maximal recordable length for our readout and thus the most conservative estimation possible here.
    Moving the threshold closer to the baseline, i.e. $V_\text{tresh}=\unit[-11]{\text{mV}}$, increases the accidental double-coincidence probability to nearly \unit[7]{\%}.
    With a less conservative four-fold coincidence window of \unit[100]{ns}, the fake double trigger probability is less than 1 in $10^6$.

\FloatBarrier
\subsection{Trigger Electronics}
    The electronics for the trigger system are divided into an part inside the gas and a part outside.
    The connection between inside and outside is realized with a multi-wire-gland feedthrough of 12 cables on the side's blind flange.
    Every SiPM needs to be supplied with an individual bias voltage of around $\unit[(24.45\pm0.25)]{V}$ plus over voltage of \unit[1]{V} to \unit[5]{V}.
    A temperature drift of $\unit[21.5]{mV/\degree C}$ for the bias voltage has also to be taken into account.
    In addition, the gain also has a temperature dependence of $\unit[-0.8]{\%/\degree C}$~\citep{SensLCSeries}.

\begin{figure}[h]
        \centering
        \includegraphics[width=.7\textwidth]{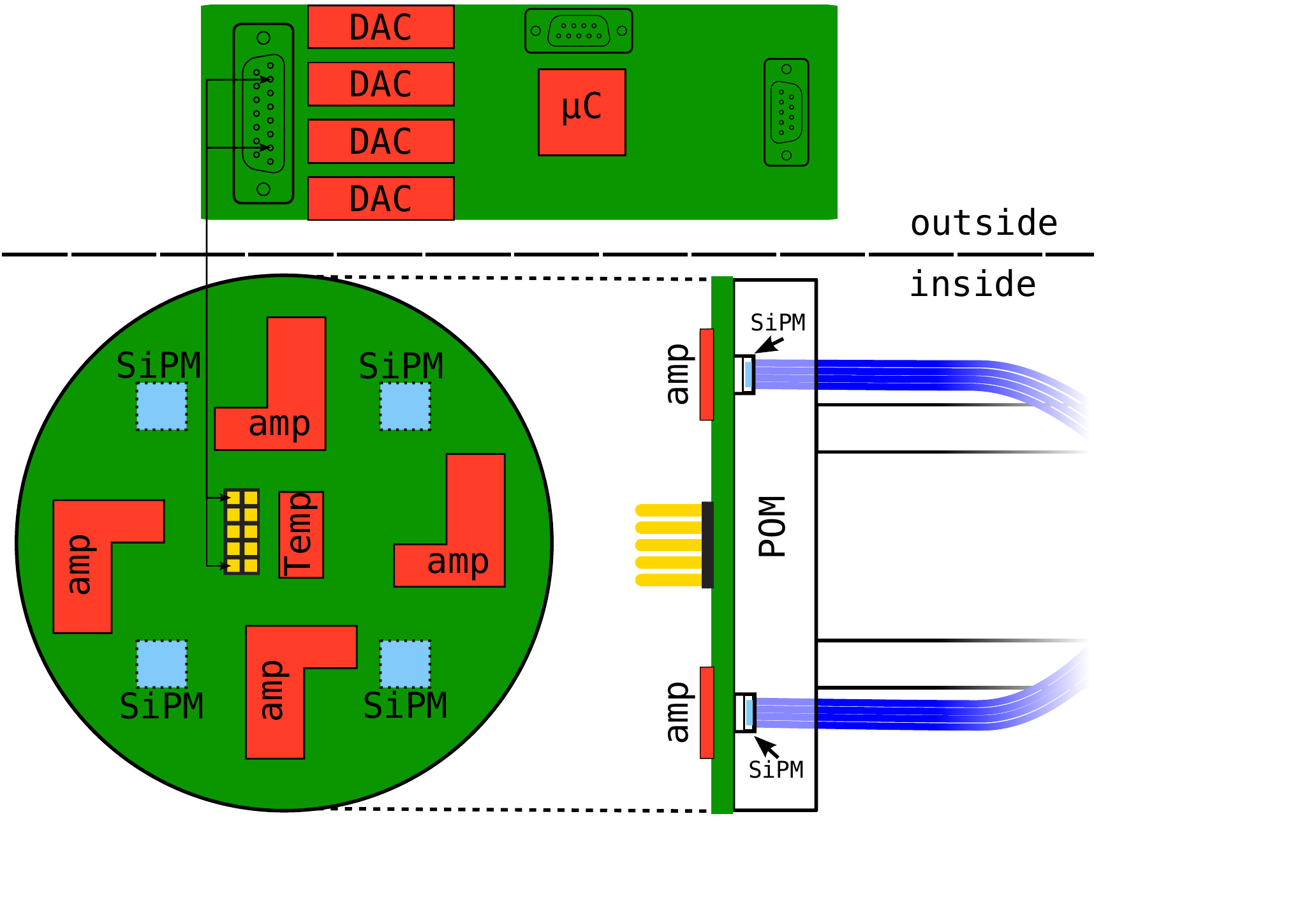}
        \caption[Overview of the trigger electronics]{
            Overview of the electronics used for generating the trigger signal.
            All digital parts are kept outside for noise reduction.
            See also figure~\ref{fig:trigger_side_scetch}.
        }
        \label{fig:trigger_electronics}
    \end{figure}

    A system that provides an individual voltage regulation specifically meant for SiPMs in detectors was adapted from~\citep{Weinstock2014}.
    SiPMs are constantly supplied with a higher-than-needed voltage a priori, but the reference potential is supplied by a programmable digital to analog converter (DAC) with an offset voltage of \unit[0]{V} to \unit[5]{V}.
    The advantage of this kind of circuit is, that every SiPM can be individually regulated for temperature drift, gain or even shut off by undercutting the minimum bias voltage.
    To increase resolution of the offset voltage, the full range of \unit[5]{V} is scaled down to \unit[3.5]{V} for every DAC.
    Common ground is given through a contact to the pressure vessel itself.
    A temperature sensitive IC sits on the gas-side PCB where it measures the temperature directly where the SiPMs are located for temperature stabilisation of the SiPMs, see figure~\ref{fig:trigger_electronics}.

    Only the amplifiers for the SiPMs, the SiPMs and the temperature sensor are located on the inside and the control electronics on the outside.
    An advantage of this is, that the outside electronics can be exchanged very easily.
    Furthermore, the inside is kept free from electro-magnetic (EM) radiation produced by active components such as the microcontroller that controls the DACs.
    Any EM signals are picked up by the amplification razor like an antenna, overlaying the output signal from gas amplification thus decreasing signal to noise ratio.

\FloatBarrier
\section{The Internals of the assembled \hpgmc}
\begin{figure}[h]
        \centering
        \includegraphics[width=\textwidth]{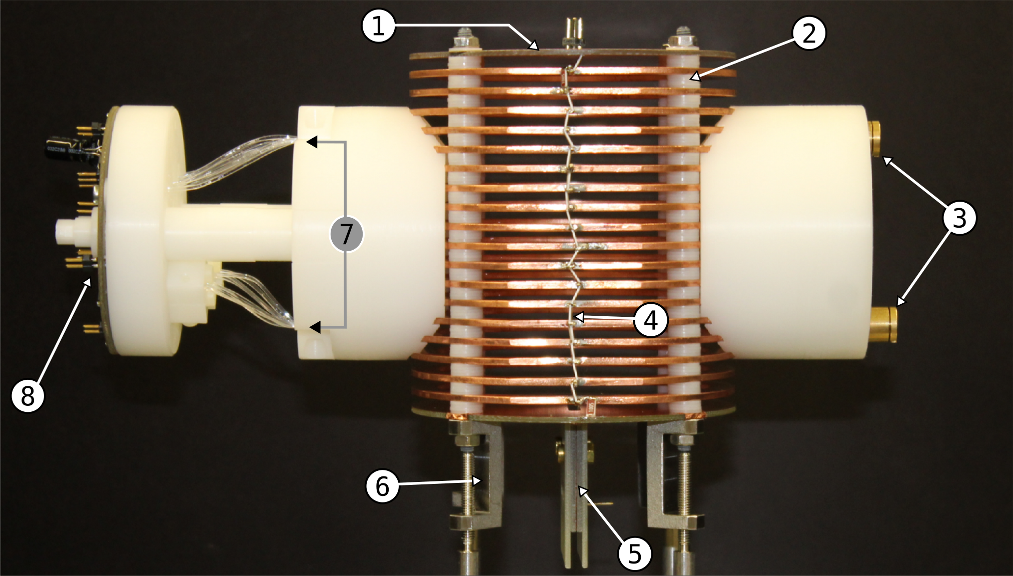}
        \caption[Fully assembled field cage with trigger and \Sr modules attached]{
            Fully assembled field cage with trigger and \Sr modules attached.
            1:~cathode, 2:~gas~gap~spacer, 3:~\Sr~capsules, 4:~HV~resistor, 5:~razor, 6:~cage~feet,~adjustable, 7:~scint.~fibres, 8:~trigger~electronics
        }
        \label{fig:the_heart}
        ~\\
        \includegraphics[angle=90, width=\textwidth]{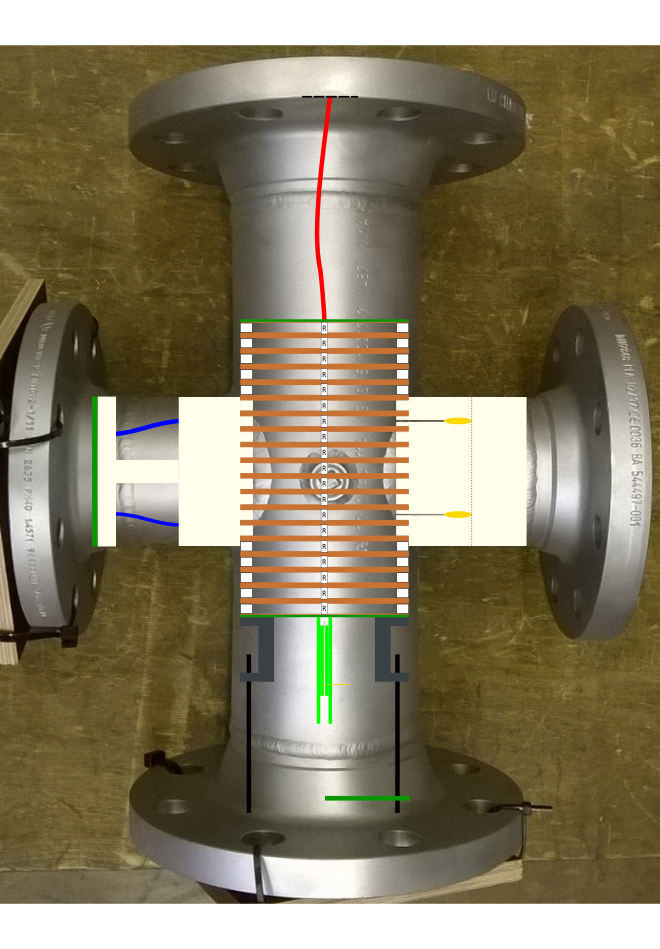}
        \caption[Location of components inside pressure vessel]{
            At the crossing of the double-tee, the field cage is located.
            Attached to it and protruding into the reduces tees are \Sr- and trigger-blocks.
        }
        \label{fig:xray_vessel}
    \end{figure}

    Figure~\ref{fig:the_heart} shows a side view of the field cage with attached source holder on the right and trigger system on the left.
    The cross-shape of the cage with the two POM pieces is assembled inside the chamber, see figure~\ref{fig:xray_vessel}.
    For assembly, the field cage is inserted into the piping through a DN100 side first.
    After the cathode is connected, the cage can be brought into the final position.
    Through the DN80 sides, it is now possible to wedge the POM pieces into place.
    All four sides can be accessed independently by removing the corresponding flanges.

    \paragraph{Trigger system and source holder} are kept in place purely by friction of a comb structure that is wedged between the field rings.
    This also guarantees a predictable and repeatable alignment of both sides.
    The outlets on the \Sr side are \unit[1]{mm} diameter bores that point across the drift space towards a $\unit[8\times1]{mm}$ horizontal line of scintillating fibres, effectively forming a collimator.
    The uncertainty on the drift distance is thus \unit[1]{mm}.

    \paragraph{Breakdown tests} were done to determine the final, experimentally verified, maximum voltage.
    Over a period of at least \unit[24]{h} no breakdown has been observed with a cathode voltage of \unit[-23]{kV} with ambient pressure and air.
    Higher voltages do not necessarily cause breakdown immediately, e.g. with \unit[-24]{kV}, breakdown occurred between \unit[30]{min} and \unit[1]{h} for repeated testing.
    Testing without a field cage, i.e. a free cathode connector, shows breakdown at similar voltages, so a more conservatively insulated HV feedthrough could increase the maximum cathode voltage.

    \paragraph{The field cage} has a total series resistance of $R_\text{tot}=19\cdot\unit[10]{M\Ohm}=\unit[190]{M\Ohm}$ from cathode to ground, so a current of at least \unit[160]{$\mu$A} needs to be provided by the HV source at maximum voltage.
    Dark currents draining through parasitic influences need to be added on top of this value.
    During breakdown testing, the current was about \unit[100]{$\mu$A} above the current expectation solely from the resistor chain path.
    A collection of the measured single resistor values can be found in figure~\ref{fig:single_R_values}.
    The homogeneity of the voltage steps is at an accuracy level of \unit[1]{\%}.

\FloatBarrier
\section{Auxiliary Gas-System}
\begin{figure}[h]
        \centering
        \includegraphics[width=\textwidth]{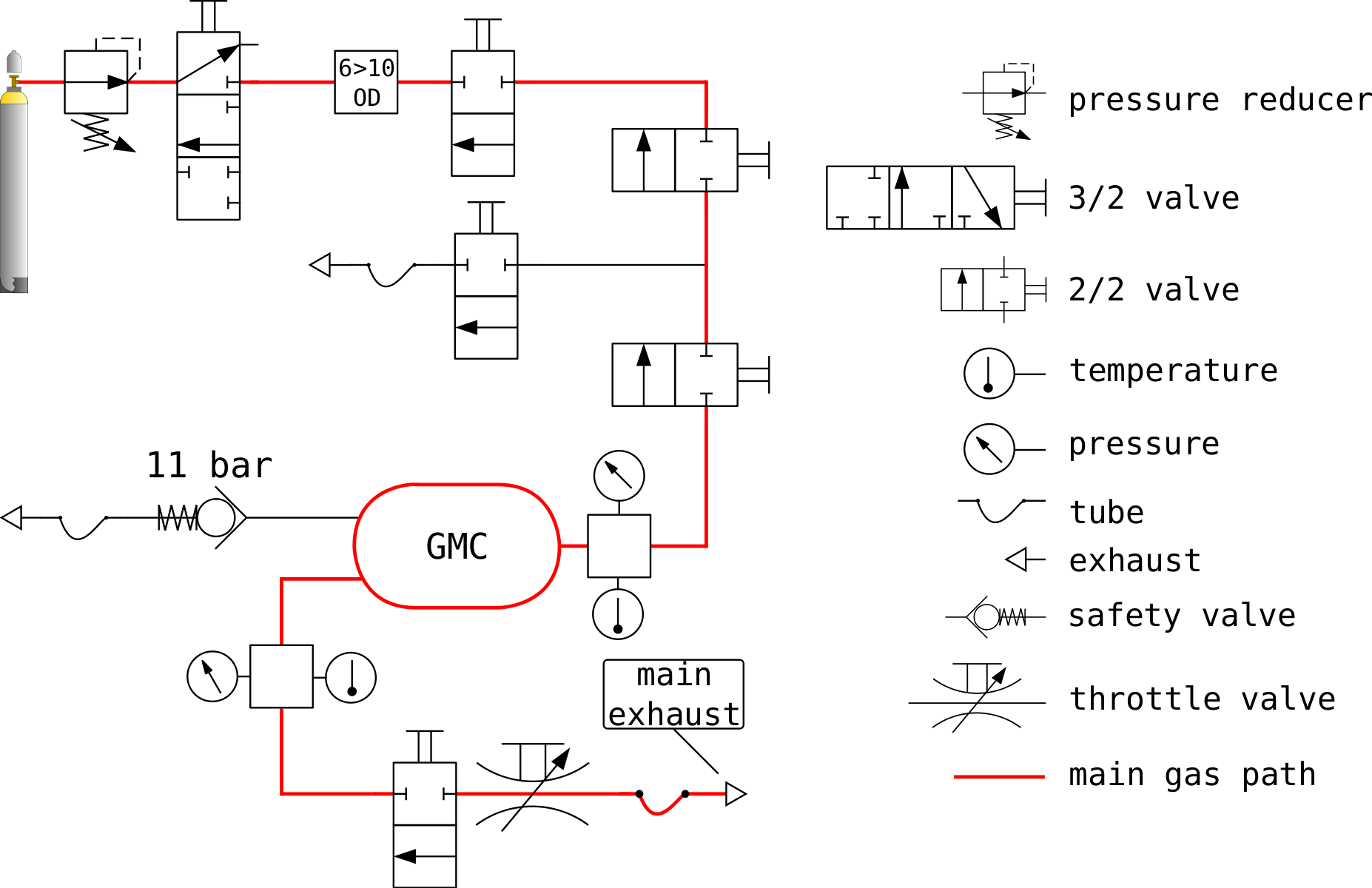}
        \caption[Auxiliary gas system supplying the \hpgmc]{
            Schematic of the supply line for the \hpgmc.
            Subsections can be closed off and modified without depressurizing the chamber.
        }
        \label{fig:gas_system}
    \end{figure}

    The gas supply of the \hpgmc comes directly from a pressurized bottle outside the laboratory.
    A pressure regulator adjusts the pressure to the desired value before passing the gas on into the laboratory.
    In figure~\ref{fig:gas_system} the full gas-system is shown.
    Temperature and pressure of inflowing and outflowing gas of the chamber is measured.
    The main exhaust has an upstream throttle valve that is used to set the throughput of gas through the \hpgmc.
    Two additional exhausts exist: One for the safety valve and one upstream of the main chamber.
    All exhausts vent to air outside of the building.

    Pressure tests have been conducted to test the tightness of the self-made flanges and the responding characteristics of the used safety valve.
    It was found, that the safety valve starts to leak starting from about \unit[10.2]{barg} until full opening at \unit[11]{barg}.
    Below \unit[10.1]{barg}, the gas loss was on average \unit[7]{ml/h}.
    The self-made flanges withheld the pressure without incident.

\begin{figure}[h]
        \centering
        \includegraphics[width=\linewidth]{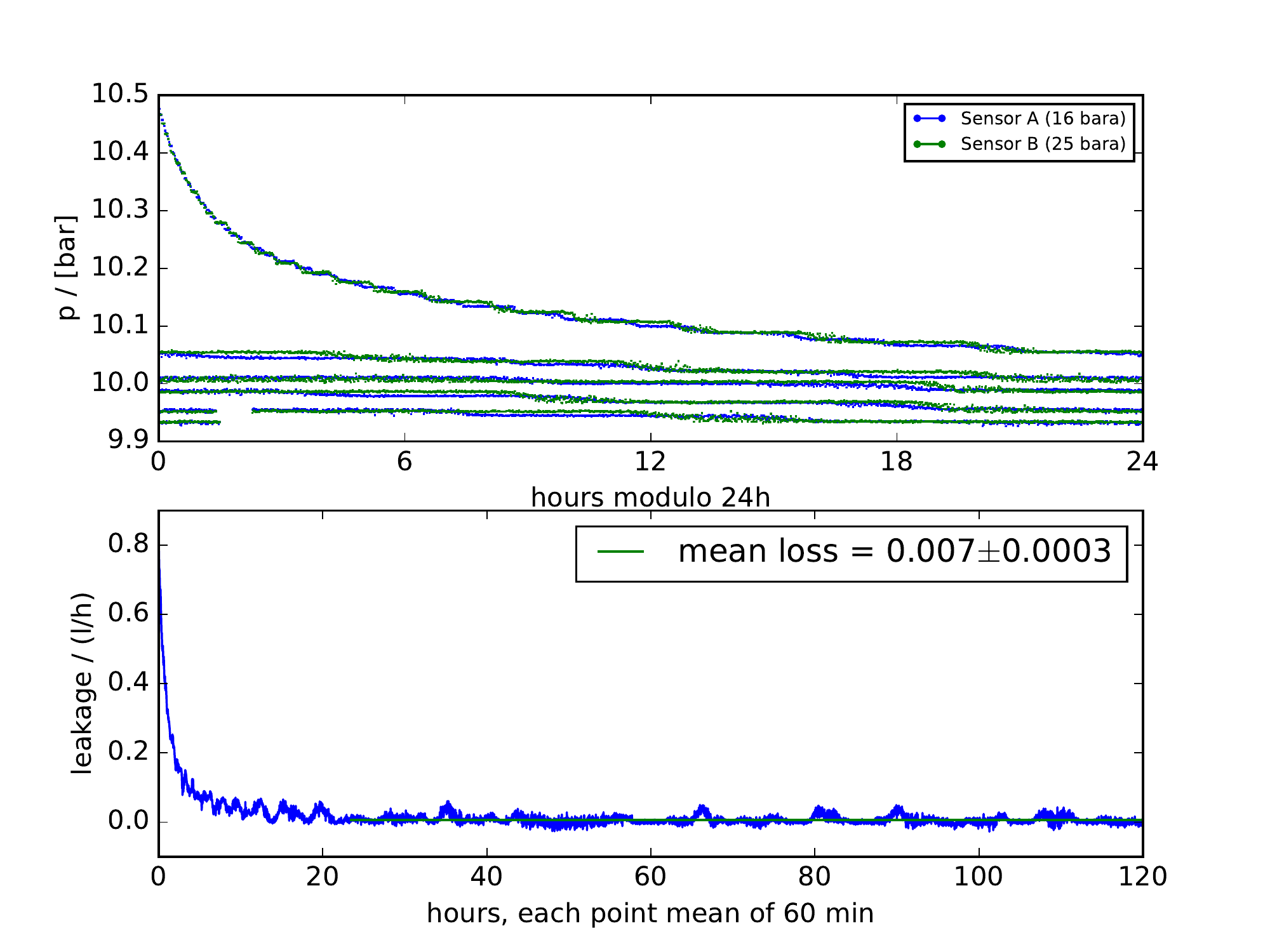}
        \caption[Testing chamber tightness and of the safety valve]{
                At the design operation pressure of \unit[10]{barg}, a low leakage rate of a few millilitres per hour were observed using the double-tee with the custom cathode and a flange with gas inlet.
                For increasing pressures approaching the threshold of the safety valve, a steep increase in leakage rate is observed, until opening of the valve at \unit[11]{barg}.
                The steps are due to the sensors internal digitizers.
        }
        \label{fig:leakage_no_refil}
    \end{figure}

    A tightness measurement over five days can be found in figure~\ref{fig:leakage_no_refil}.
    The steps are produced by the DAC used by the pressure transducers that produces the output current.
    Resolution of one DAC step is \unit[0.6]{\permil} of the full sensor range, i.e. \unit[16]{bara} and \unit[25]{bara} for the two sensors used.
    This value does not affect overall resolution significantly in comparison to the \unit[5]{\permil} inherent non-linearity of the sensor.
\FloatBarrier

    \chapter{Conclusion and Outlook}
    A \hpgmc for measurements of drift velocity has been designed in this work.
    It is operable up to \unit[10]{barg} and \unit[-23]{kV} at atmospheric pressure.
    It is expected to reach the full, possible cathode voltage of \unit[-30]{kV} for higher pressures.

    All subsystems inside the chamber have been built and tested outside the high pressure environment.
    The test results are consistently promising for a complete assembly.
    With the current readout system, only drift velocity can be determined.
    Due to the modular design, exchange of the anode PCB with the attached razor amplification region is not difficult.
    So, various technologies for high pressure applications can be demonstrated and investigated using the \hpgmc.
    To be able to measure a precise gain in addition to the drift velocity, only the cathode needs to be modified to hold a \Fe source.
    Final commissioning will start soon, with the goal to produce and measure first signals in a high pressure drift gas.

    Currently, a \hptpc is being developed by groups in the UK with an application for beam time at CERN for 2018.
    The goal is to measure proton and possibly also pion cross-sections on argon in a low momentum regime.
    A proven full system of this GMC could be used for calibration measurements during the beamtime and contribute towards a more precise measurement.

    \chapter{Acknowledgements}
I would like to thank everyone in the institute, who made my work in the last year possible:
Stefan Roth for offering me the opportunity to build a detector from scratch.
Thomas Radermacher for his expertise on SiPMs and proof reading.
Lukas Koch for the enforced coffee breaks and Python knowledge.
Everyone in the Halle for making the time enjoyable and providing input on gaseous and scintillator based experiments.
The mechanical and electronics workshops for actually building the system and helping with debugging of unruly circuits.
Special thanks go to Jochen Steinmann for his invaluable help in all stages of the project.
    \appendix
    \chapter{Appendix}
\clearpage

\section{Sources Side}%
    \label{apx:sources}
    \begin{figure}[!h]
        \begin{center}
            \subfigure[]{
                \includegraphics[height=.45\linewidth, width=.45\linewidth]{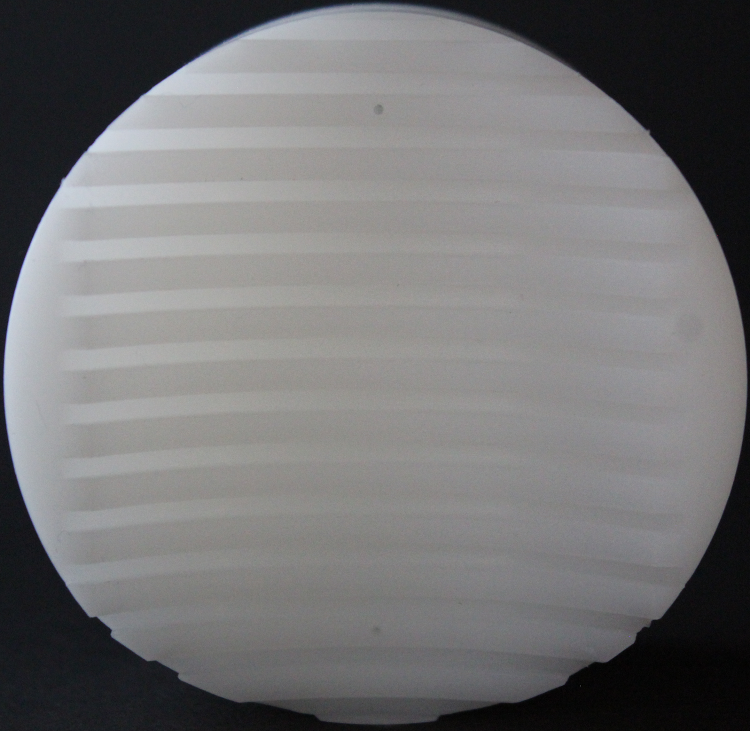}
            }%
            \subfigure[]{
                \includegraphics[height=.45\linewidth, width=.45\linewidth]{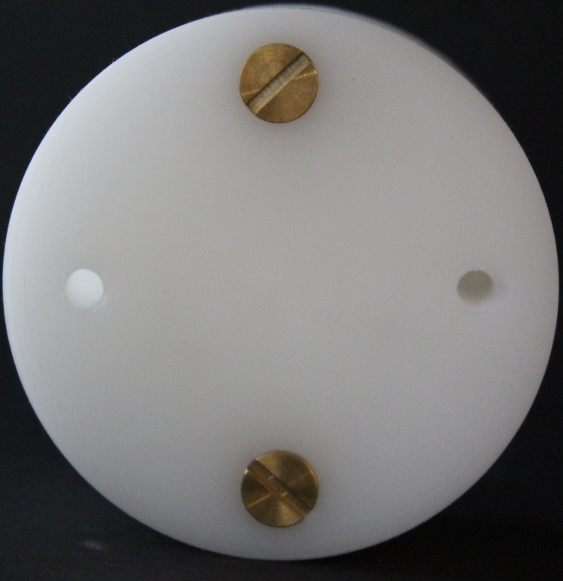}
            }
            \caption[]{
                (a)~Side facing into the drift volume.
                The comb structure is wedged between the field forming rings.
                (b)~Averted side with the screws that are locking the \Sr inside the capsules.
            }
            \label{fig:sr90_views}
            \vspace{1cm}
            \includegraphics[width=\linewidth]{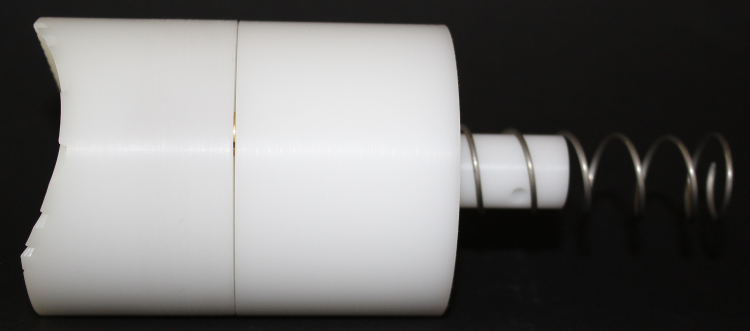}
            \caption[]{
                Fully assembled POM block holding the \Sr sources inside.
                A spring pushes the block into place.
                The field cage is attached from the left.
            }
        \end{center}
    \end{figure}

    \begin{figure}[h]
        \centering
        \includegraphics[width=\linewidth]{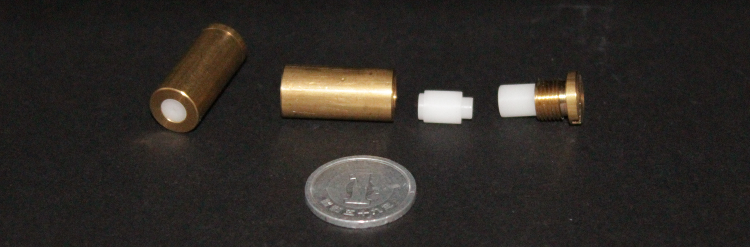}
        \caption[]{
            The two capsules housing the \Sr sources with a \unit[1]{\yen} coin for size comparison.
            On the left capsule, the \unit[1]{mm} exit bore is visible.
            The inside is filled with POM, with a recess for the smaller \Sr sources in the center.
        }
        \label{fig:capsule_explosion}
    \end{figure}

\FloatBarrier
\clearpage
\section{Trigger Side}%
    \vspace{-1em}%
    \label{apx:trigger}
    \begin{figure}[!h]
        \begin{center}
            \subfigure[]{
                \includegraphics[height=.45\linewidth, width=.45\linewidth]{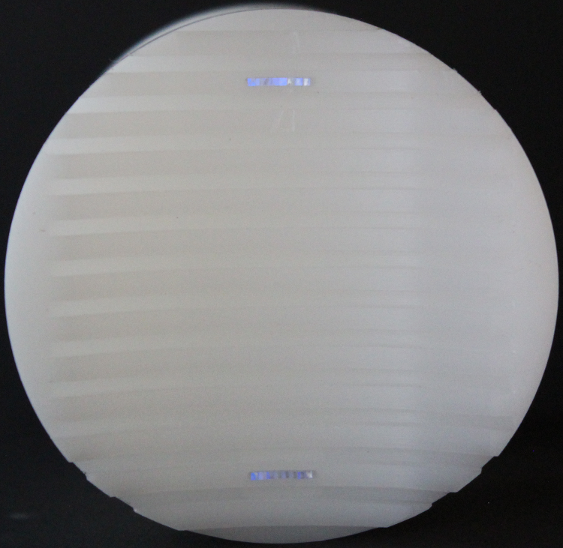}
            }%
            \subfigure[]{
                \includegraphics[height=.45\linewidth, width=.45\linewidth]{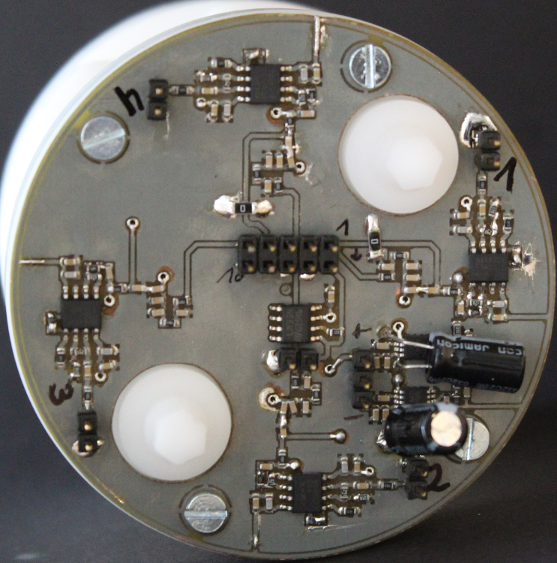}
            }
            \caption[]{
                (a)~Side facing into the drift volume.
                The $8\times1$ scintillating fibre windows are perpendicular to the drift direction.
                (b)~Trigger electronics board view.
            }
            \label{fig:trigger_views}
            \vspace{1cm}
            \includegraphics[width=\linewidth]{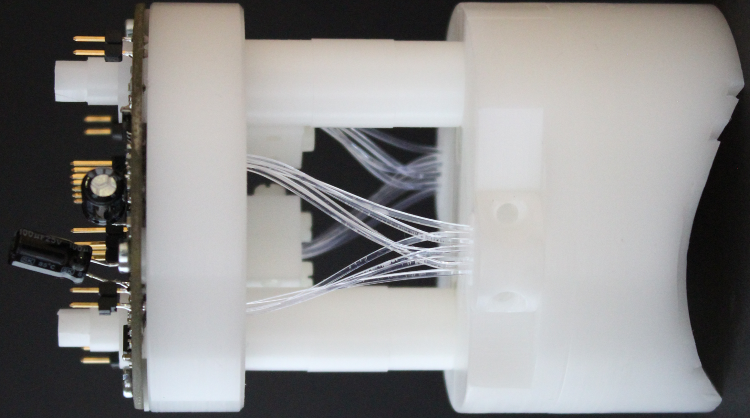}
            \caption[]{
                View along the fully assembled POM block on the trigger side.
                The field cage is attached from the right side.
            }
            \label{fig:trigger_full_side}
        \end{center}
    \end{figure}

\FloatBarrier
\clearpage
\section{Misc}
    \begin{figure}[h]
        \begin{center}
            \subfigure[]{
                \includegraphics[width=.45\linewidth]{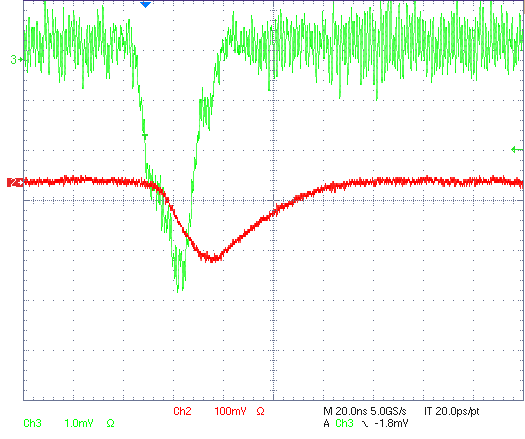}
                \label{fig:ampreg_min_signal}
            }
            \subfigure[]{
                \includegraphics[width=.45\linewidth]{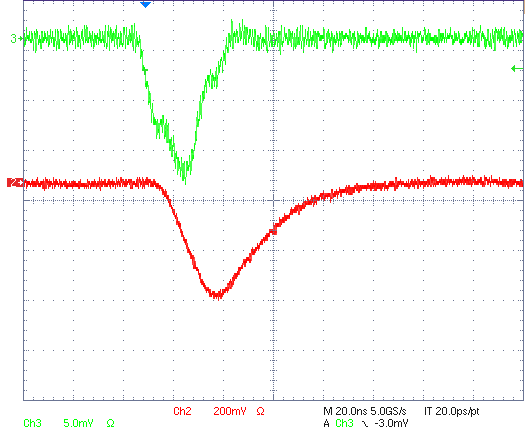}
                \label{fig:ampreg_regular_signal}
            }
        \end{center}
        \caption[]{
            Testing of the amplifier for the amplification region using the test input signal path connected to a pulse generator.
            (a) Few mV signals (green) can be amplified reliably (red).
            (b) During operation expected signal height of \unit[-15]{mV} produces a $\approx\unit[-400]{mV}$ output.
        }
        \label{fig:razoramp_signals_generator}
    \end{figure}
    \begin{figure}[!h]
        \centering
        \includegraphics[width=.7\linewidth]{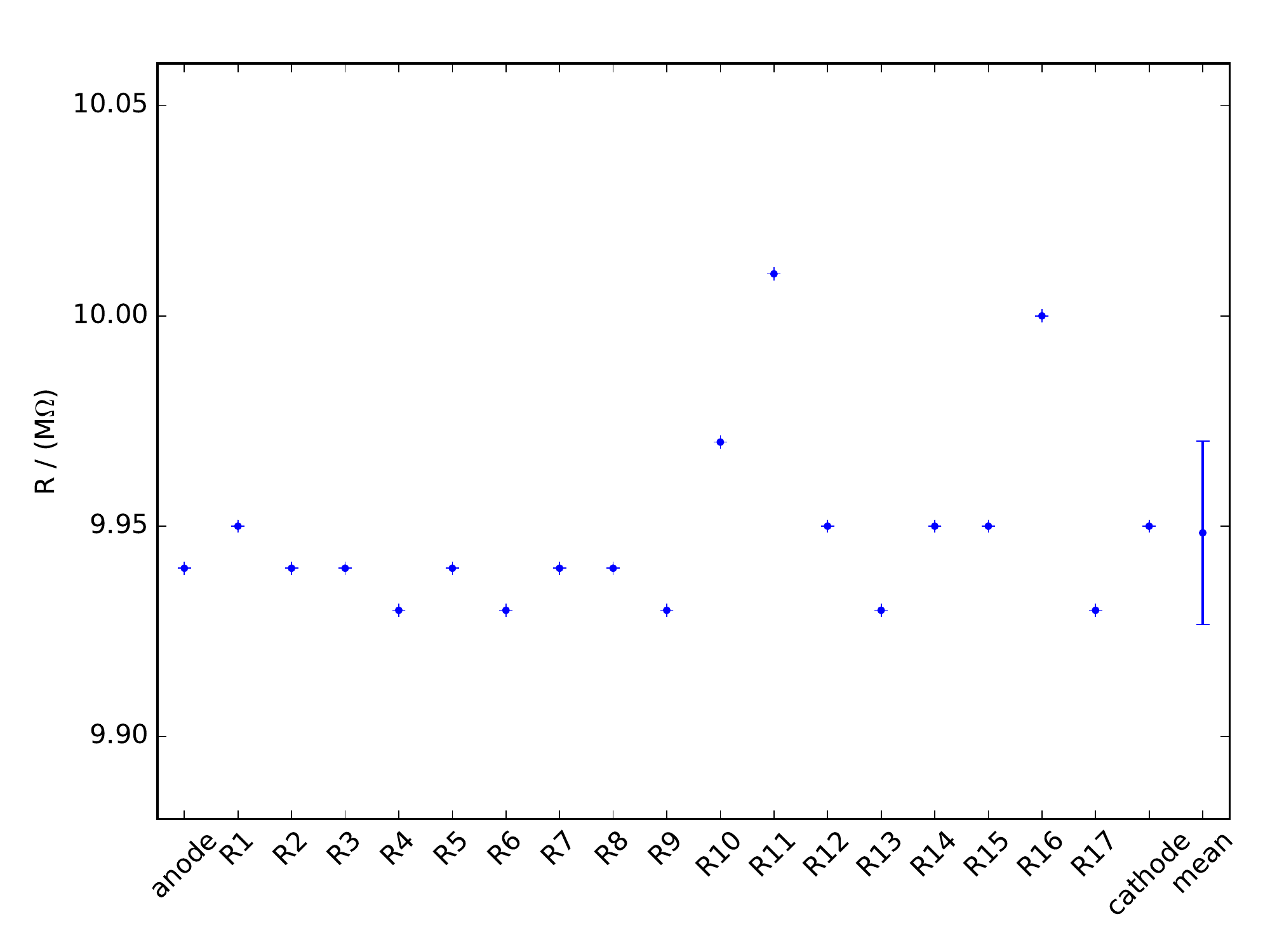}
        \caption[]{
            The values of the resistors in the resistor chain of the field cage.
        }
        \label{fig:single_R_values}
    \end{figure}

    \listoffigures
    \listoftables

    \bibliographystyle{plainnat}
    \bibliography{../citations/mixed}
\end{document}